\documentclass[12pt]{article}
\usepackage{tikz}
\usetikzlibrary{decorations.pathreplacing}
\usepackage{amsmath}
\usepackage{amssymb}
\usepackage[authoryear,round]{natbib}
\usepackage{graphicx}
\usepackage{setspace}


\usepackage{subcaption}
\usepackage{amsfonts}
\usepackage{marginnote}
\usepackage{enumerate}
\usepackage[utf8]{inputenc}
\usepackage[T1]{fontenc}
\usepackage{newtxtext,newtxmath}
\usepackage{url}

\usepackage{xcolor}

\usepackage{booktabs}
\usepackage{adjustbox}
\usepackage{float}
\usepackage{hyperref}
\definecolor{maroon}{RGB}{128,0,0}
\hypersetup{
	colorlinks=true,
	linkcolor=maroon,
	citecolor=maroon,
	urlcolor=maroon,
	linktoc=all
}
\usepackage[nameinlink,noabbrev]{cleveref}
\usepackage{setspace}

\newcommand{\blind}{0}

\def\spacingset#1{\renewcommand{\baselinestretch}%
{#1}\small\normalsize}

\providecommand{\U}[1]{\protect\rule{.1in}{.1in}}
\newtheorem{theorem}{Theorem}[section]

\newtheorem{proposition}[theorem]{Proposition}

\newfont{\bbf}{cmbx12 scaled 1435}

\makeatletter
\renewcommand\@biblabel[1]{}
\makeatother

\title{\bf Aggregating many estimators using estimated weights}
\author{Emmanuel Guerre\thanks{The authors thank participants at the 2026 Tsinghua Symposium for High-Dimensional Econometrics and Machine Learning, 5th Workshop of Econometrics Rio - S\~{a}o Paulo, 2026 Annual Research Symposium in Economics and Finance at QMUL, as well as audiences of various seminars. Thanks to Magdalena Pietrzykowska for excellent research assistance.} \\
	School of Economics and Finance\\
	Queen Mary, University of London\\
	United Kingdom \and Yuting Wang \\ CESAER \& GAEL\\ CNRS, INRAE, Institut Agro Dijon \& Universit\'e Grenoble Alpes \\ France}
\date{\today}

\begin{document}

\spacingset{1}

\if0\blind
{
	\maketitle
} \fi

\if1\blind
{
	\bigskip
	\bigskip
	\bigskip
	\begin{center}
		{\LARGE\bf Aggregating many estimators using estimated weights}
	\end{center}
	\medskip
} \fi

\bigskip
\begin{abstract}
Consider an increasing number of consistent estimators to be averaged when only estimated weights are available. The underlying parameter of interest can be identical across estimators (homogeneity) or not (heterogeneity). The contribution of the paper is threefold. First, it is shown that the interaction of the estimated weights with the estimators can generate  specific bias terms. This constrains the number of estimators that can be aggregated when weight estimation is ignored in inference. Second, the paper proposes estimated adaptive weights, which allow for standard Gaussian inference in a uniform manner and  are asymptotically optimal both under homogeneity and heterogeneity. Third, conditions ensuring the validity of the \citet{cochran1937} Q test of homogeneity with estimated variance are given.
\end{abstract}

\noindent%
\vfill

\newpage
\spacingset{1.3} 
\raggedbottom
\setlength{\bibsep}{0pt}
\setlength{\abovedisplayskip}{4pt plus 1pt minus 1pt}
\setlength{\belowdisplayskip}{4pt plus 1pt minus 1pt}
\setlength{\abovedisplayshortskip}{2pt plus 1pt minus 1pt}
\setlength{\belowdisplayshortskip}{2pt plus 1pt minus 1pt}
\setlength{\jot}{2pt}
	
\section{Introduction}
\label{sec:introduction}

Models with many agents, goods, choices, or interactions may involve very large amounts of data for each observation; see \citet{varian2014}, \citet{einav2014}, and \citet{ng2017} for general discussions of the resulting applications, challenges, and methods. Optimising, or even evaluating, a likelihood or another objective function can then become computationally difficult. We consider settings in which the initial sample can be divided into meaningful subsamples, such as geographic or institutional units, within which the parameter of interest can be estimated feasibly. The considered class of subsample estimators is broad and includes standard econometric estimators. Final estimates are obtained by aggregating the subsample estimates using estimated weights. However, usual optimal weights are generally infeasible in full generality, especially when the estimator variance depends on the parameter in an unknown way which can vary across subsamples. 
We propose a feasible adaptive  weighting scheme, which is asymptotically optimal whether the underlying parameter is constant (homogeneous) or not (heterogeneous) across subsamples. Under suitable conditions linking the number of estimators to subsample size, the procedure allows for optimal inference across this class of aggregators. We further propose a test of parameter homogeneity, a property in line with external validity. Under homogeneity, the consistency rate of our aggregation procedure is given by the square-root of the total sample size, and falls down to the square-root of the number of subsamples under heterogeneity.

These results connect to several strands of the computer science, econometrics, and statistics literatures, particularly distributed estimation, meta-analysis, and panel data.
The distributed estimation literature has developed various methods which can be useful for econometrics in a rich data environment or with complex models. In this approach, several computers can work jointly to compute a gradient and a Hessian over subsamples, which are then summed to produce full sample counterparts used in the next iteration of descent or Newton-Raphson optimisation algorithms \citep{tsitsiklis1986,chu2006,duan2022}. However, the partial gradient and Hessian must be centralized at each iteration, which can be costly if there are many subsamples or if many iterations are needed to achieve convergence. It is also ineffective in the case of subsample-specific parameters, such as in the location choice analysis of \citet{wang2025}.  In this case, it may be faster to run estimation on each computer and to aggregate the results to produce an average estimation. In the homogeneous context, this simple distributed method has been considered by \citet{mcdonald2010}, \citet{zhang2013,zhang2015}, \citet{rosenblatt2016}, and \citet{volgushev2019}, among others. While this branch of the literature typically assumes that the observations are i.i.d., \citet{gu2023} consider aggregation of estimators whose variance can vary across subsamples. They use optimal inverse-variance weights and a (sub)sample-splitting procedure to ensure that the weights are independent of the aggregated estimators.

However, the computer science literature typically assumes parameter homogeneity, i.e., that the parameter of interest remains constant across subsamples. This concern is addressed in the meta-analysis literature.  Many meta-analyses from the economic literature rely on  a Bayesian analysis of the \citet{rubin1981} hierarchical model which assumes that the heterogeneous parameter of interest is drawn from a normal distribution. See the review section of \citet{vivalt2020} who relates heterogeneity with the important issue of external validity.  Some papers such as \citet{meager2019,meager2022} and \citet{bandiera2021} consider a small number of studies. Other works are closer to the rich-data environment of interest here. For instance, \citet{brown2024} and \citet{wolfson2019} consider between 500 and 700 estimates, while \citet{xue2020} analyse more than 12,000 ones. While papers consider the possibility of publication bias (see \citealp{andrews2019} and the references therein), the baseline model assumes that the estimators are unbiased. The estimator variance is regarded as known, justified by the belief that the subsample sizes are large enough ``to guarantee a good estimate of the true within-study variance, with little or no impact on the results'' as noted in \citet{bellio2016}. Among the few exceptions is \citet{liu2015} who consider estimated inverse-variance weights under parameter homogeneity and for a finite number of estimators, a framework inappropriate for distributed estimation but for which variance estimation does not affect limit distributions. By contrast, \citet{cochran1937} and \citet{cochran1953} consider a Gaussian framework where weight estimations affect finite-sample distribution.
This model is also estimated in a frequentist way in the statistical or medical literature. However, \citet{kulinskaya2014} point to a lack of well-grounded knowledge, noting that the ``[ubiquitous related Maximum Likelihood procedure] has been applied to thousands of data sets, despite the lack of theoretical or simulation studies to confirm when it does work, if ever''. While \citet{zeng2015} provide a theoretical study of the MLE based on an increasing number of estimators with known distribution up to the parameter of interest and the heterogeneity parameter variance, they restrict to the case where the parameter heterogeneity variance decreases with the average subsample size.

A third  related area is panel-data econometrics. Estimation of nonlinear panel-data models with individual and time effects has attracted considerable interest, see e.g., \citet{fernandezval2016} and the references therein among many others. For linear panel data models, \citet{lu2023} 
propose a Bootstrap approach for average individual and time parameters which is uniformly valid under both homogeneity and heterogeneity. For a nonlinear regression distribution specification,  \citet{fernandezval2026} consider the standard mean of individual parameters, which are bootstrapped to obtain uniformly valid confidence intervals under both homogeneity and heterogeneity, without weights that would capture estimator variance heterogeneity.   
As noted in \citet{vivalt2020} among others, parameter heterogeneity poses a threat to external validity. Testing for homogeneity was first considered by \citet{cochran1937}, in a Gaussian framework limited to a finite number of univariate estimators.  \citet{swamy1970} applied a similar approach for testing homogeneity of a panel data linear regression slope with a fixed number of individuals and growing time dimension, while \citet{pesaran2008} allow both to grow.  See also \citet{ritz2008} for related results.

Our results extend the literature in several directions. First, many references are tied to specific models and estimation methods, mostly related to linear regression. Our framework, presented in Section \ref{Frmwrk}, allows for general estimation methods which can be biased, thereby ensuring wide applicability in a large data framework. Second, potential heterogeneity in both the parameter and its variance is considered, calling for estimated weights when the variance is unknown. Indeed, averaging estimators without weighting becomes inefficient under homogeneity. The paper introduces \emph{adaptive}
weights in \eqref{Hatwstar}, which are asymptotically optimal under homogeneity and heterogeneity. Theorem \ref{Opt} establishes that these weights allow for standard Gaussian inference which is uniformly valid, producing optimal confidence intervals in the class of weighted estimators without resorting to the Bootstrap. This requires using a parameter-heterogeneity variance which vanishes fast enough under homogeneity. The proposed one \eqref{Checksig} extends   \citet{cochran1954}, \citet{swamy1970} and \citet{hedges1985} to our general setup. Third, Theorem \ref{Homtest} shows the validity of the proposed homogeneity test. Fourth, all our results, starting with Proposition \ref{Expansion}, account for the interaction bias terms created by weight estimation, as well as a correlation bias term caused by the dependence of the population weights on the subsample parameters, see Proposition \ref{Randweightshet}.

The rest of the paper is organized as follows. Section~\ref{sec:framework} introduces the framework. Section~\ref{sec:main-assumptions} groups and discusses our main Assumptions while Section~\ref{sec:main-results} gathers our main results.  Section~\ref{sec:simulation-application} reports the results of simulation experiments for instrument regression and discrete choice specifications together with an application to racial differences in mortgage denial across U.S. metropolitan areas. Section~\ref{sec:concluding-remarks} concludes. Appendix~\ref{app:additional-simulation-results} provides additional simulation results and Appendix~\ref{app:proof-section} gathers the proofs of our main results.

\section{Heterogeneity, estimation and weights \label{Frmwrk}}
\label{sec:framework}

\paragraph{Sample and Subsamples.}

Consider a sample $X_i$, $i=1,\ldots,N$, which is divided into disjoint groups of observations, $\left\{X_i,i \text{ in }\mathcal{G}_g\right\}$, indexed by $g$, $g=1,\ldots,G_N$, where the  number  of groups $G_N$   increases with the total sample size $N$. Each group $\mathcal{G}_g$ gathers $n_g$ observations, with $\sum_{g=1}^{G_N} n_g = N$.
This division into subsamples can be arbitrarily decided by the researcher, as in the standard distributed estimation scenario where the total sample size $N$ is large and using the full sample leads to computational difficulties when estimating a model of interest. Alternatively, the subsample structure can come from the data collection process. Each subsample may correspond to a different data source. For example, the subsample index $g$ may denote a geographic location, such as a city or a country. 
The proposed framework also covers cases where subsamples are not necessarily observed. In meta-analysis, the subsample index $g$ can, for instance, identify a study based on the subsample $\left\{X_i,i \text{ in }\mathcal{G}_g\right\}$. Each study may report subsample estimation results, without disclosing all the underlying data.

The results in this paper are for the case where all the groups grow, i.e., $\inf_{1 \leq g \leq G_N} n_g$ diverges with $N$ in an (approximately) balanced way, with each subsample containing a similar proportion of the observations. That is, for some $C>0$ large enough and when $N$ grows,
\begin{align*}
	\frac{1}{C} \leq \min_{1\leq g \leq G_N} \frac{n_g}{N/G_N} \leq   \max_{1\leq g \leq G_N} \frac{n_g}{N/G_N} \leq C,
\end{align*}
meaning that all the  $ n_g$'s are asymptotically of exact order $N/G_N$.

\paragraph{Parameter heterogeneity.} 
The distribution of the observations can be identical or differ across groups. In the latter case, this may affect the multidimensional parameter of interest. As a consequence, the parameter of interest will be denoted from now on as $\theta_g$, $g=1,\ldots,G_N$, to allow for subsample variations. Our framework considers the following two scenarios:
\begin{equation}
	\begin{array}{lll}
		\theta_g = \theta & \text{for all $g=1,\ldots,G_N$,} &  \text{(Homogeneity)} \\
		\theta_g = \theta +\delta_g & \text{for all $g=1,\ldots,G_N$, }  & \text{(Heterogeneity)} \\
	\end{array}
	\label{Par}
\end{equation}
where the deviations $\delta_g$ are centered, independent, and identically distributed random vectors with variance $\Sigma$.  The homogeneity case is often referred to as the common-effect or fixed-effect model\footnote{With a singular ``effect'', as $\theta_g=\theta$ does not vary across studies.}, while the heterogeneity case corresponds to the random effects model in the meta-analysis literature. The homogeneity case is ubiquitous in distributed estimation papers, especially in those where numerical estimator computation proceeds by summing subsample gradients to compute a sample one. The heterogeneity scenario considered here is common in  meta-analysis and panel-data econometrics.  The parameter $\theta$ in \eqref{Par} then becomes the parameter of interest. Estimation of $\Sigma$ will also be considered.


\paragraph{Estimation of group parameters.}
Each $\theta_g$ can be estimated using an estimator $\widehat{\theta}_g$ based on the subsample
$\left\{X_i,i\in \mathcal{G}_g\right\}$ with $n_g$ observations. 
We assume that $\widehat{\theta}_g$, $g=1,\ldots,G_N$ are independent given the $\theta_g$'s. However, we do not assume that the estimators are identical, as economic parameters can be estimated differently across studies in a meta-analysis. 
The paper focuses on estimators that converge at the usual $\sqrt{n_g}$ parametric rates, assuming that the bias and variance of the $\widehat{\theta}_g$'s satisfy
\begin{align*}
	\mathrm{Var} \left(\left. \widehat{\theta}_g \right| \theta_g \right)
	=
	\frac{\Omega_{n_{g},g} \left( \theta_g\right)}{n_g}
	,
	\qquad 
	\mathbb{E} \left[\left. \widehat{\theta}_g \right| \theta_g\right] = \theta_g + \frac{B_{n_g,g} \left(\theta_g\right)}{n_g^b}, \quad b>\frac{1}{2},
\end{align*}
where $\Omega_{n_{g},g} \left( \cdot \right)$ and $B_{n_g,g} \left(\cdot\right)$ respectively converge to bounded and non-vanishing $\Omega_{g} \left( \cdot \right)$ and $B_{g} \left(\cdot\right)$. 
As a consequence, $\widehat{\theta}_g$ can be represented as
\begin{align}
	\widehat{\theta}_g = \theta + \delta_g + \widehat{\varepsilon}_g + \frac{B_{n_g,g} \left(\theta_g\right)}{n_g^b},
	\text{ with }
	\left\{
	\begin{array}{l}
		\mathbb{E} \left[\delta_g\right] = 0 \text{ and } \mathrm{Var} \left(\delta_g\right) = \Sigma,
		\\
		\mathbb{E} \left[\widehat{\varepsilon}_g \left|\theta_g\right.\right] = 0 \text{ and } \mathrm{Var} \left[\widehat{\varepsilon}_g \left|\theta_g\right.\right] = \frac{\Omega_{n_{g},g} \left( \theta_g\right)}{n_g},
		\\
		\text{ implying in particular } \mathrm{Cov} \left(\delta_g,\widehat{\varepsilon}_g \right)=0,
	\end{array}
	\right.
	\label{Representation}
\end{align}
where $\widehat{\varepsilon}_g = \widehat{\theta}_g - \theta_g - B_{n_g,g} \left(\theta_g\right)/n_g^b$, the pair $\left(\delta_g,\widehat{\varepsilon}_g\right)$ being independent across subsamples but not necessarily identically distributed. 

For brevity, the bias rate exponent $b$ is common to all estimators. The restriction $b>1/2$ ensures that the  estimator Mean Squared Errors (MSE hereafter) are dominated by the variance term when $n_g$ grows, as expected for asymptotically normal and centered $\widehat{\theta}_g$'s. 
In addition to $\widehat{\theta}_g$, we assume that a consistent estimator $\widehat{\Omega}_g = \widehat{\Omega}_g \left(\widehat{\theta}_g\right)$ of $\Omega_{n_{g},g} (\theta_g)$ is available. 

\paragraph{Weighted estimators.}
The weighted estimators studied in this paper are of the form:  
\begin{align}
	\widehat{\theta} \left(\widehat{W}\right)
	=
	\left[
	\sum_{g=1}^{G_N} \widehat{W}_g 
	\right]^{-1}
	\sum_{g=1}^{G_N} \widehat{W}_g \widehat{\theta}_g,
	\quad
	& \widehat{W}_g = \widehat{W}_g\left(\widehat{\theta}_g;X_i, 1  \leq i \leq N \right)= \widehat{W}_g\left(\widehat{\theta}_g\right), \quad g=1,\ldots G_N,
	\label{Weighted}
\end{align}
where $\widehat{W}_g$ is a conformable square matrix estimating some population weight $W_g = W_g \left(\theta_g\right)$. Multiplying all $\widehat{W}_g$'s by an invertible matrix does not change the weighted estimator \eqref{Weighted}. Hence, a sequence of weights $\left\{\widehat{W}_g, 1\leq g \leq G_N \right\}$ can be identified with $\left\{\varPhi_N \cdot \widehat{W}_g, 1\leq g \leq G_N \right\}$, where $\left\{\varPhi_N,N\geq 1\right\}$ is a sequence of deterministic or stochastic invertible matrices. In particular, the constant weights $\widehat{W}_g = \widehat{W}$ correspond to equal weighting.

Note that \eqref{Representation} gives the decomposition:
\begin{align*}
	\widehat{\theta} \left(\widehat{W}\right)
	=
	\theta +
	\left[
	\sum_{g=1}^{G_N} \widehat{W}_g 
	\right]^{-1}
	\sum_{g=1}^{G_N} \widehat{W}_g \left(\delta_g + \widehat{\varepsilon}_g \right)
	+
	\left[
	\sum_{g=1}^{G_N} \widehat{W}_g 
	\right]^{-1}
	\sum_{g=1}^{G_N} \widehat{W}_g 
	\frac{B_{n_g,g} (\theta_g)}{n_g^b}.
\end{align*}
If $b$ is strictly larger than 1 or if the $B_{n_g,g} (\theta_g)$'s cannot be removed via a suitable bias correction, the weighted average of the bias components $B_{n_g,g} (\theta_g)/n_g^b$ can be viewed as a negligible remainder term with respect to the weighted average of the stochastic terms $\delta_g + \widehat{\varepsilon}_g$, at the price of some restrictions to be specified later on. Associated constant optimal weights  are proportional to the inverse of the variance of $\delta_g + \widehat{\varepsilon}_g$, which is equal to
$
	\Sigma
	+
	\mathbb{E} \left[\Omega_{n_{g},g} (\theta_g)\right]/n_g
$,
suggesting the optimal population weights:
$
	W_g^{\sharp}=\left( \Sigma + n_g^{-1} \cdot \mathbb{E} \left[\Omega_{n_{g},g} \left(\theta_g \right)\right]\right)^{-1}
$.
The difficulty is that the expectation $\mathbb{E} \left[\Omega_{n_{g},g} (\theta_g)\right]$ can be difficult to estimate in full generality, so that implementing empirical versions of the weights is not feasible. For instance, if the i.i.d. $X_i$ of subsample $\mathcal{G}_g$ have the distribution $\mathcal{N} (\theta_g, \Omega_g (\theta_g))$ given $\theta_g$ where $\Omega_g (\cdot)$ is unknown and varies with $g$ due to unobserved heterogeneity, it is possible to estimate $\Omega_g (\theta_g)$ using the subsample variance. But recovering the full  $\Omega_g (\cdot)$ as necessary to estimate $\mathbb{E} \left[\Omega_{n_{g},g} (\theta_g)\right]$ may be out of reach due to  subsample dependence. Another example uses a discrete choice model $Y_i = \mathbb{I} \left(U_i \leq X_i \theta_g\right)$ for all $i$ in $\mathcal{G}_g$, where the $U_i$ are i.i.d. with a logistic distribution, independent of the $X_i$'s and $\theta_g$'s. In this model, the asymptotic variance of the MLE of $\theta_g$ is, $\Lambda(\cdot)$ being the Logistic function, 
\begin{align*}
	&\Omega_g (\theta_g)
	= \mathbb{E} ^{-1}\left[ \left. \Lambda \left(X_i \theta_g\right) \left(1- \Lambda \left(X_i \theta_g\right) \right)X_i^{\prime} X_i\right| \theta_g,i \in \mathcal{G}_g\right],
	\text{ estimated by }
	\widehat{\Omega}_g (\widehat{\theta}_g),
	\\
	&
	\quad
	\widehat{\Omega}_g (\tau)
	=
	\left[\frac{1}{n_g}
	\sum_{i \in \mathcal{G}_g}
	\Lambda \left(X_i \tau \right) \left(1-\Lambda \left(X_i \tau \right)\right)X_i^{\prime} X_i
	\right]^{-1}.
\end{align*} 
If the $X_i$'s, $i$ in $\mathcal{G}_g$ were independent of $\theta_g$, the sample mean 
$G_N^{-1}  \sum_{j=1}^{G_N} \widehat{\Omega}_g (\widehat{\theta}_j)$ would be a consistent estimator of $\mathbb{E} \left[\Omega_{n_{g},g} \left(\theta_g \right)\right]$ under reasonable conditions. But it would not be the case if the $X_i$'s, $i$ in $\mathcal{G}_g$ depended on $\theta_g$, as  those $X_i$'s, $i$ in $\mathcal{G}_g$, are independent of $\widehat{\theta}_j$ when $j$ differs from $g$ and would bias the proposed estimator. The situation where $X_i$'s, $i$ in $\mathcal{G}_g$,  depend on $\theta_g$ can arise in some econometric applications, as in \citet{fernandezval2026} who consider a second-stage regression model where $\theta_g = Z_g \beta+ \eta_g$,  with possible dependence between $X_i$'s, $i$ in $\mathcal{G}_g$, and $Z_g$.

However, it is possible to match the asymptotic behavior of $W_g^{\sharp}$. If $\Omega_{n_{g},g} (\cdot)$ has a bounded derivative, the Taylor Inequality gives for all $\Sigma$, 
\begin{align*}
	\mathbb{E} \left[\Omega_{n_{g},g} (\theta_g)\right] =  \Omega_{n_{g},g} (\theta) + O \left( \|\Sigma\|^{\frac{1}{2}}\right), 
	\quad
	\Omega_{n_{g},g} (\theta_g)
	=  \Omega_{n_{g},g} (\theta) + O_{\mathbb{P}} \left( \|\Sigma\|^{\frac{1}{2}}\right), 
\end{align*}
so that $\mathbb{E} \left[\Omega_{n_{g},g} (\theta_g)\right] = \Omega_{n_{g},g} (\theta_g) + O_{\mathbb{P}} \left( \|\Sigma\|^{\frac{1}{2}}\right)$, noting that the available
$\widehat{\Omega}_g \left(\widehat{\theta}_g\right)$ is an  estimator of $\Omega_{n_{g},g} (\theta_g)$.  
It then follows, uniformly in $\Sigma$ and when $n_g$ diverges:\footnote{Using $o_{\mathbb{P}} (a_N b_N) = o_{\mathbb{P}} (a_N^2+ b_N^2)$
since $|a_N b_N| \leq (a_N^2+b_N^2)/2$.} 
\begin{align*}
	W_g^{\sharp} & =n_g \left( n_g \Sigma + \mathbb{E} \left[\Omega_{n_{g},g} \left(\theta_g \right)\right]\right)^{-1}
	=
	n_g \left( n_g \Sigma + \Omega_{n_{g},g} \left(\theta_g \right)
	+
	o_{\mathbb{P}} \left(n_g^{-1/4} \left\|n_g \Sigma\right\|^{1/2}\right)
	\right)^{-1}
	\\
	&
	=
	n_g \left( n_g \Sigma + \Omega_{n_{g},g} \left(\theta_g \right)
	+
	o_{\mathbb{P}} \left(n_g^{-1/2} + \left\|n_g \Sigma\right\|\right)
	\right)^{-1}
	\\
	& 
	=
	(1+ o_{\mathbb{P}} (1))
	n_g \left( n_g \Sigma + \Omega_{n_{g},g} \left(\theta_g \right)
	\right)^{-1},
\end{align*}
assuming the eigenvalues of $\Omega_{n_{g},g} \left(\theta_g \right)$ stay bounded away from $0$.
This suggests the more feasible population weights
\begin{align*}
	W_g^{\star} (\theta_g,\Sigma) = \left(\Sigma + \frac{\Omega_{g} \left(\theta_g\right)}{n_g}\right)^{-1}.
\end{align*}
Note however that these weights depend on $\theta_g$, at the difference of $W_g^{\sharp}$, so that the mean and variance of $\sum_{g=1}^{G_N} W_g^{\star} (\theta_g,\Sigma) \delta_g$ are not explicit, unlike when using the weights $W_g^{\sharp}$.\footnote{This issue is addressed in Proposition \ref{Randweightshet}.  Using $G_N^{-1} \sum_{j=1}^{G_N} \widehat{\Omega}_g \left(\widehat{\theta}_j\right)$ instead of $\widehat{\Omega}_g \left(\widehat{\theta}_g\right)$ is possible but may be subject to the same issue, because the mapping $\Omega_g (\cdot)$ may still depend on $\theta_g$. Using $\widehat{\Omega}_g \left(\widehat{\theta}_g\right)$ is simpler and can help to discriminate the subsamples with a poor estimation of $\theta_g$.
}

 Recall that $\Omega_{g} \left(\theta_g\right)$ in $W^{\star} (\theta_g,\Sigma)$ can  be estimated by $\widehat{\Omega}_g$.  The key point is that $\Sigma$ can be estimated with an error that is negligible with respect to $1/n_g$ under or near homogeneity. Indeed, as in standard analysis of variance, the variance of the estimators $\widehat{\theta}_g$ decomposes into an inter-group variance plus an intra-group one given by the average of the 
$\Omega_{n_{g},g} \left(\theta_g\right)/n_g$'s. See \citet{cochran1954}, \citet{swamy1970} or \citet{hedges1985}, who propose in a first step to correct the sample inter-group variance of the $\widehat{\theta}_g$ with an intra-group term. This suggests the variance estimator $\check{\Sigma}$:
\begin{align}
	\check{\Sigma} = \widehat{\Sigma} - 
	\widehat{\Sigma}_w, 
	\text{ where } 
	\widehat{\Sigma} = \frac{1}{G_N}
	\sum_{g=1}^{G_N}
	\left(
	\widehat{\theta}_g
	-
	\overline{\widehat{\theta}}
	\right)
	\left(
	\widehat{\theta}_g
	-
	\overline{\widehat{\theta}}
	\right)^{\prime}
	\text{ and }
	\widehat{\Sigma}_w=\frac{1}{G_N} \sum_{g=1}^{G_N}  \frac{ \widehat{\Omega}_g \left(\widehat{\theta}_g\right)}{n_g},
	\label{Checksig}
\end{align}
which is studied more specifically in
Proposition \ref{Varconsistency}.
Alternative estimators can involve the estimator variances $\Omega_{n_{g},g} \left(\theta_g\right)/n_g$ to weight the $\widehat{\theta}_g$'s in $\overline{\widehat{\theta}}$ and $\widehat{\Sigma}$, as in \citet{dersimonian1986}.  See also \citet{veroniki2016} for a review of heterogeneity variance estimators and simulation experiments.
The associated candidate weights are therefore:
\begin{align}
	\widehat{W}_g^{\star} \left(\widehat{\theta}_g\right)
	&
	=
	\left(\check{\Sigma} + \frac{\widehat{\Omega}_{g} \left(\widehat{\theta}_g\right)}{n_g}\right)^{-1},
	\label{Hatwstar}
\end{align}
with a suitable redefinition when the inverse does not exist.
These weights are termed \emph{adaptive} since they deliver asymptotically optimal variance performance under both homogeneity and heterogeneity, as shown in Theorem \ref{Opt}. The paper also covers more general sorts of weights, including the inverse-variance weights $n_g \widehat{\Omega}_g^{-1}$, which are asymptotically optimal under homogeneity.

We now introduce variance estimators for $\widehat{\theta} \left(\widehat{W}\right)$, which can be used for the construction of confidence intervals or hypothesis testing. Our estimation of  the asymptotic variance of  $\widehat{\theta} \left(\widehat{W}\right)$ is:
\begin{align}
	& 
	\left(
	\sum_{g=1}^{G_N}
	\widehat{W}_g \left(\widehat{\theta}_g\right)
	\right)^{-1}
	\widehat{\boldsymbol{V}}_{\Sigma} \left(\widehat{W}\right)
	\left(
	\sum_{g=1}^{G_N}
	\widehat{W}_g^{\prime}\left(\widehat{\theta}_g\right)
	\right)^{-1},
	\widehat{\boldsymbol{V}}_{\Sigma} \left(\widehat{W}\right)
	=
	\sum_{g=1}^{G_N} \widehat{W}_g \left(\widehat{\theta}_g\right)\left( \check{\Sigma} + \frac{\widehat{\Omega}_g\left(\widehat{\theta}_g\right)}{n_g}\right)\widehat{W}_g^{\prime}\left(\widehat{\theta}_g\right).
	\label{Varest}
\end{align}

\paragraph{Homogeneity test.} Homogeneity is important for external validity in meta-analysis and for assessing the reliability of aggregated estimators. Indeed, their convergence rate is $1/\sqrt{N}$ under homogeneity, but drops down to $1/\sqrt{G_N}$ under heterogeneity, which can be much smaller. It is also related to external validity as previously mentioned.
We consider a multivariate version of the Cochran Q-test of homogeneity, see  \citet{cochran1937}, \citet{pesaran2008}, \citet{ritz2008} among others for specific models.  Let $\widehat{\theta}_{IV}$ be the estimator of $\theta$ optimally weighted under homogeneity,
\begin{align}
	\widehat{\theta}_{IV} = \overline{\widehat{\theta}} \left(\widehat{\Omega}^{-1}\right)
	=
	\left[
	\sum_{g=1}^{G_N}
	n_g \cdot \widehat{\Omega}_{g}^{-1} \left(\widehat{\theta}_g\right)
	\right]^{-1}
	\sum_{g=1}^{G_N}
	n_g \cdot \widehat{\Omega}_{g}^{-1} \left(\widehat{\theta}_g\right) 
	\widehat{\theta}_g,
	\label{IV}
\end{align}
and for $D= \mathrm{Dim} (\theta)$, define
$
	\widehat{\xi}_g
	=
	n_g
	\left(
	\widehat{\theta}_g - \widehat{\theta}_{IV}
	\right)^{\prime}
	\widehat{\Omega}_{g}^{-1} \left(\widehat{\theta}_g\right)
	\left(
	\widehat{\theta}_g - \widehat{\theta}_{IV}
	\right)
	-
	D
$.
The homogeneity test-statistic is
\begin{align}
	\widehat{Q}
	=
	\sqrt{G_N}
	\frac{\overline{\widehat{\xi}}}{\widehat{s}},
	\quad
	\widehat{s}^2 = \frac{1}{G_N} \sum_{g=1}^{G_N} \widehat{\xi}_g^2 -\left(\overline{\widehat{\xi}}\right)^{2}.
	\label{Homstat}
\end{align}
Let $q_{\alpha}$ be the $1-\alpha$ quantile of the standard normal distribution, i.e., $\mathbb{P} \left(\mathcal{N} (0,1) \geq q_{\alpha} \right) =\alpha$. 
The homogeneity test rejects the null of homogeneity if $\widehat{Q} \geq q_{\alpha}$. Alternative rejection regions and critical values can be considered when the $\widehat{\theta}_g$'s are asymptotically normal. In this case, a candidate approximation for the distribution of $\overline{\widehat{\xi}}$ is a standardized  Chi-Square with $(G_N-1) \times D$ degrees of freedom under the null of homogeneity. The homogeneity test considered in the simulation experiment rejects homogeneity if $\overline{\widehat{\xi}}$  is larger than the $1-\alpha$ quantile of this distribution.

\section{Main assumptions}
\label{sec:main-assumptions}

When applied to a vector, $\left\| \cdot \right\|$ is the Euclidean norm, while for a matrix, $\left\| W \right\| = \max_{u:\left\| u \right\|=1} \left\| Wu \right\|$ is the associated operator norm. 
The set $\mathcal{T}$ stands for the possible values of all the  $\theta_g$. 
In the main text and Appendix~\ref{app:proof-section}, $C$ stands for a positive constant that may vary from line to line.    The expectations in Assumptions E and W are with respect to the sample, and then with respect to the parameter deviations $\delta_g$'s, which are random under heterogeneity but collapse to $0$ under homogeneity.
In these assumptions, $\kappa$ is a positive constant used for some inequalities. The estimators $\widehat{\theta}_g$ and data generating processes, weights and variance estimators $\widehat{\Omega}_g \left(\widehat{\theta}_g\right)$, which satisfy these inequalities define a class $\mathcal{D}_{\kappa}$, over which our results hold uniformly. Note that for any $\kappa>0$, the class $\mathcal{D}_{\kappa}$ allows for heterogeneous alternatives depending on $N$ converging to homogeneity.

\subparagraph{Assumption He.}
 \emph{Under heterogeneity, the i.i.d. deviation vectors $\delta_g$ in \eqref{Par} are centered,  with variance $\Sigma$ and $\mathbb{E}^{\frac{1}{8}} \left[ \left\|\Sigma^{-1/2} \delta_g\right\|^8\right] < \kappa$. The matrix $\Sigma$ stays in a class of matrices satisfying $\|\Sigma\|\leq \kappa$ and $\| \Sigma^{-1} \|  \| \Sigma\|\leq \kappa$, ensuring that the highest and smallest eigenvalues of $\Sigma$ remain of the same order but can go to $0$.  }

\subparagraph{Assumption E.} \emph{
$
	\mathbb{E}^{\frac{1}{8}}
	\left[\left\|\sqrt{n_g} \left(\widehat{\theta}_g - \theta_g \right)\right\|^8\right]
$,
$
\sup_{\tau \in \mathcal{T}}  \left\| B_{n_g,g} (\tau)\right\| $,
$ \sup_{\tau \in \mathcal{T}} \left\| \Omega_{n_g,g}^{-1} (\tau)\right\| $ are smaller or equal to $\kappa$ for all $1 \leq g \leq G_N$ and $N$.
It also holds
$
	\max_{1 \leq g \leq G_N} \sup_{\tau \in \mathcal{T}} \left\|B_{n_{g},g} (\tau)- B_g (\tau) \right\| = o(1)
$ and
$
	\max_{1 \leq g \leq G_N}\sup_{\tau \in \mathcal{T}}\left\|\Omega_{n_{g},g} (\tau)- \Omega_g (\tau)\right\| =o(1)$.
}

\subparagraph{Assumption V.}
\emph{There is a $v>0$ such that, for all $N$ and $1 \leq g \leq G_N$,
$
		\sup_{\tau \in \mathcal{T}}
		n_g^{v}
		\left\|
		\Omega_{n_g,g} \left(\tau \right)
		-
		\Omega_{g} \left(\tau \right)
		\right\|
		$
is smaller or equal to $\kappa$, as
		$\sup_{\tau \in \mathcal{T}} \left\| \Omega_{g}^{(1)} \left(\tau \right)\right\|
		$
		,
		$
		\sup_{\tau \in \mathcal{T}} \left\| \Omega_{n_g,g}^{(1)} \left(\tau \right)\right\|
		$,
		$ \sup_{\tau \in \mathcal{T}}  \left\|\Omega_{g} \left(\tau \right)\right\|$ and $ \sup_{\tau \in \mathcal{T}} \left\|\Omega_{g}^{-1} \left(\tau \right)\right\|$. 
	   For all $N$ and $1 \leq g \leq G_N$,
		$ 
		\mathbb{E}^{\frac{1}{8}}
		\left[
		\left\|
		\sqrt{n_g}
		\left(	\widehat{\Omega}_g \left(\theta_g\right) -\Omega_{g} \left(\theta_g\right) \right)\right\|^8
		\right]
		$
		and
		$ 
		\mathbb{E}^{\frac{1}{8}}
		\left[
		\sup_{\tau \in \mathcal{T}}
		\left\|   \widehat{\Omega}_g^{(1)} (\tau)\right\|^8 \right] 
		$  are lesser or equal to $\kappa$.}

\subparagraph{Assumption W.}
\emph{The population weight functions $W_g (\cdot)$ are deterministic.
The population weights are well-conditioned in the sense that, for all $N$:
  	\begin{align*}
  		\max_{1 \leq g \leq G_N} \sup_{\tau \in \mathcal{T}} \left\| W_g  (\tau )\right\|\leq \kappa,
  		\quad
  		\max_{1 \leq g \leq G_N} \sup_{\tau \in \mathcal{T}} \left\| W_g ^{-1} (\tau )\right\|\leq \kappa,
  		\quad
  		\sup_{\tau \in \mathcal{T}} 
  		\frac{\left\| \left(\sum_{g=1}^{G_N} W_g (\tau )\right)^{-1} \right\|}{ \left(\sum_{g=1}^{G_N} \left\|W_g(\tau)\right\|\right)^{-1} } 
  		\leq 
  		\kappa.
  	\end{align*}
   $W_g \left(\cdot \right)$ is continuously differentiable with, for all $N$: 
  \begin{align*}
  	\max_{1 \leq g \leq G_N} \sup_{\tau \in \mathcal{T}} \left\| \left( \frac{N}{G_N} \|\Sigma\| +1\right) W_g^{(1)} (\tau) \right\| \leq \kappa.
\end{align*}}
  
 \emph{
	The weight matrices $\widehat{W}_g \left(\cdot \right)$ are continuously differentiable, and 
	 it holds, for all $N$,
\begin{align*}
	&
	\max_{1 \leq g \leq G_N} 
	\mathbb{E}^{\frac{1}{8}}
	\left[
		\left\|
		\left( n_g \|\Sigma\| +1 \right) 
		\sqrt{n_g}
		\left(	\widehat{W}_g \left(\theta_g\right) -W_g \left(\theta_g\right) \right)\right\|^8
	\right]
	\leq \kappa,
	\\
	&
	\max_{1 \leq g \leq G_N} 
	\mathbb{E}^{\frac{1}{8}}
	\left[
	\sup_{\tau \in \mathcal{T}}
	\left\|   \left( n_g \|\Sigma\| +1 \right) \widehat{W}_g^{(1)} (\tau)\right\|^8 \right] 
	\leq \kappa .
\end{align*}	
}

\subparagraph{Heterogeneity.} As with the other assumptions, Assumption He requests  finite  eighth moments. This moment condition allows us to avoid imposing dependence assumptions, which would not hold for weights like \eqref{Hatwstar}.

\subparagraph{Subsample estimation.}  Assumption E deals with the estimation procedures, which may differ across subsamples but should all be consistent.  It imposes a bias term negligible with respect to the $1/\sqrt{n_g}$ rate, as standard for estimators with a centered normal limit distribution. As mentioned above, a usual value for the bias exponent $b$ is $b=1$ (see \citet{rilstone1996} for usual econometric estimators), with higher values of $b$ possibly achieved through bias correction \citep{newey2004,yang2015,kim2016,gu2023}. 
The case $b<1$ arises, for instance, in simulation-based estimation using few simulations (\citealp{gourieroux1996}), or in semiparametric estimation involving a nonparametric step (\citealp{ichimura2007}). Moments finiteness may be an issue for econometric estimators such as instrument ones, which have a ratio expression with a denominator that can go to $0$. Assumption E can be weakened by showing that the $\widehat{\theta}_g$'s coincide with a truncation satisfying Assumption E.\footnote{See the proof of Theorem \ref{Opt} for an example of such a technique applied to weights. Suppose for instance that the decomposition $\widehat{\theta}_g = \theta_g + \frac{\widehat{N}_g}{\widehat{D}_g}$ holds, with $\max_{1 \leq g \leq G_N} \mathbb{E} \left[ \left|\sqrt{n_g}\left(\widehat{D}_g -D_g\right)\right|^8\right] = O(1)$ and $\inf_{g} |D_g|>2 \epsilon>0$, $\max_{1 \leq g \leq G_N} \mathbb{E} \left[ \left|\sqrt{n_g}\widehat{N}_g\right|^8\right] = O(1)$. Then  $\widetilde{\theta}_g = \theta_g + \frac{\widehat{N}_g}{\widehat{D}_g}\mathbb{I} \left[|\widehat{D}_g-D_g|< \epsilon\right]$ will satisfy Assumption E, with
\begin{align*}
	\mathbb{P} \left[ \widehat{\theta}_g \neq \widetilde{\theta}_g \text{ for some $1 \leq g \leq G_N$}\right]
	\leq
	\mathbb{P} \left( \max_{1 \leq g \leq G_N} \left| \widehat{D}_g - D_g \right| \geq \epsilon \right)
	\leq \frac{1}{\epsilon^8} \sum_{g=1}^{G_N} \frac{\mathbb{E} \left[ \left|\sqrt{n_g}\left(\widehat{D}_g -D_g\right)\right|^8\right]}{n_g^4}
	=
	O\left( \frac{G_N^5}{N^4}\right),
\end{align*}
which vanishes asymptotically if $G_N = o(N^{4/5})$. Therefore results for the aggregated $\widetilde{\theta}_g$ transfer to corresponding weighted means of $\widehat{\theta}_g$.
} Note also that Assumption E imposes well-conditioned asymptotic variances. 

\subparagraph{Variance estimation.}
 Assumption V deals in particular with weights given by the estimator variances which are related to the optimal ones under homogeneity. Various examples of variance estimators $\widehat{\Omega}_g$ are provided in  \citet{newey1994} for usual estimators; see also  Gouri\'eroux and Monfort (1996) and \citet{ichimura2007} for some of the extensions mentioned above.  Such variance estimators  typically target the asymptotic variance $\Omega_g \left(\theta_g\right)$. Consequently, the finite-sample to asymptotic difference $\Omega_{n_g,g} \left(\theta_g\right)-\Omega_g \left(\theta_g\right)$ can be interpreted as a bias term that can affect our results, see Theorem \ref{Homtest} on our modified \citet{cochran1937} homogeneity Q-test.

\subparagraph{Weights.} Assumption W is designed for bounded versions of population weights. For instance, instead of $\widehat{W}_g^{\star}  = n_g \left(n_g \check{\Sigma} + \widehat{\Omega}_g\right)^{-1}$ of \eqref{Hatwstar}, Assumption W would consider the equivalent weight:
\begin{align*}
	& \widehat{W}^{\ast}_g \left(\widehat{\theta}_g,\check{\Sigma}\right)
	=
	\frac{G_N n_g}{N}
	\left(\frac{N}{G_N} \|\Sigma \| + 1\right)
	\left(n_g \check{\Sigma} + \widehat{\Omega}_g \left(\widehat{\theta}_g\right)\right)^{-1},
	\text{ with population counterpart}
	\\
	&
	W^{\ast}_g \left(\theta_g,\Sigma\right)
	=
	\frac{G_N n_g}{N}
	\left(\frac{N}{G_N} \|\Sigma \| + 1\right)
	\left(n_g \Sigma + \Omega_g (\theta_g) \right)^{-1}.
\end{align*}
That these weights are not feasible in practice as they depend on the unknown $\Sigma$ is not an issue, as these weights are equivalent to the feasible ones issued from \eqref{Hatwstar}. These bounded weights are only used for the purpose of the proof. In other words, Assumption W implicitly supposes that there is an equivalent bounded version of the considered weights that satisfies the listed statements. 

Restricting to well-conditioned weights is reasonable as the subsample estimator variances are also well-conditioned. The condition on the population weight derivative $W_g^{(1)} (\cdot)$ requests that the dependence of $W_g$ on $\theta_g$ decreases when $\Sigma$ increases, and is key to controlling a bias term $\mathbb{E} \left[ W_g (\theta_g) \delta_g \right]$, see \eqref{CLT_esp} in Proposition \ref{Randweightshet}.

 Moment finiteness up to order eight echoes Assumption E. In practice, it could be difficult to verify this condition for weights involving estimated inverse variance. However, for theoretical purposes, the original weights can often be replaced by truncated weights that yield asymptotically an identical estimator and satisfy the required moment conditions. See, for instance,  the proof of Theorem \ref{Opt} for an example of this approach.

\section{Main results}
\label{sec:main-results}

Our main results are organised as follows. Perhaps the most useful ones are Theorem \ref{Opt}, which establishes the adaptive optimality of the weights \eqref{Hatwstar} and the asymptotic normality of the related aggregated estimator, and Theorem \ref{Homtest} on the modified \citet{cochran1937} homogeneity Q-test.  Theorem \ref{Opt} crucially builds on Proposition \ref{Varconsistency}, which derives the consistency rate of $\check{\Sigma}$ under heterogeneity and homogeneity, and is fast enough near homogeneity to ensure that inference with the adaptive weight estimator is valid uniformly over $\mathcal{D}_{\kappa}$.

Theorem \ref{Opt} relies on some preliminary results, which hold in particular for weights where $\Sigma$ is considered as known, such as $\left(\Sigma + \widehat{\Omega}_g \left(\widehat{\theta}_g\right)/n_g\right)^{-1}$. Weight estimation aims to produce an aggregation procedure that achieves a performance close to the one using the corresponding population weights. However, the weight estimation error can be correlated with the group estimators, thereby generating additional bias for the final estimator as described in Proposition \ref{Expansion}. A second issue is specific to heterogeneity: as the asymptotic population weights $W_g (\theta_g)$ may depend on the parameter values $\theta_g$, they are potentially correlated with these parameter values and the $\delta_g$. This may create another bias term as $\widehat{\theta}_g$ depends on $\delta_g$ by \eqref{Representation}, and also affect the computation of the variance of the aggregated estimator, as illustrated in Proposition \ref{Randweightshet}. Proposition \ref{Inference} shows that uniform validity of standard Gaussian inference can be ensured by restricting the number $G_N$ of subsamples. Proposition  \ref{Randweightshet} is also a key tool to  establish Theorem \ref{Opt} .

Following \citet{lu2023} and \citet{fernandezval2026}, our results hold uniformly over the class $\mathcal{D}_{\kappa}$, allowing in particular for alternatives with a heterogeneity variance $\Sigma_N$ depending upon $N$. This covers the three cases of alternatives discussed in \citet{fernandezval2026}, i.e., strong, moderate or no heterogeneity where $(N/G_N) \Sigma_N$ diverges, has a (positive definite) limit, or asymptotically vanishes, conditioning the convergence rate and asymptotic variance of aggregated estimators. Under strong heterogeneity, $\theta$ can be estimated with the sometimes slow $\sqrt{G_N}$ rate, which improves to $\sqrt{N}$ under moderate heterogeneity or homogeneity.

\subsection{The impact of weight estimation}
\label{sec:impact-weight-estimation}

 The next proposition expands the aggregated estimator $\widehat{\theta} \left(\widehat{W} \right)$ around $\widehat{\theta} \left(W\right)$. As
\begin{align*}
 \widehat{\theta} \left(\widehat{W} \right)
 & =\theta+ \widehat{B} + \widehat{E}_{He} + \widehat{E}_{\theta} \text{ where }
  \widehat{B} = \left[\sum_{g=1}^{G_N} \widehat{W}_g \left(\widehat{\theta}_g \right) \right]^{-1} \sum_{g=1}^{G_N} \widehat{W}_g\left(\widehat{\theta}_g \right)  \frac{B_{n_g,g} (\theta_g)}{n_g^b},
 \\
 & 
 \quad
 \widehat{E}_{He}=
 \left[\sum_{g=1}^{G_N} \widehat{W}_g \left(\widehat{\theta}_g \right) \right]^{-1} \sum_{g=1}^{G_N} \widehat{W}_g \left(\widehat{\theta}_g \right) \delta_g,
 \text{ and }
 \widehat{E}_{\theta} = \left[\sum_{g=1}^{G_N} \widehat{W}_g \left(\widehat{\theta}_g \right) \right]^{-1} \sum_{g=1}^{G_N} \widehat{W}_g \left(\widehat{\theta}_g \right) 
 \widehat{\varepsilon}_g,
\end{align*}
this amounts to studying the errors induced by changing $\widehat{W}_g$ to $W_g$ in the bias, heterogeneity, estimation terms $\widehat{B}$, $\widehat{E}_{He}$, and $\widehat{E}_{\theta}$, respectively. Proposition \ref{Expansion} below gives a rate for the approximation of each of these items.

\begin{proposition}
	Suppose that the subsamples are balanced.
	Under Assumptions E, He, and W, $\widehat{\theta} \left(\widehat{W} \right)$ in \eqref{Weighted}  satisfies, uniformly over $\mathcal{D}_{\kappa}$,
	$
		\widehat{\theta} \left(\widehat{W}  \right) 
		= 
		\theta 
		+ 
		\left(
		1
		+
		O_{\mathbb{P}}
		\left[
		\left(\frac{G_N}{N}\right)^{\frac{1}{2}}
		\right]
		\right)
		\left[
		\widetilde{B}
		+
		\widetilde{E}_{He} + \widetilde{E}_{\theta}
		\right]
	$ where
\begin{align*}
		&\widetilde{B}
		=
		\left[\sum_{g=1}^{G_N} W_g \left(\theta_g\right) \right]^{-1}
		\sum_{g=1}^{G_N} W_g\left(\theta_g\right) \frac{B_{n_g,g} \left(\theta_g\right)}{n_g^{b}},
		\quad
		 \widetilde{E}_{He} = \left[\sum_{g=1}^{G_N} W_g \left(\theta_g\right)\right]^{-1} \sum_{g=1}^{G_N} W_g \left(\theta_g\right) \delta_g ,
		\\
		& \widetilde{E}_{\theta} = \left[\sum_{g=1}^{G_N} W_g\left(\theta_g\right) \right]^{-1} \sum_{g=1}^{G_N} W_g \left(\theta_g\right)
		\widehat{\varepsilon}_g,
		\quad r_{\theta} = \frac{G_N}{N}\frac{1}{\frac{N}{G_N} \| \Sigma\| +1 },
		 \quad
		 r_{He}
		 = 
		 \frac{G_N}{N}
		 \frac{\left(\frac{N}{G_N} \left\|\Sigma\right\|\right)^{\frac{1}{2}}}{\frac{N}{G_N} \| \Sigma\| +1 }
		 .
	 \end{align*}
	\label{Expansion}
\end{proposition}

The bias term $\widetilde{B}$, which collects the bias of all estimators $\widehat{\theta}_g$, is common to all aggregation procedures. More specific items come from the interactions of the aggregated estimators and the estimated weights.
The proof of Proposition \ref{Expansion} works by expanding $\widehat{W}_g \left(\widehat{\theta}_g\right)$ around $\widehat{W}_g \left(\theta_g\right)$, which in turn approaches $W_g (\theta_g)$.  The numerators in $\widehat{E}_{He}$ and $\widehat{E}_{\theta}$ differ from those of $\widetilde{E}_{He}$ and $\widetilde{E}_{\theta}$ due to these expansion terms. In the univariate case, they are given by:
\begin{align*}
	& \sum_{g=1}^{G_N} \widehat{W}_g^{(1)} \left(\theta_g\right) \left(\widehat{\theta}_g - \theta_g \right) \delta_g, 
	\qquad
	\sum_{g=1}^{G_N} \left[\widehat{W}_g \left(\theta_g\right) - W_g \left(\theta_g\right)\right]\delta_g,&
	\text{ for }\widetilde{E}_{He},
	\\
	& \sum_{g=1}^{G_N} \widehat{W}_g^{(1)} \left(\theta_g\right) \left(\widehat{\theta}_g - \theta_g \right)^2, 
	\qquad
	 \sum_{g=1}^{G_N} \left[\widehat{W}_g \left(\theta_g\right) - W_g \left(\theta_g\right)\right] \left(\widehat{\theta}_g - \theta_g \right), 
	 & \text{ for }\widetilde{E}_{\theta}.
\end{align*}
Bounds for these items are collected in $r_{He}$ and $r_{\theta}$, respectively.

 
As a consequence of the proposition, estimating weights induces additional bias terms. Under homogeneity, i.e., $\Sigma = 0$, the order of $r_{\theta}$ is $G_N/N$, which is identical to the rate of the estimator bias term $\widetilde{B}$ when $b=1$. The magnitude of these bias terms decreases with heterogeneity, i.e., when $\| \Sigma\|$ increases.

\subsection{Random population weights and heterogeneity}
\label{sec:random-population-weights}

Compared to \citet{gu2023} who consider inverse-variance weighting under homogeneity of the parameter, our framework raises new challenges as the numerator $ \sum_{g=1}^{G_N} W_g (\theta_g) \delta_g$ of the linearized term $\widetilde{E}_{He}$ is not necessarily centered, typically when $W_g (\theta_g)$ and $\delta_g$ are correlated. Hence weighted estimators may face a heterogeneity correlation bias term. As well, the variance of $ \sum_{g=1}^{G_N} W_g (\theta_g) \delta_g$ may be difficult to estimate.
Define:
\begin{align}
	\begin{array}{l}
		\boldsymbol{V} \left(W\right)
		=
		\mathrm{Var}
		\left(
		\sum_{g=1}^{G_N}
		W_g \left(\theta_g\right)
		\left(\delta_g+\widehat{\varepsilon}_g\right)
		\right),
		\\
		\widetilde{\boldsymbol{V}} \left(W\right)
		=
		\sum_{g=1}^{G_N} W_g \left(\theta_g\right) \left(\ \Sigma + n_g^{-1} \cdot \Omega_{n_g,g} \left(\theta_g\right)\right) W_g^{\prime} \left(\theta_g\right),
		\label{Var}
	\end{array}
\end{align}
The next proposition gives the order of the heterogeneity correlation bias and shows that the tractable $\widetilde{\boldsymbol{V}} \left(W\right)$ is a suitable approximation for the variance of $\sum_{g=1}^{G_N}
W_g \left(\theta_g\right)
\left(\delta_g+\widehat{\varepsilon}_g\right)$.
\begin{proposition}
	Suppose that the subsamples are balanced.
	Under Assumptions E, He, and W, it holds, uniformly over $\mathcal{D}_{\kappa}$,
	\begin{align}
		&
		\mathbb{E}
		\left[
		\sum_{g=1}^{G_N}
		W_g \left(\theta_g\right) \left(\delta_g + \widehat{\varepsilon}_g\right)
		\right]
		=
		O
		\left(
		\frac{G_N^2}{N}
		\frac{\frac{N}{G_N}\| \Sigma\|}{\frac{N}{G_N} \| \Sigma\| +1}\right),
		\label{CLT_esp}
		\\
		&
		\widetilde{\boldsymbol{V}}^{-1/2 } \left(W\right) \boldsymbol{V} \left(W\right)\widetilde{\boldsymbol{V}}^{-1/2 } \left(W\right)
		=
		\mathrm{Id}
		+O_{\mathbb{P}} 
		\left[
		\left(\frac{G_N}{N}\right)^{\frac{1}{2}}
		+
		\frac{1}{\sqrt{G_N}}
		\right],
			\label{Vareq}	
	\end{align}
    implying in particular that the exact order of $\boldsymbol{V} \left(W\right)$ and $\widetilde{\boldsymbol{V}} \left(W\right)$ is
    $\frac{G_N^2}{N} \left(\frac{N}{G_N} \|\Sigma\|+1\right)$ if $G_N=o(N)$ diverges. In this case, it also holds:
    \begin{align}
    	\boldsymbol{V}^{-\frac{1}{2}}\left(W\right) 
    	\mathbb{E}
    	\left[
    	\sum_{g=1}^{G_N}
    	W_g \left(\theta_g\right) \left(\delta_g + \widehat{\varepsilon}_g\right)
    	\right]
    	=
    	O_{\mathbb{P}}
    	\left[	
    	\sqrt{\frac{G_N^2}{N}}
    	\frac{\frac{N}{G_N}\| \Sigma\|}{\left(\frac{N}{G_N} \| \Sigma\| +1\right)^{\frac{3}{2}}}\right],
    	\text{ uniformly over $\mathcal{D}_{\kappa}$.}
    	\label{CLT_Esp}
    \end{align}  
\label{Randweightshet}
\end{proposition}

\paragraph{Order of the variance.} The exact order of $\frac{G_N^2}{N} \left(\frac{N}{G_N} \|\Sigma\|+1\right)$ of the variance $\boldsymbol{V} \left(W\right)$ of $\sum_{g=1}^{G_N}
W_g \left(\theta_g\right)
\left(\delta_g+\widehat{\varepsilon}_g\right)$ exhibits different behaviors under heterogeneity and homogeneity. For a given $\Sigma \neq 0$, $\boldsymbol{V} \left(W\right)$ diverges with the exact order $G_N$, while, if $\Sigma =0 $, it  goes to $0$ if $G_N = o (N^{1/2})$. As  $\sum_{g=1}^{G_N} \widehat{W}_g (\widehat{\theta}_g)$ is of exact order $G_N$, the order of $\left[\sum_{g=1}^{G_N} \widehat{W}_g (\widehat{\theta}_g)\right]^{-1} \sum_{g=1}^{G_N}
W_g \left(\theta_g\right)
\left(\delta_g+\widehat{\varepsilon}_g\right)$ is $ O_{\mathbb{P}} \left[ N^{-1/2} \left(\frac{N}{G_N} \| \Sigma\| +1\right)^{1/2}\right]$, which boils down to $O_{\mathbb{P}}  \left(N^{-1/2}\right)$ under homogeneity and $O_{\mathbb{P}}  \left(G_{N}^{-1/2}\right)$ under heterogeneity.

\paragraph{Weight correlation and estimation bias under heterogeneity.} Under heterogeneity, i.e., if $\Sigma \neq 0$,  the order $r_{He}= \left(G_N/N\right)^{3/2}$ of the estimation bias from Proposition \ref{Expansion} is dominated by $G_N^2/N$, the one of the correlation bias stated  in \eqref{CLT_esp}.  The corresponding quantities $r_{He} / \left\| \boldsymbol{V}^{1/2} \left(W\right)\right\|$ and \eqref{CLT_Esp} must asymptotically vanish to ensure that these bias terms do not affect the Central-Limit Theorem of Proposition \ref{Inference}. While $r_{He} / \left\| \boldsymbol{V}^{1/2} \left(W\right)\right\| =o(1)$ when $N$ diverges, \eqref{CLT_Esp} vanishes asymptotically when the stronger condition $G_N= o( N^{2/3})$ holds. This suggests that the weight correlation bias plays the strongest role under heterogeneity.

\paragraph{Consistency.}
When $G_N$ and $N/G_N$ both diverge, $r_{\theta}$, $r_{He}$, $\widetilde{B}$ and $\left[\sum_{g=1}^{G_N} \widehat{W}_g (\widehat{\theta}_g)\right]^{-1} \sum_{g=1}^{G_N}
W_g \left(\theta_g\right)
\left(\delta_g+\widehat{\varepsilon}_g\right)= O_{\mathbb{P}} \left[ N^{-1/2} \left(\frac{N}{G_N} \| \Sigma\| +1\right)^{1/2}\right]$
all vanish uniformly over $\mathcal{D}_{\kappa} $ as discussed above. Therefore consistency of $\widehat{\theta} (\widehat{W})$ holds provided that the correlation bias term in \eqref{CLT_esp} is negligible with respect to $G_N$, which holds if $N/G_N$ diverges. Hence  $\widehat{\theta} (\widehat{W})$ is consistent uniformly over $\mathcal{D}_{\kappa}$, for any $\kappa>0$, provided that $G_N$ and $N/G_N$ both diverge.

\subsection{Variance estimation}
\label{sec:variance-estimation}

Proposition \ref{Varconsistency} studies the heterogeneity variance estimator $\check{\Sigma}$ from \eqref{Checksig}, and the estimator $\widehat{\boldsymbol{V}}_{\Sigma} \left(\widehat{W}\right)$ in \eqref{Varest}
of the variance term $\boldsymbol{V} \left(W\right)$. 

\begin{proposition}
	Suppose that the subsamples are balanced and that Assumptions E, He and V hold. Let  $\underline{v} = \min \left(v,\frac{1}{2}\right)$. 
	Then, uniformly over $\mathcal{D}_{\kappa}$,
	\begin{align}
		\check{\Sigma}
		& =
		\Sigma
		+
		\left\|\Sigma\right\|^{\frac{1}{2}}
		O_{\mathbb{P}} 
		\left[
		\frac{\left\|\Sigma\right\|^{\frac{1}{2}}}{\sqrt{G_N}}
		+
		\frac{1}{\sqrt{N}}
		+
		\left(\frac{G_N}{N}\right)^{b}
		\right]
		+
		O_{\mathbb{P}} \left[\left( \frac{G_N}{N} \right)^{1+\underline{v}}+ 
		\left(\frac{G_N}{N}\right)^{2b}
		+
		\frac{\sqrt{G_N}}{N} 
		\right]
		\nonumber \\
		&=
		\Sigma
		\left[
		1
		+
		O_{\mathbb{P}}
		\left(
		\frac{1}{\sqrt{G_N}}
		\right)
		\right]
		+
		\left\|\Sigma\right\|^{\frac{1}{2}}
		O_{\mathbb{P}} 
		\left[
		\frac{1}{\sqrt{N}}
		+
		\left(\frac{G_N}{N}\right)^{b}
		\right]
		+
		o_{\mathbb{P}}
		\left( \frac{G_N}{N} \right)
		.
		\label{Checksigwrtng}
	\end{align}
	
	Moreover, if $G_N=o(N)$ diverges and under  Assumption W, it also holds uniformly over $\mathcal{D}_{\kappa}$:
	\begin{align*}
		\widetilde{\boldsymbol{V}}^{-1/2} \left(W\right)\widehat{\boldsymbol{V}}_{\Sigma} \left(\widehat{W}\right)\widetilde{\boldsymbol{V}}^{-1/2} \left(W\right) = \mathrm{Id} + o_{\mathbb{P}} (1).
\end{align*}
	\label{Varconsistency}
\end{proposition}
\vspace{-2\baselineskip}

\paragraph{Heterogeneity variance estimation.}
A key feature of $\check{\Sigma}$, which is due to the bias correction $\widehat{\Sigma}_w$ in \eqref{Checksig}, is rate adaptation which can be illustrated with the two polarised cases of heterogeneity and homogeneity, assuming that $b$ is large enough or $G_N$ small enough. Under heterogeneity, \eqref{Checksigwrtng} yields that $\check{\Sigma} = \Sigma + O_{\mathbb{P}} \left(G_N^{-1/2}\right)$, as would be feasible if the $\theta_g$ were directly observed. Under homogeneity so that $\Sigma=0$, this improves to $\check{\Sigma} = o_{\mathbb{P}} (G_N/N)$. 

\paragraph{Estimation of $\widetilde{\boldsymbol{V}} \left(W\right)$.} The rate adaptation property described above is key for obtaining an estimator of $\widetilde{\boldsymbol{V}} \left(W\right)$ as in Proposition \ref{Varconsistency}.  Indeed, $\widehat{\boldsymbol{V}}_{\Sigma} \left(\widehat{W}\right)$ involves the item $\check{\Sigma} + \widehat{\Omega}_g \left(\widehat{\theta}_g\right)/n_g$, which is equal to $(1+o_{\mathbb{P}}(1)) \Omega_{n_g,g} \left(\theta_g\right)/n_g$ under homogeneity and $(1+o_{\mathbb{P}}(1))\Sigma$ under heterogeneity, uniformly in $g$. This is crucial to capture the different rate behaviors of $\widetilde{\boldsymbol{V}} \left(W\right)$. Note also that Proposition \ref{Varconsistency} and \eqref{Vareq} yield that
$
\widehat{\boldsymbol{V}}_{\Sigma} \left(\widehat{W}\right) = (1+o_{\mathbb{P}}(1)) \mathrm{Var} \left(\sum_{g=1}^{G_N} W_g (\theta_g) \left(\delta_g + \widehat{\varepsilon}_g\right)\right)
$,
which is needed for inference.

\subsection{Inference with estimated weights}
\label{sec:inference-estimated-weights}

We now study how estimated weights affect inference. While \eqref{Vareq} and Proposition \ref{Varconsistency} show that consistent variance estimation is feasible under the weak restrictions that both $G_N$ and $N/G_N$ diverge, Propositions \ref{Expansion} and \ref{Randweightshet} show that, together with the baseline estimators bias, the weight estimation and correlation between population weights and heterogeneity deviations matter. The next proposition gives conditions on the growth of the number of groups $G_N$ ensuring that standard Gaussian inference is feasible. 
\begin{proposition}
	Suppose that the subsamples are balanced, that $G_N$ diverges, and that Assumptions E, He, and W hold. Let $\underline{b}=\min(b,1)$. Then the distribution of the standardized estimator,
	\begin{align*}
		\widehat{\boldsymbol{V}}_{\Sigma}^{-\frac{1}{2}} \left(\widehat{W}\right)
		\left[
		\sum_{g=1}^{G_N} \widehat{W}_g  \left(\widehat{\theta}_g\right)
		\right]  \left( \widehat{\theta}\left(\widehat{W}\right)-\theta\right)
	\end{align*}
	converges to a standard multivariate normal distribution, uniformly over 
	$\mathcal{D}_{\kappa}$ if $G_N = o \left( N^{1-\frac{1}{2\underline{b}}}\right)$, and uniformly over $\mathcal{D}_{\kappa} \cap \left\{ \frac{1}{\kappa} \leq \left\|\Sigma\right\| \right\}$ if $G_N = o \left( N^{1-\frac{1}{2\underline{b}+1}}\right)$, for any $\kappa>0$.
	\label{Inference}
\end{proposition}

Under Assumption V, it is easily seen that the weights $\left(\frac{N}{G_N} \| \Sigma \| + 1\right) \frac{G_N n_g}{N}\left(n_g \Sigma + \widehat{\Omega}_g \left(\widehat{\theta}_g\right)\right)^{-1}$ satisfy Assumption W, so that Proposition \ref{Inference} applies. Specialising this result to homogeneity implies that the inverse-variance weighted estimator $\widehat{\theta}_{IV}
	=
	\left[
	\sum_{g=1}^{G_N}
	n_g \widehat{\Omega}_g^{-1} \left(\widehat{\theta}_g\right)
	\right]^{-1}
	\sum_{g=1}^{G_N}
	n_g \widehat{\Omega}_g^{-1} \left(\widehat{\theta}_g\right)
	\widehat{\theta}_g$ 
satisfies the Central-Limit Theorem:
\begin{align*}
	\left[
	\sum_{g=1}^{G_N}
	n_g \widehat{\Omega}_g^{-1} \left(\widehat{\theta}_g\right)
	\right]^{\frac{1}{2}}
	\left(
	\widehat{\theta}_{IV}
	-
	\theta
	\right)
	\stackrel{d}{\rightarrow}
	\mathcal{N}
	\left(0,\mathrm{Id}\right),
\end{align*} 
if $\Sigma=0$ and $G_N = o \left( N^{1-\frac{1}{2\underline{b}}}\right)$,
a result which parallels \citet[Theorems 6 and 7]{gu2023} but holds without sample-splitting, at the price of a stronger growth condition for $G_N$ compared to \citet{gu2023}. Indeed, sample-splitting removes the weight estimation bias and these authors only assume that $G_N = o \left(N^{2/3}\right)$. Note however that the considered weights become infeasible under heterogeneity when $\Sigma$ is unknown, and that sample-splitting may be difficult to implement in meta-analysis.

Restricting to the class of DGP $\mathcal{D}_{\kappa} \cap \left\{ \frac{1}{\kappa} \leq \left\|\Sigma\right\| \right\}$ is mostly of theoretical interest, but useful to illustrate that heterogeneity allows for a better growth condition for $G_N$. The intuition is that $\widetilde{\boldsymbol{V}} (W)$ diverges faster under heterogeneity so that the bias-variance condition which drives the rate of admissible $G_N$ can be slightly relaxed.

\subsection{Optimal inference}
\label{sec:optimal-inference}

The next theorem complements Proposition \ref{Inference} by considering aggregation with the candidate optimal weights \eqref{Hatwstar}. It works using a second-order Taylor Inequality for this weighted estimator around the one using weights based on $\Sigma$ instead of $\check{\Sigma}$.  Applying Proposition \ref{Inference} then gives the Central-Limit Theorem stated below, while the optimality statement of the aggregated  estimator using the adaptive weights follows from standard arguments.

\begin{theorem}
	Suppose that the subsamples are balanced, and that Assumptions E, He, and V hold.  Assume that $G_N$ diverges with $G_N = o \left(N^{1-\frac{1}{2\underline{b}}}\right)$, $\underline{b} = \min (b,1)$.
	
	Then, both
	\begin{align*}
		\left[\sum_{g=1}^{G_N}W_g^{\star} \left(\theta_g,\Sigma\right)\right]^{\frac{1}{2}}
		\left(
		\widehat{\theta} \left(\widehat{W}_g^{\star}\right) - \theta
		\right)
		\text{ and }
			\left[\sum_{g=1}^{G_N} \widehat{W}_g^{\star} \left(\widehat{\theta}_g\right)\right]^{\frac{1}{2}}
		\left(
		\widehat{\theta} \left(\widehat{W}_g^{\star}\right) - \theta
			\right),
	\end{align*}  
	converge in distribution to a standard multivariate normal uniformly over $\mathcal{D}_{\kappa}$.
	
	Moreover, $	\widehat{\theta} \left(\widehat{W}_g^{\star}\right)$ is asymptotically optimal among the class of weighted estimators $\widehat{\theta} (W)$ with asymptotic variance $\left(\sum_{g=1}^{G_N}W_g \right)^{-1} \sum_{g=1}^{G_N} W_g \left(\Sigma + \frac{\Omega_{n_g,g} (\theta_g)}{n_g}\right) W_g^{\prime} \left(\sum_{g=1}^{G_N}W_g^{\prime} \right)^{-1}$, which is asymptotically larger than the one of $\widehat{\theta} \left(\widehat{W}_g^{\star}\right)$ uniformly in $\Sigma$,
	\begin{align*}
		\left(\sum_{g=1}^{G_N}W_g \right)^{-1} \sum_{g=1}^{G_N} W_g \left(\Sigma + \frac{\Omega_{n_g,g} (\theta_g)}{n_g}\right) W_g^{\prime} \left(\sum_{g=1}^{G_N}W_g^{\prime} \right)^{-1}
		\succcurlyeq
		(1+o_{\mathbb{P}} (1))\left[\sum_{g=1}^{G_N} \left(\Sigma + \frac{\Omega_{n_g,g} (\theta_g)}{n_g}\right)^{-1}\right]^{-1},
	\end{align*}
	where $\succcurlyeq$ is the usual ordering of symmetric matrices.
	\label{Opt}
\end{theorem}

Theorem \ref{Opt} is valid uniformly over the class $\mathcal{D}_{\kappa}$ of DGP, including homogeneity and heterogeneity, as well as heterogeneous alternatives with $\Sigma = \Sigma_N$ such as in \citet{fernandezval2026} who consider strong (diverging $\frac{N}{G_N} \| \Sigma\|$), moderate (bounded $\frac{N}{G_N} \| \Sigma\|$) or no heterogeneity (asymptotically vanishing $\frac{N}{G_N} \| \Sigma\|$). Under strong heterogeneity, Theorem \ref{Opt} implies that $\widehat{\theta} \left(\widehat{W}_g^{\star}\right)$ converges to $\theta$ with the rate $\sqrt{G_N}$, which improves to $\sqrt{N}$ under moderate or no heterogeneity. Moderate heterogeneity was already considered in \citet{zeng2015} for known $\Omega_{n_g,g} (\cdot)$. Their Theorem 2 states, for a one-dimensional parameter with $\Sigma_N = \frac{G_N}{N} \sigma^2>0$, a Central-Limit Theorem identical to the one above, showing in addition that $\widehat{\theta} \left(\widehat{W}_g^{\star}\right)$ can improve on the Maximum Likelihood estimator. 

Theorem \ref{Opt} therefore extends \citet{zeng2015} and \citet{gu2023} by covering a wider range of heterogeneity scenarios. Besides considering a more general class of estimators, it also extends \citet{lu2023} and \citet{fernandezval2026} due to its optimality focus. These authors indeed use for inference purposes the equal-weights estimator
\begin{align}
	\widehat{\theta}_{EW}
	=
	\frac{1}{G_N} \sum_{g=1}^{G_N} \widehat{\theta}_g,
	\label{EW}
\end{align}
which is asymptotically normal by Proposition \ref{Inference} with the asymptotic variance $\frac{1}{G_N^2}\sum_{g=1}^{G_N} \left(\Sigma + \frac{\Omega_{n_g,g} (\theta_g)}{n_g}\right)$. The latter variance is improved by the one of $\widehat{\theta} \left(\widehat{W}_g^{\star}\right)$ by the asymptotic optimality statement of Theorem \ref{Opt}, meaning in particular that confidence intervals relying on $\widehat{\theta} \left(\widehat{W}_g^{\star}\right)$ are asymptotically shorter than counterparts based on $\widehat{\theta}_{EW}$.  \citet{lu2023} and \citet{fernandezval2026} also rely on bootstrap procedures to achieve rate adaptation. This is computationally more involved than a straightforward application of the Central-Limit Theorem proposed here. 

The growth rate condition $G_N = o \left(N^{1-1/(2\underline{b})}\right)$, which gives at best $G_N = o \left(N^{1/2}\right)$, is in line with the ones used in \citet{zeng2015},  \citet{lu2023} and \citet{fernandezval2026} who assume  $G_N = o \left(N^{1/2}\right)$, corresponding to the case of subsample estimators with a bias of order $1/n_g$, so that $\underline{b}=1$.  

In applications, it is often more convenient to relate the number $G_N$ of groups to the average subsample size $n=N/G_N$ instead of the total sample size. The condition $G_N = o \left(N^{1-1/(2\underline{b})}\right)$ then becomes $G_N = o \left(n^{2 \underline{b}-1}\right)$, which gives at best $G_N = o(n)$ when $b=1$.

\subsection{Testing for homogeneity \label{Qtest}}
\label{sec:testing-homogeneity}

Rate-adaptive inference procedures derived from Theorem \ref{Opt} do not need to involve a homogeneity testing step as they automatically adjust to the data DGP heterogeneity. However, as mentioned earlier, testing for homogeneity is interesting independently of parameter  inference, as heterogeneity is a threat to external validity as noted in \citet{vivalt2020}. The next result shows that the homogeneity test based on the statistic \eqref{Homstat} has the correct asymptotic level and is asymptotically consistent.

\begin{theorem}
	Suppose that Assumptions He, E, and V hold. In addition, assume that the subsamples are balanced, that, for $\underline{vb}=\min \left(2b-1,v,\frac{1}{2}\right)$, $G_N$ diverges with $G_N= o \left(N^{1-\frac{1}{2\underline{vb}+1}}\right)$, and that\\ $\min_{1 \leq g \leq G_N} \mathrm{Var} \left(n_g \widehat{\varepsilon}_g^{\prime }\Omega^{-1}_{n_g,g} (\theta_g) \widehat{\varepsilon}_g\right) \geq C>0$ for $N$ large enough.
	
	Then  $\widehat{Q}$ converges in distribution to a standard normal under homogeneity, while it diverges for any given $\Sigma \neq 0$. As a consequence,
	the test which rejects the null of homogeneity if $\widehat{Q} \geq q_{\alpha}$ is an asymptotic $\alpha$-level and consistent test.
\label{Homtest}
\end{theorem}

 Notably, Theorem \ref{Homtest} makes no assumption on the asymptotic distribution of the estimators $\widehat{\theta}_g$, which are not associated with any specific model. 
In this sense, Theorem \ref{Homtest}  extends the seminal paper of \citet{cochran1937} who considered a mean parameter under Gaussianity, as well as \citet{swamy1970} or \citet{pesaran2008} who considered panel data linear regression models. \citet{ritz2008} considered the case of general estimators but assumed that their variances are known. The condition on the number of groups $G_N$ used in Theorem \ref{Homtest} differs from the one used above for estimation because the variance estimator $\widehat{\Omega}_{g} \left(\widehat{\theta}_g\right)$ plays a crucial role in the construction of the  statistic $\widehat{Q}$, especially under the null of homogeneity considered here. 


\section{Simulation illustrations and application}
\label{sec:simulation-application}

\subsection{Simulation exercises}
\label{sec:simulation-exercises}
This section considers two specifications: a discrete-choice model in which the correlation between the covariates varies across groups and, reported in Appendix~\ref{app:iv-regression-specification}, an instrumental variable regression specification in which the second-stage variance depends on the considered subsamples. The subsample size is constant across groups and set to $n=1,000$. The number of groups, $G_N$, varies from $1$ to $600$, so that the total sample size $N$ ranges from $1,000$ to $600,000$. Each simulation experiment uses $100,000$ replications.

The simulation illustration first evaluates three weighting schemes, which are compared with a full-sample estimator. The three schemes are equal ($W_g=1/G_N$ as in \eqref{EW}), inverse-variance ($\widehat{W}_g \left(\widehat{\theta}_g\right) = n_g \cdot \widehat{\Omega}_g^{-1} \left( \widehat{\theta}_g \right)$), and adaptive  \eqref{Hatwstar} weights. 
The adaptive weights are computed using $\max \left(0,\check{\Sigma} \right)$ instead of $\check{\Sigma}$, which is sometimes negative.
The confidence intervals computed in the simulation experiments are based on the variance estimator $\widehat{\boldsymbol{V}}_{\Sigma} \left(\widehat{W} \right)$ in \eqref{Varest}.  Chi-squared critical values are used for the homogeneity test.

The simulation illustration investigates the square-root Mean Squared Error (RMSE), the bias, and the confidence interval coverage of each estimation procedure. For the sake of brevity, the simulation experiments focus on the polar cases of  homogeneity and heterogeneity. The level and power of the homogeneity test are also considered.

In each specification, the focus is on a univariate parameter $\theta_{1g}\sim \mathcal{N} \left(1,\sigma_{\delta}^2\right)$, with $\Sigma=0$ under homogeneity. The parameter of interest is $\mathbb{E} \left[\theta_{1g}\right]=1$. The RMSE is computed using the following average of the simulation estimates $\widehat{\theta}_{1}^s$, which accounts for the consistency rates $\sqrt{N}$ and
$\sqrt{G_N}$ under homogeneity and heterogeneity respectively: 
\begin{align}
	\left[
	\frac{1}{100,000}
	\sum_{s=1}^{100,000}
	N
	\left(
	\widehat{\theta}_1^{s}- 1
	\right)^2
	\right]^{\frac{1}{2}}
	\text{ and }
	\left[
	\frac{1}{100,000}
	\sum_{s=1}^{100,000}
	G_N
	\left(
	\widehat{\theta}_1^{s}- 1
	\right)^2
	\right]^{\frac{1}{2}}
	.
	\label{RMSE}
\end{align}
The reported bias from the simulation experiments is computed as a proportion of the RMSE, that is, under homogeneity and heterogeneity respectively:
\begin{align}
	\frac{\frac{\sqrt{N}}{100,000}
		\sum_{s=1}^{100,000}
		\left(\widehat{\theta}_1^s-1\right)
	}{RMSE}
	\text { and }
		\frac{\frac{\sqrt{G_N}}{100,000}
			\sum_{s=1}^{100,000}
			\left(\widehat{\theta}_1^s-1\right)
		}{RMSE}
		.
	\label{Biassim}
\end{align}

For the discrete choice model considered here, it holds for all $i$ in group $\mathcal{G}_g$:
\begin{align}
\left\{
\begin{array}{ll}
	y_i = \mathbb{I} \left( y_i^{\star} \geq 0\right), 
	\quad
	y_i^{\star} = \theta_{1g} x_{1i} + \theta_2 x_{2i} - U_i,
	& U_i \stackrel{i.i.d.}{\sim} \mathrm{ Logistic},
	\\
	\left[
	\begin{array}{c}
		x_{1i}
		\\
		x_{2i}
	\end{array}
	\right]
	\sim
	\mathcal{N}
	\left(
	\left[
	\begin{array}{c}
		0
		\\
		0
	\end{array}
	\right]
	,
	\left[
	\begin{array}{cc}
		1 & \rho_g
		\\
		\rho_g & 1
	\end{array}
	\right]
	\right)
	,
	&
	\\
	\theta_{1g} \sim \mathcal{N} \left(1,\sigma_{\delta}^2 \right),
	\quad
	\rho_g
	\sim
	\left\{
	\begin{array}{cc}
		\mathcal{U}_{[.9,.95]}
		&
		\text{ with prob. $.55$},
		\\
		\mathcal{U}_{[0,.1]}
		&
		\text{ with prob. $.45$},
	\end{array}
	\right.
\end{array}
\right.,
&
\label{DCM}
\end{align}
where $\theta_g$ and $\rho_g$ are independent. $x_{i}=\left[x_{1i},x_{2i}\right]$ and $U_i$ are also independent given  $\theta_g$ and $\rho_g$. The standard deviation $\sigma_{\delta}$ is set to $0.3$ under heterogeneity. The group correlation parameter $\rho_g$ lies in $[0.9,0.95]$ with probability $0.55$, in which case the variance of $x_i$ is close to non-invertible, making $\theta_g = \left[\theta_{1g},\theta_2\right]^{\prime}$  difficult to estimate. By contrast, $\theta_g$ is better identified when $0 \leq \rho_g \leq 0.1$, which occurs with probability $0.45$.

 $\theta_g$ is estimated using a couple of Newton-Raphson iterations  based on the first-order condition of the maximum likelihood estimator (MLE) initialized at $\theta_g$, so that the resulting $\widehat{\theta}_g$ is equivalent to the MLE up to an $o_{\mathbb{P}} \left(n^{-1/2}\right)$ error term. An estimator $\widehat{\theta}$ is similarly computed over the whole sample using the initial value $1$, so that $\widehat{\theta}$ is equivalent to the MLE up to an order $o_{\mathbb{P}} \left((G_N n)^{-1/2}\right)$. The asymptotic variance of $\widehat{\theta}_{1g}$ is obtained using the Hessian formula.

An important difference between the discrete choice model and the IV specification \eqref{IVmodel} is that this variance asymptotically depends on $\theta_g$. Hence, under heterogeneity, the inverse-variance weighted estimator will not estimate $\mathbb{E} \left[\widehat{\theta}_{1g}\right]$ but $\mathbb{E} \left[	\widehat{\Omega}_{g}
\left(\widehat{\theta}_g\right)^{-1} \widehat{\theta}_{1g}\right]/\mathbb{E} \left[	\widehat{\Omega}_{g}
\left(\widehat{\theta}_g\right)^{-1} \right]$. Averaging over simulations, the inverse-variance weighted estimator that aggregates $\widehat{\theta}_{1g}$, $g=1,\ldots,NG_{N}$, yields an approximate value of $0.9385$ for    $\mathbb{E} \left[	\widehat{\Omega}_{g}
\left(\widehat{\theta}_g\right)^{-1} \widehat{\theta}_{1g}\right]/\mathbb{E} \left[	\widehat{\Omega}_{g}
\left(\widehat{\theta}_g\right)^{-1} \right]$. The simulation results below therefore use $0.9385$ to center the inverse-variance estimator under heterogeneity. Appendix~\ref{app:dcm-additional-results} reports simulation results when centering this estimator at $1$. As expected, the bias dominates, leading to a deterioration in both the RMSE and the confidence interval coverage as $G_N$ increases under heterogeneity with the centering of Appendix~\ref{app:dcm-additional-results}.

\begin{figure}[!htbp]
	\centering
	\begin{subfigure}[t]{0.4\textwidth}
		\includegraphics[width=\textwidth]{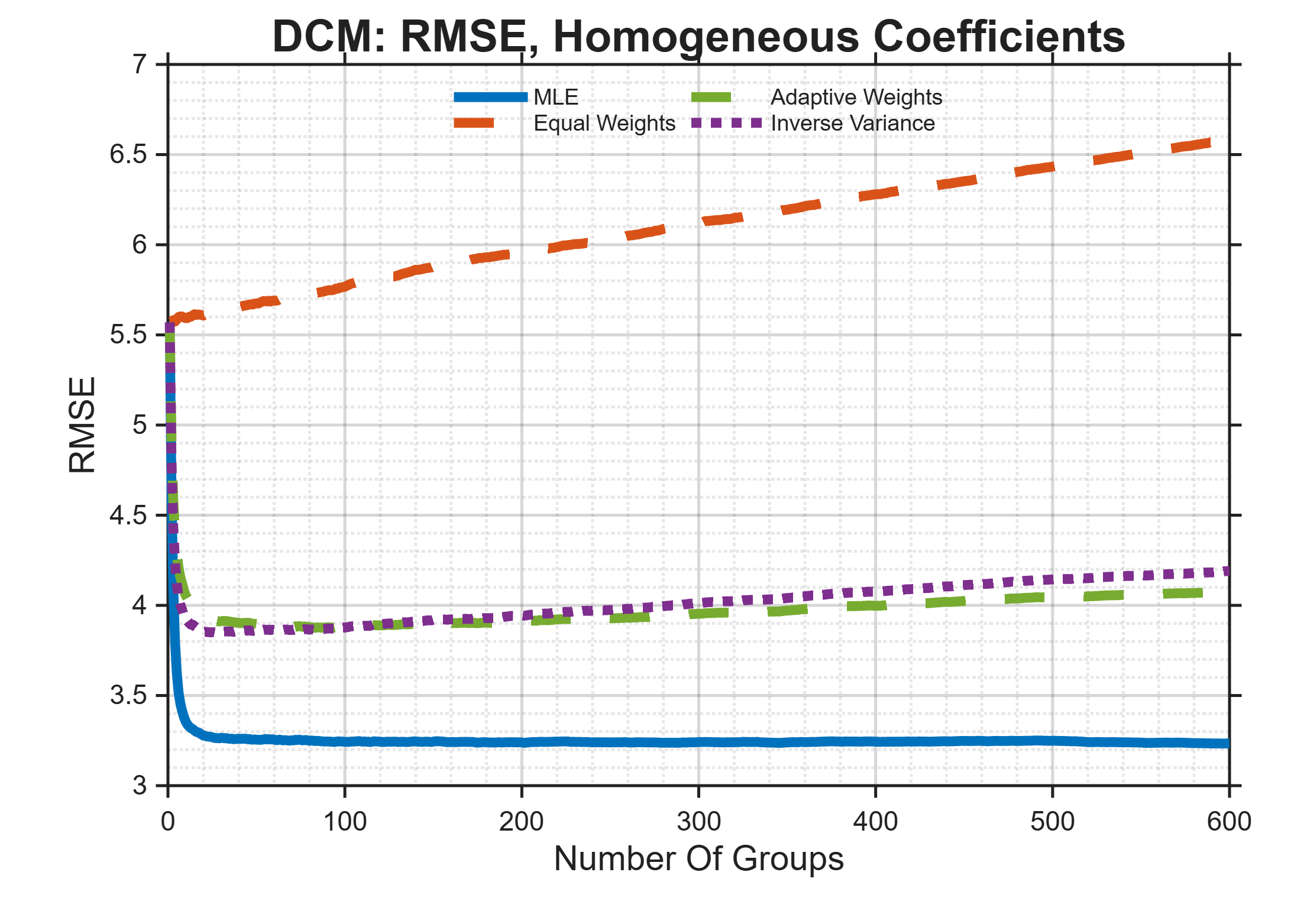}
		\caption{RMSE under homogeneity, $\theta_1=1$}
		\label{fig:RMSEhomoDCM}
	\end{subfigure}%
	~
	\begin{subfigure}[t]{0.4\textwidth}
		\includegraphics[width=\textwidth]{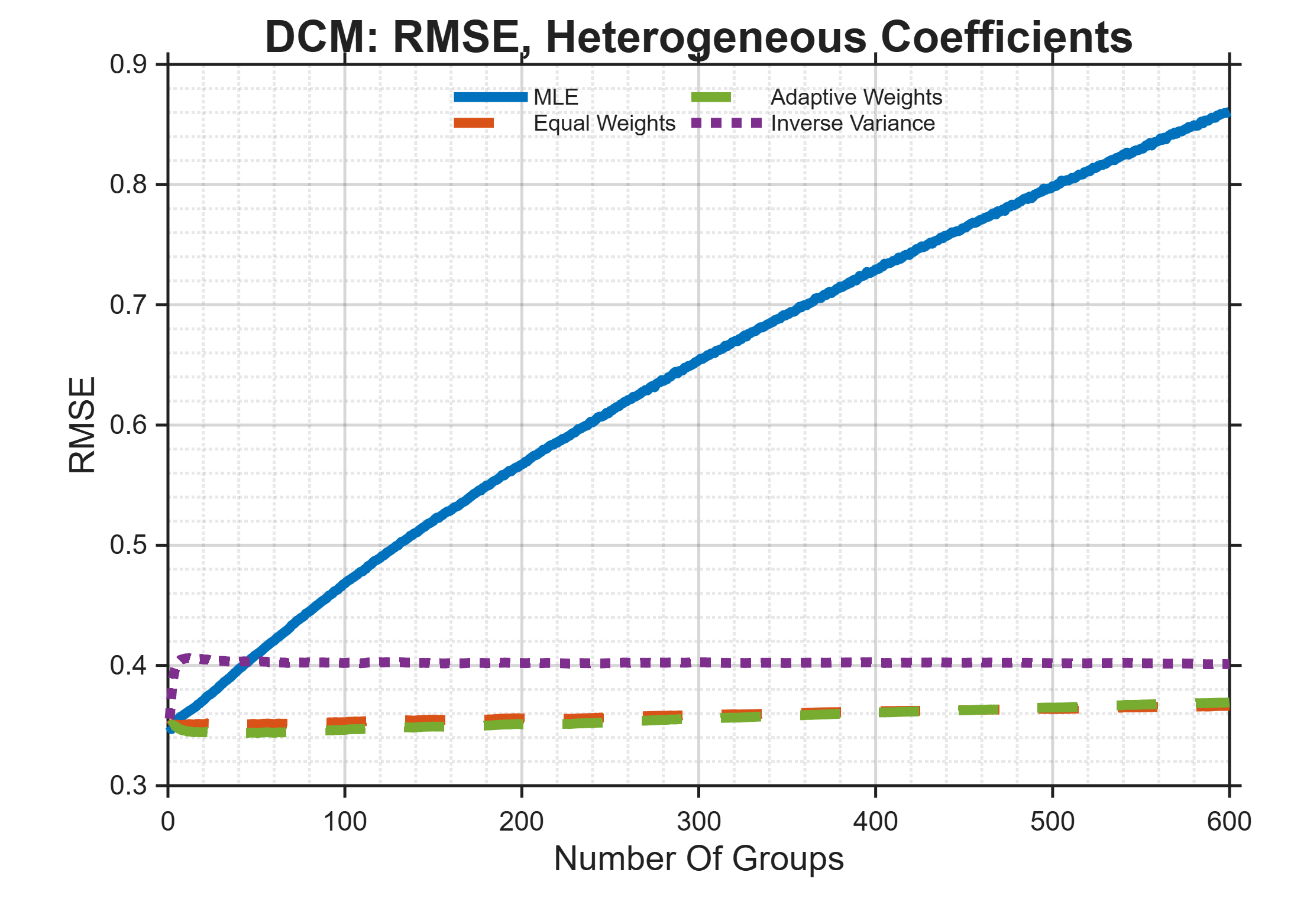}
		\caption{RMSE under heterogeneity, $\theta_1=.9385$ for inverse-variance weights}
		\label{fig:RMSEhetero2DCM}
	\end{subfigure}
	\caption{RMSE for the DCM specification \eqref{DCM}. The RMSE is standardized as in \eqref{RMSE}. Homogeneity in the left panel and heterogeneity in the right panel. RMSE on $y$-axis with number of groups $G_N$ on $x$-axis.}
	\label{fig:RMSE2DCM}
\end{figure}

\subparagraph{RMSE.} Figure \ref{fig:RMSE2DCM} displays the RMSE results for this specification. Under homogeneity, the MLE asymptotically achieves the Cram\'{e}r--Rao bound. The inverse-variance and adaptive-weight estimators should also attain this bound asymptotically, yet their RMSEs remain slightly above that of the MLE. They nevertheless outperform the equal-weight estimator, with improvements of at least $30\%$, as shown in Figure \ref{fig:RMSEhomoDCM}.  Under heterogeneity, Figure \ref{fig:RMSEhetero2DCM} shows that the RMSE of the MLE deteriorates when $G_N$ increases. The loss of the inverse-variance weighting procedure relative to equal and adaptive weighting is small, around $10\%$ in magnitude.

\subparagraph{Coverage.} 
Under homogeneity, the inverse-variance and adaptive-weight procedures deliver higher coverage than expected for $G_N$ below $100$ for equal weights, $300$ for inverse-variance, and $350$ for adaptive weights, respectively. However, coverage never exceeds $97\%$. For larger $G_N$, the coverage of the equal-weight procedure falls to about $91\%$ at $G_N=600$, whereas the random-weight procedures remain above $94\%$, compared with the nominal coverage of $95\%$. Under heterogeneity, the coverage of the three weighting procedures is nearly identical, bearing in mind that a different parameter value is used for the inverse-variance weights (see Figure~\ref{fig:CovheteroDCM} for coverage when $\theta$ is set to $1$ for all procedures). The empirical coverage of each weighted procedure is very close to the nominal $95\%$ level as soon as $G_N \geq 30$.

\begin{figure}[!htbp]
	\centering
	\begin{subfigure}[t]{0.4\textwidth}
		\includegraphics[width=\textwidth]{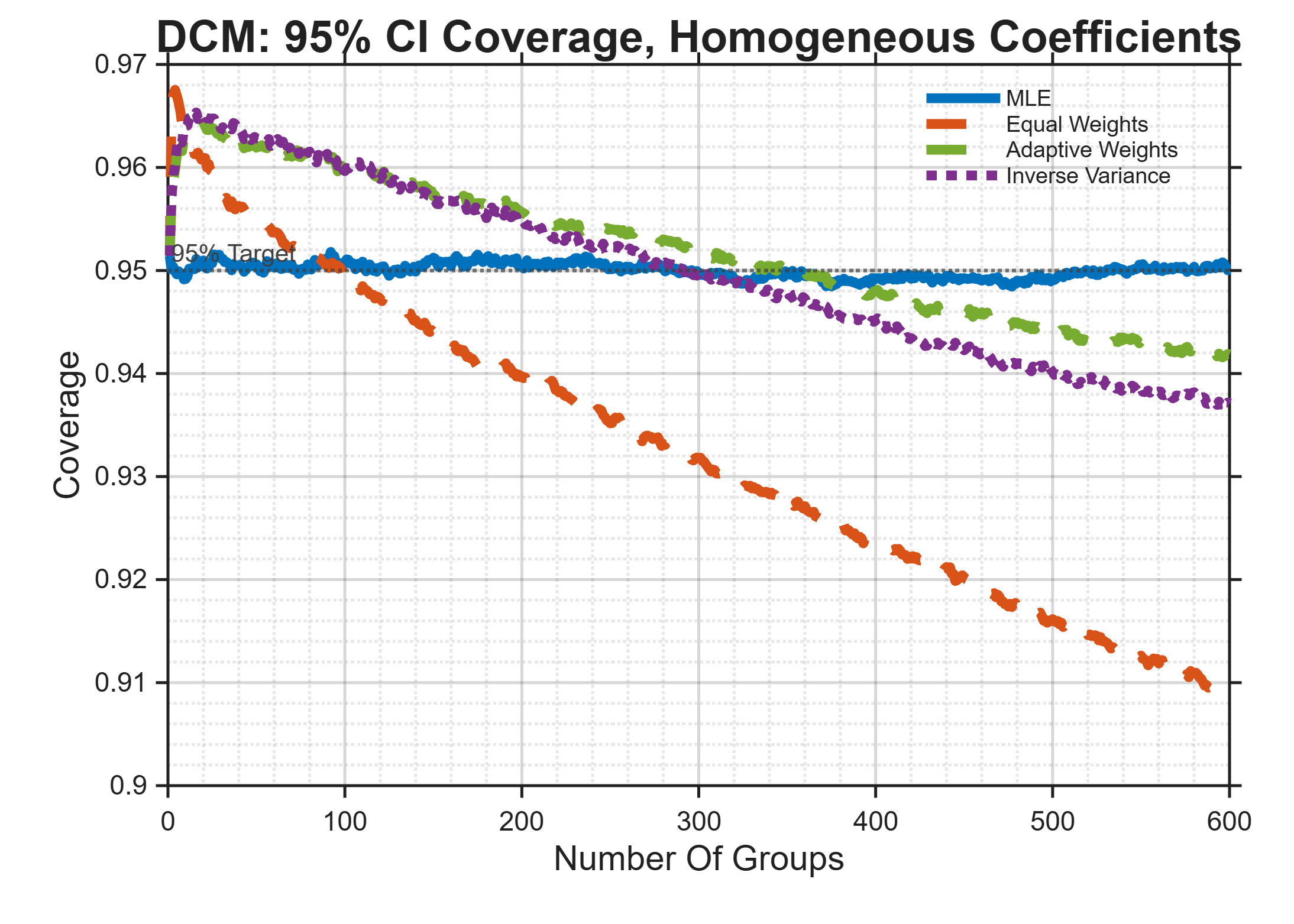}
		\caption{Coverage under homogeneity, $\theta_1=1$ for all weights}
		\label{fig:CovhomoDCM}
	\end{subfigure}
	~
	\begin{subfigure}[t]{0.4\textwidth}
		\includegraphics[width=\textwidth]{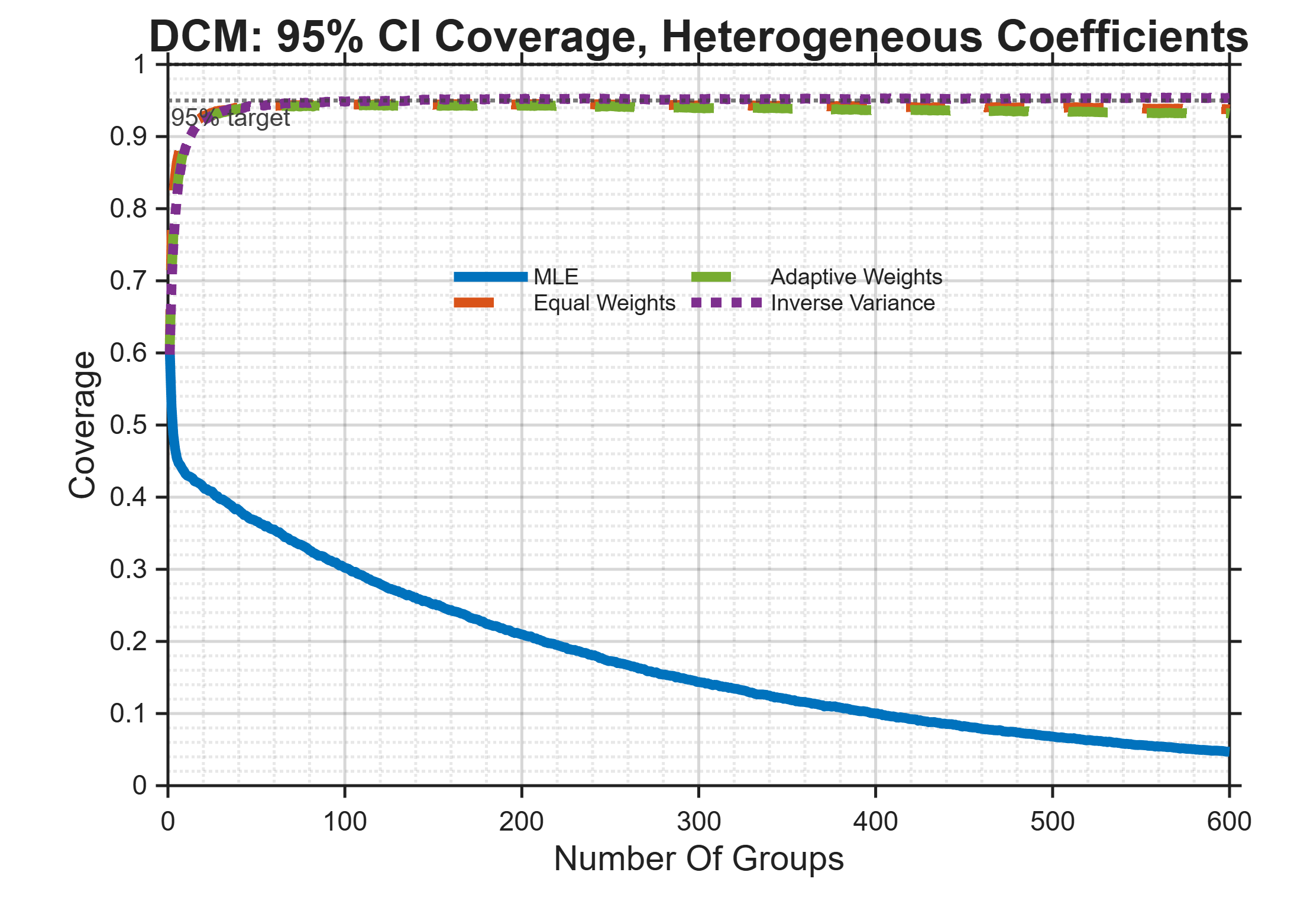}
		\caption{Coverage under heterogeneity, $\theta_1=.9385$ for inverse-variance weights}
		\label{fig:Cov2heteroDCM}
	\end{subfigure}
	\caption{Coverage of a $95\%$ confidence interval for the DCM specification \eqref{DCM}.  Homogeneity in the left panel and heterogeneity in the right panel. Coverage on $y$-axis with number of groups $G_N$ on $x$-axis.}
	\label{fig:Cov2DCM}
\end{figure}

\subparagraph{Homogeneity Q test.} The simulation experiment considers a nominal level of $5\%$, for $G_N=5,10,50,100$. The $y$-axis gives power as a function of the slope heterogeneity standard deviation $\sigma_{\delta}$ in \eqref{IVmodel}. The empirical level, achieved for $\sigma_{\delta}=0$, is close to the nominal level. As expected, the power increases with $G_N$. 
\begin{figure}[!htbp]
	\centering
	\includegraphics[scale=.37]{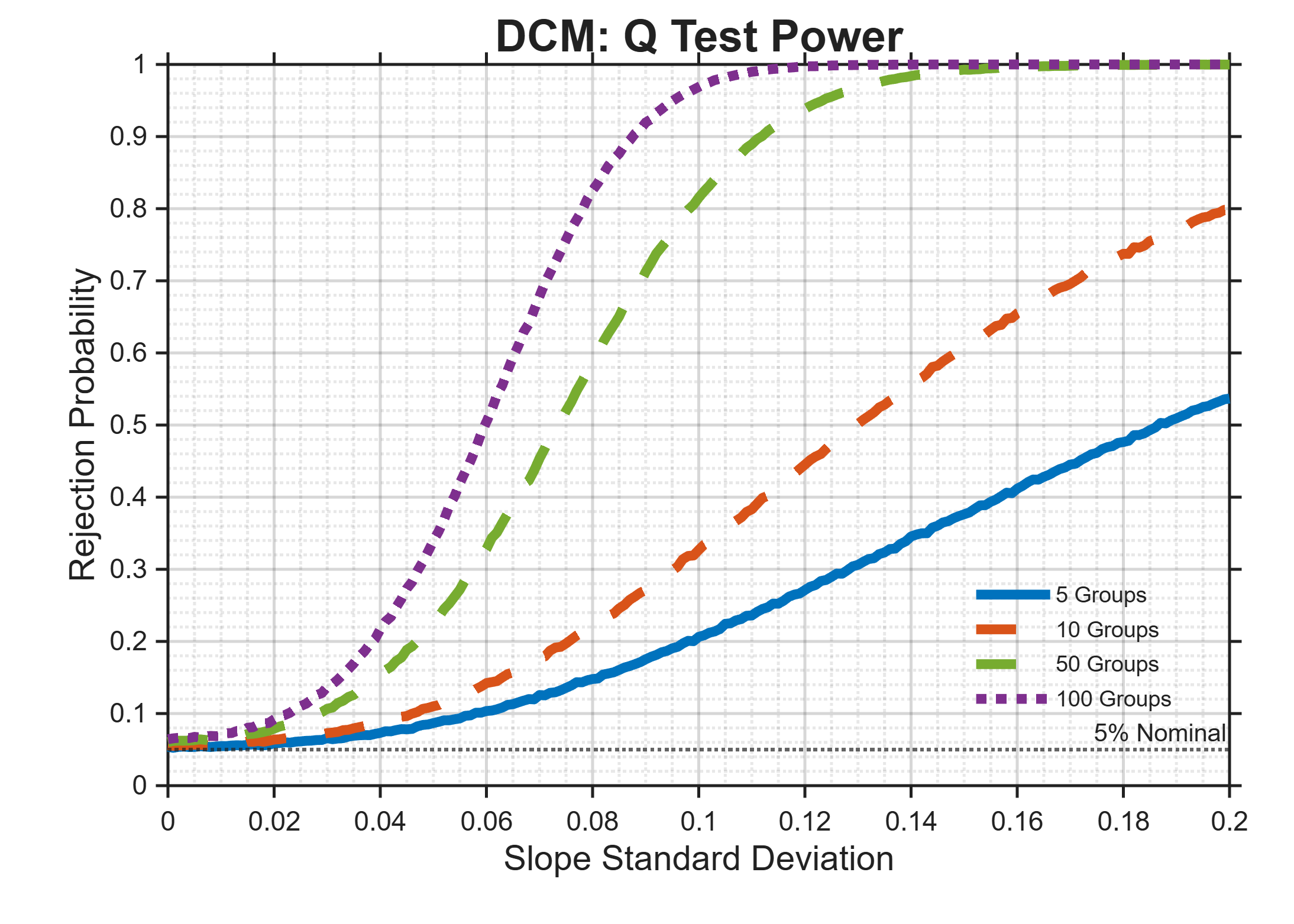}
	\caption{Homogeneity Q-test for the  model \eqref{DCM}, nominal $5\%$ level.  Power on the $y$-axis, heterogeneity standard deviation $\sigma_{\delta}$ on the $x$-axis.}
	\label{fig:QDCM}
\end{figure}
\subsection{Racial disparities in mortgage denial}
\label{sec:hmda}

We illustrate the aggregation problem using mortgage applications from the U.S. Home Mortgage Disclosure Act (HMDA), a widely used source in empirical mortgage research on lending technology, socioeconomic disparities in mortgage outcomes, and geographic heterogeneity \citep{fuster2019,bhutta2021,bartlett2022,gerardi2023,dursundeneef2023,garcia2026}. 
We focus on the conditional difference in mortgage denial between Black and White applicants within each Metropolitan Statistical Area (MSA), viewed as a specific mortgage market.  A discrete-choice model as in  \eqref{DCM} is estimated within each MSA subsample. Different aggregation methods are then compared and homogeneity is tested.

\paragraph{Data and MSA-specific estimates.}
We use  2020--2024 HMDA application-level data on originated and denied applications from non-Hispanic Black and non-Hispanic White applicants. We retain MSAs with at least 1,000 applications and positive counts in each of the four race-by-outcome cells defined by Black/White and denial/origination status. The resulting sample contains \(G=384\) MSAs and approximately 4.65 million applications. Table~\ref{tab:desc-msa-summary} summarizes the variables and the retained sample.

\begin{table}[!htbp]
	\centering
	\caption{Descriptive statistics and MSA sample support}
	\label{tab:desc-msa-summary}
	\scriptsize
	\begin{adjustbox}{max width=\textwidth}
		\begin{tabular}{lrrr}
\toprule
quantity & observations & mean & sd \\
\midrule
\multicolumn{4}{l}{\textbf{Panel A. Application-level continuous variables}} \\
Log income & 4647610 & 4.5874 & 0.7331 \\
Log loan amount & 4647610 & 12.4171 & 0.7869 \\
CLTV ratio & 4647610 & 0.8120 & 0.1618 \\
DTI score & 4647610 & 1.6754 & 1.2608 \\
\addlinespace
\multicolumn{4}{l}{\textbf{Panel B. MSA-level sample support}} \\
MSA sample size &  & 12103.1510 & 17597.8156 \\
MSA denial rate &  & 0.0828 & 0.0279 \\
\bottomrule
\end{tabular}

	\end{adjustbox}
\end{table}

For application \(i\) in MSA \(g\), we estimate the Logit specification,
\begin{align*}
	\mathbb{P} (Denied_{ig}=1\mid X_{ig})
	=
	\Lambda\left(
	\alpha_g+\theta_{Black,g}Black_{ig}+X_{ig}\theta_{X,g}
	\right),
\end{align*}
$\Lambda (\cdot)$ being the Logistic transformation, and
where the dependent variable \(Denied_{ig}\) equals one for a denied application and zero for an originated application. The  regressor \(Black_{ig}\) equals one for a non-Hispanic Black applicant, with non-Hispanic White applicants as the reference group. The controls \(X_{ig}\) include Female, log income, log loan amount, the combined loan-to-value (CLTV) ratio, and an ordinal DTI score constructed from the public HMDA debt-to-income categories. The  variance estimators $\widehat{\Omega}_{g} \left(\widehat{\theta}_g\right)$ are the standard sandwich ones.

\begin{figure}[!htbp]
	\centering
	\includegraphics[width=\textwidth]{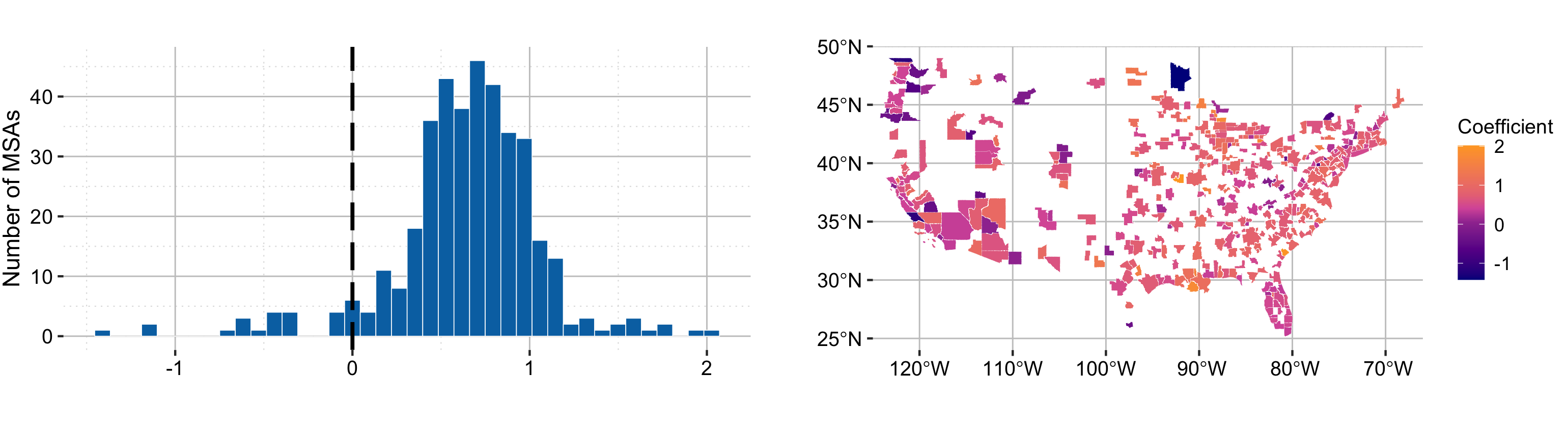}
	\caption{Black coefficient by MSA}
	\label{fig:trimmed-blackmap}
	\vspace{.5em}
	\begin{minipage}{.95\textwidth}
		\scriptsize \textit{Note:} The left panel plots the distribution of MSA-specific Black logit coefficients. The right panel maps the same coefficients for MSA geographies.
	\end{minipage}
\end{figure}

Table~\ref{tab:trimmed-dist} shows substantial cross-MSA variation. The mean Black coefficient is \(0.639\). Figure~\ref{fig:trimmed-blackmap} shows that most estimates are positive but geographically dispersed.

\begin{table}[!htbp]
	\centering
	\caption{Cross-MSA coefficient distributions}
	\label{tab:trimmed-dist}
	\scriptsize
	\begin{adjustbox}{max width=\textwidth}
		\begin{tabular}{lrrrr}
\toprule
variable & G & mean & sd  \\
\midrule
\multicolumn{5}{l}{\textbf{Panel A. Black coefficient}} \\
Black applicant & 384 & 0.6392 & 0.4168  \\
\addlinespace
\multicolumn{5}{l}{\textbf{Panel B. Other non-intercept coefficients}} \\
DTI score & 384 & 0.6800 & 0.1398  \\
CLTV ratio & 384 & 0.1168 & 0.8738  \\
Log income & 384 & 0.0131 & 0.2577  \\
Log loan amount & 384 & -0.4959 & 0.2451  \\
Female applicant & 384 & -0.1725 & 0.1511  \\
\bottomrule
\end{tabular}

	\end{adjustbox}
\end{table}

\vspace{-25pt} 
\paragraph{Aggregation and heterogeneity.}
We aggregate the MSA-specific  coefficients using adaptive (AD), equal (EQ), and inverse-variance (IV) weights as in \eqref{Hatwstar}, \eqref{EW} and \eqref{IV} respectively. While the coefficients of the Panel B variables of Table \ref{tab:trimmed-agg} are aggregated in a vectorial manner, a scalar aggregation is used for the Black applicant variable. While vectorial aggregation is supposed to produce a smaller asymptotic variance by Theorem \ref{Opt}, it may also combine more bias terms, so that the resulting estimator may have a larger bias. 

\begin{table}[!htbp]
	\centering
	\caption{Aggregation of MSA-specific coefficients with $95\%$ confidence intervals}
	\label{tab:trimmed-agg}
	\scriptsize
	\begin{adjustbox}{max width=\textwidth}
		\begin{tabular}{lrlll}
\toprule
variable & G & AD & EQ & IV \\
\midrule
\multicolumn{5}{l}{\textbf{Panel A. Black coefficient}} \\
Black applicant & 384 & 0.6969 [0.6646, 0.7292] & 0.6392 [0.5976, 0.6808] & 0.7190 [0.6613, 0.7768] \\
\addlinespace
\multicolumn{5}{l}{\textbf{Panel B. Other non-intercept coefficients}} \\
DTI score & 384 & 0.6770 [0.6634, 0.6906] & 0.6800 [0.6660, 0.6939] & 0.6891 [0.6668, 0.7115] \\
CLTV ratio & 384 & 0.0463 [-0.0390, 0.1316] & 0.1168 [0.0295, 0.2041] & -0.2453 [-0.3866, -0.1040] \\
Log income & 384 & 0.0242 [-0.0007, 0.0492] & 0.0131 [-0.0126, 0.0388] & 0.1204 [0.0749, 0.1659] \\
Log loan amount & 384 & -0.4796 [-0.5037, -0.4554] & -0.4959 [-0.5203, -0.4714] & -0.3904 [-0.4336, -0.3471] \\
Female applicant & 384 & -0.1731 [-0.1855, -0.1606] & -0.1725 [-0.1876, -0.1574] & -0.1900 [-0.2071, -0.1730] \\
\bottomrule
\end{tabular}

	\end{adjustbox}
\end{table}

For the Black coefficient, adaptive weighting  gives \(0.697\),  equal weighting \(0.639\), and inverse-variance weighting \(0.719\).  While these values are economically similar, the random-weights estimators seem substantially higher than the equal weight ones. In particular, the AD and IV based confidence intervals do not contain the EQ point estimation. The standard deviation of the AD procedure is around $30\%$ lower than those of the EQ and IV estimators.

For the Panel B coefficients which are aggregated in a vectorial manner, the AD procedure also produces a smaller estimated standard deviation, but marginally compared to the EQ estimator. The IV one has a larger estimated standard deviation than its two competitors. The AD and EQ procedures are in agreement for the significance of all coefficients except for the CLTV ratio variable, which is viewed as non-significant with AD inference. Note also that the EQ and IV estimates of this coefficient have opposite signs, which may be due to bias.

Consider now homogeneity of the Black applicant coefficient. The left panel of Figure \ref{fig:trimmed-blackmap} and its standard deviation of 0.4168 across MSAs in Table \ref{tab:trimmed-dist} both suggest some heterogeneity. The corrected standard deviation computed from $\check{\Sigma}$ in \eqref{Checksig} yields a smaller 0.2498, suggesting that estimation uncertainty is an important source of spurious heterogeneity. 
However, computation of the modified Cochran homogeneity statistic $\widehat{Q}$ in \eqref{Homstat} provides strong evidence of parameter heterogeneity, with a value of 6.1598 for the Black coefficient alone and of  14.0173 for the variables of Panel B, leading to a rejection of the null of homogeneity at any usual statistical level.

\section{Concluding remarks}
\label{sec:concluding-remarks}

Using estimated weights to aggregate subsample estimators is common in meta-analysis and is also relevant when applying distributed estimation to batches of observations from different origins, which may affect the parameter of interest and/or the variances of the estimators. The equal weighting procedure often considered in the distributed estimation literature may be inappropriate in many economic datasets where heterogeneity affects the dispersion of the estimators. 
However, optimal deterministic weights based on the inverse of the sum of the heterogeneity variance and expected estimation variance can be difficult to apply. In full generality, the estimation variance may depend on the parameter of interest in an unknown way, so that its expectation may be difficult to estimate under parameter heterogeneity. This paper proposes an adaptive weighting scheme which addresses this issue in a uniform manner, at the price of using estimated weights which asymptotically depend on the subsample parameter. This creates an estimation bias and, under heterogeneity, a correlation bias due to the dependence of the weights on the subsample parameter. The latter also implies that the variance of the weighted estimator is difficult to compute.

We show that, under suitable conditions, these issues can be addressed when the number of groups $G_N$ is negligible with respect to the (average) subsample size $n= N/G_N$, where $N$ is the total sample size. Under such conditions, standard Gaussian inference applies to the adaptive weighting scheme, which is also optimal. This uses a heterogeneity-variance estimator which asymptotically vanishes with a fast rate under homogeneity, as in \citet{hedges1985}.  As our framework covers a large range of estimation methods,
these results extend \citet{gu2023}, who restrict to the homogeneous case, and extend \citet{lu2023}, \citet{fernandezval2026}, who both consider equal-weight averages which may be inefficient for confidence intervals.
As heterogeneity is a threat to external validity as noted in \citet{vivalt2020},  the paper also considers a homogeneity test which extends \citet{cochran1937}, \citet{pesaran2008}, \citet{ritz2008}. 
Our simulations show that our theoretical results seem applicable when $G_N$  is negligible with respect to $n$, typically less than $ n/10$, $n$ being in the range of a thousand. An application to Black--White differences in mortgage denial across U.S. metropolitan areas illustrates our methodology.

\begingroup
\setstretch{1}
\bibliographystyle{agsm}
\bibliography{GW260830}

@article{andrews2019,
  author  = {Andrews, Isaiah and Kasy, Maximilian},
  title   = {Identification of and Correction for Publication Bias},
  journal = {American Economic Review},
  year    = {2019},
  volume  = {109},
  pages   = {2766--2794}
}

@article{veroniki2016,
  author  = {Veroniki, Areti Angeliki and Jackson, Dan and Viechtbauer, Wolfgang and Bender, Ralf and Bowden, Jack and Knapp, Guido and Kuss, Oliver and Higgins, Julian P. T. and Langan, Dean and Salanti, Georgia},
  title   = {Methods to Estimate the Between-Study Variance and Its Uncertainty in Meta-Analysis},
  journal = {Research Synthesis Methods},
  year    = {2016},
  volume  = {7},
  pages   = {55--79}
}

@article{bandiera2021,
  author  = {Bandiera, Oriana and Fisher, Greg and Prat, Andrea and Ytsma, Erina},
  title   = {Do Women Respond Less to Performance Pay? Building Evidence from Multiple Experiments},
  journal = {American Economic Review: Insights},
  year    = {2021},
  volume  = {3},
  pages   = {435--454}
}

@article{bartlett2022,
  author  = {Bartlett, Robert and Morse, Adair and Stanton, Richard and Wallace, Nancy},
  title   = {Consumer-Lending Discrimination in the FinTech Era},
  journal = {Journal of Financial Economics},
  year    = {2022},
  volume  = {143},
  pages   = {30--56}
}

@article{bhutta2021,
  author  = {Bhutta, Neil and Hizmo, Aurel},
  title   = {Do Minorities Pay More for Mortgages?},
  journal = {The Review of Financial Studies},
  year    = {2021},
  volume  = {34},
  pages   = {763--789}
}

@article{bellio2016,
  author  = {Bellio, Ruggero and Guolo, Annamaria},
  title   = {Integrated Likelihood Inference in Small-Sample Meta-Analysis for Continuous Outcomes},
  journal = {Scandinavian Journal of Statistics},
  year    = {2016},
  volume  = {43},
  pages   = {191--201}
}

@article{brown2024,
  author  = {Brown, Alexander L. and Imai, Taisuke and Vieider, Ferdinand M. and Camerer, Colin F.},
  title   = {Meta-Analysis of Empirical Estimates of Loss Aversion},
  journal = {Journal of Economic Literature},
  year    = {2024},
  volume  = {62},
  pages   = {485--516}
}

@inproceedings{chu2006,
  author    = {Chu, Cheng-Tao and Kim, Sang Kyun and Lin, Yi-An and Yu, Yuanyuan and Bradski, Gary and Ng, Andrew Y. and Olukotun, Kunle},
  title     = {Map-Reduce for Machine Learning on Multicore},
  booktitle = {Advances in Neural Information Processing Systems 19},
  year      = {2006},
  pages     = {281--288},
  publisher = {MIT Press}
}

@article{cochran1937,
  author  = {Cochran, William G.},
  title   = {Problems Arising in the Analysis of Similar Experiments},
  journal = {Supplement to the Journal of the Royal Statistical Society},
  year    = {1937},
  volume  = {4},
  pages   = {102--118}
}

@article{cochran1954,
  author  = {Cochran, William G.},
  title   = {The Combination of Estimates from Different Experiments},
  journal = {Biometrics},
  year    = {1954},
  volume  = {10},
  pages   = {101--129}
}

@article{cochran1953,
  author  = {Cochran, William G. and Carroll, Stephen G.},
  title   = {A Sampling Investigation of the Efficiency of Weighting Inversely as the Estimated Variance},
  journal = {Biometrics},
  year    = {1953},
  volume  = {9},
  pages   = {447--459}
}

@article{dersimonian1986,
  author  = {DerSimonian, Rebecca and Laird, Nan},
  title   = {Meta-Analysis in Clinical Trials},
  journal = {Controlled Clinical Trials},
  year    = {1986},
  volume  = {7},
  pages   = {177--188}
}

@article{duan2022,
  author  = {Duan, Ruixue and Ning, Yang and Chen, Yong},
  title   = {Heterogeneity-Aware and Communication-Efficient Distributed Statistical Inference},
  journal = {Biometrika},
  year    = {2022},
  volume  = {109},
  pages   = {67--83}
}

@article{dursundeneef2023,
  author  = {Dursun-de Neef, Hande Ozden},
  title   = {Bank Specialization, Mortgage Lending and House Prices},
  journal = {Journal of Banking and Finance},
  year    = {2023},
  volume  = {151},
  pages   = {106836}
}

@article{einav2014,
  author  = {Einav, Liran and Levin, Jonathan},
  title   = {Economics in the Age of Big Data},
  journal = {Science},
  year    = {2014},
  volume  = {346},
  pages   = {1243089}
}

@article{fernandezval2016,
  author  = {Fernandez-Val, Ivan and Weidner, Martin},
  title   = {Individual and Time Effects in Nonlinear Panel Models with Large N, T},
  journal = {Journal of Econometrics},
  year    = {2016},
  volume  = {192},
  pages   = {291--312}
}

@misc{fernandezval2026,
  author = {Fernandez-Val, Ivan and Gao, Wayne Yuan and Liao, Yuan and Vella, Francis},
  title  = {Dynamic Heterogeneous Distribution Regression Panel Models, with an Application to Labor Income Processes},
  year   = {2026},
  note   = {arXiv}
}

@article{fuster2019,
  author  = {Fuster, Andreas and Plosser, Matthew and Schnabl, Philipp and Vickery, James},
  title   = {The Role of Technology in Mortgage Lending},
  journal = {The Review of Financial Studies},
  year    = {2019},
  volume  = {32},
  pages   = {1854--1899}
}

@article{garcia2026,
  author  = {Garcia, Manuel and Garriga, Carlos},
  title   = {The Determinants of Mortgage Denial},
  journal = {Federal Reserve Bank of St. Louis Review},
  year    = {2026},
  volume  = {108},
  pages   = {1--36}
}

@article{gerardi2023,
  author  = {Gerardi, Kristopher and Willen, Paul S. and Zhang, David Hao},
  title   = {Mortgage Prepayment, Race and Monetary Policy},
  journal = {Journal of Financial Economics},
  year    = {2023},
  volume  = {147},
  pages   = {498--524}
}

@book{gourieroux1996,
  author    = {Gourieroux, Christian and Monfort, Alain},
  title     = {Simulation-Based Econometric Methods},
  publisher = {Oxford University Press},
  year      = {1996}
}

@article{gu2023,
  author  = {Gu, J. and Chen, Song Xi},
  title   = {Distributed Statistical Inference under Heterogeneity},
  journal = {Journal of Machine Learning Research},
  year    = {2023},
  volume  = {24},
  number  = {387},
  pages   = {1--57}
}

@book{hedges1985,
  author    = {Hedges, Larry V. and Olkin, Ingram},
  title     = {Statistical Methods for Meta-Analysis},
  publisher = {Academic Press},
  address   = {New York},
  year      = {1985}
}

@book{ibragimov1981,
  author    = {Ibragimov, Ildar A. and Has'minskii, Rafail Z.},
  title     = {Statistical Estimation: Asymptotic Theory},
  publisher = {Springer-Verlag},
  year      = {1981}
}

@incollection{ichimura2007,
  author    = {Ichimura, Hidehiko and Todd, Petra E.},
  title     = {Implementing Nonparametric and Semiparametric Estimators},
  booktitle = {Handbook of Econometrics},
  year      = {2007},
  volume    = {6B},
  pages     = {5369--5468}
}

@article{kulinskaya2014,
  author  = {Kulinskaya, Elena and Morgenthaler, Stephan and Staudte, Robert G.},
  title   = {Combining Statistical Evidence},
  journal = {International Statistical Review},
  year    = {2014},
  volume  = {82},
  pages   = {214--242}
}

@article{kim2016,
  author  = {Kim, Kyoo Il},
  title   = {Higher Order Bias Correcting Moment Equations for M-Estimation and Its Higher Order Efficiency},
  journal = {Econometrics},
  year    = {2016},
  volume  = {4},
  pages   = {48}
}

@article{liu2015,
  author  = {Liu, Dungang and Liu, Regina Y. and Xie, Minge},
  title   = {Multivariate Meta-Analysis of Heterogeneous Studies Using Only Summary Statistics: Efficiency and Robustness},
  journal = {Journal of the American Statistical Association},
  year    = {2015},
  volume  = {110},
  pages   = {326--340}
}

@article{lu2023,
  author  = {Lu, Xun and Su, Liangjun},
  title   = {Uniform Inference in Linear Panel Data Models with Two-Dimensional Heterogeneity},
  journal = {Journal of Econometrics},
  year    = {2023},
  volume  = {235},
  number  = {2},
  pages   = {694--719},
  doi     = {10.1016/j.jeconom.2022.07.002}
}

@inproceedings{mcdonald2010,
  author    = {McDonald, Ryan and Hall, Keith and Mann, Gideon},
  title     = {Distributed Training Strategies for the Structured Perceptron},
  booktitle = {Human Language Technologies: The 2010 Annual Conference of the North American Chapter of the Association for Computational Linguistics},
  year      = {2010},
  pages     = {456--464},
  address   = {Los Angeles, California},
  publisher = {Association for Computational Linguistics}
}

@article{meager2019,
  author  = {Meager, Rachael},
  title   = {Understanding the Average Impact of Microcredit Expansions: A Bayesian Hierarchical Analysis of Seven Randomized Experiments},
  journal = {American Economic Journal: Applied Economics},
  year    = {2019},
  volume  = {11},
  pages   = {57--91}
}

@article{meager2022,
  author  = {Meager, Rachael},
  title   = {Aggregating Distributional Treatment Effects: A Bayesian Hierarchical Analysis of the Microcredit Literature},
  journal = {American Economic Review},
  year    = {2022},
  volume  = {112},
  number  = {6},
  pages   = {1818--1847},
  doi     = {10.1257/aer.20181811}
}

@article{newey2004,
  author  = {Newey, Whitney K. and Smith, Richard J.},
  title   = {Higher Order Properties of GMM and Generalized Empirical Likelihood Estimators},
  journal = {Econometrica},
  year    = {2004},
  volume  = {72},
  pages   = {219--255}
}

@incollection{newey1994,
  author    = {Newey, Whitney K. and McFadden, Daniel},
  title     = {Large Sample Estimation and Hypothesis Testing},
  booktitle = {Handbook of Econometrics},
  year      = {1994},
  volume    = {4},
  pages     = {2112--2245}
}

@incollection{ng2017,
  author    = {Ng, Serena},
  title     = {Opportunities and Challenges: Lessons from Analyzing Terabytes of Scanner Data},
  booktitle = {Advances in Economics and Econometrics, Eleventh World Congress},
  publisher = {Cambridge University Press},
  year      = {2017},
  volume    = {2},
  pages     = {1--34}
}

@article{pesaran2008,
  author  = {Pesaran, M. Hashem and Yamagata, Takashi},
  title   = {Testing Slope Homogeneity in Large Panels},
  journal = {Journal of Econometrics},
  year    = {2008},
  volume  = {142},
  pages   = {50--93}
}

@book{pollard2002,
  author    = {Pollard, David},
  title     = {A User's Guide to Measure Theoretic Probability},
  publisher = {Cambridge University Press},
  year      = {2002}
}

@article{rilstone1996,
  author  = {Rilstone, Paul and Srivastava, V. K. and Ullah, Aman},
  title   = {The Second-Order Bias and Mean Squared Error of Nonlinear Estimators},
  journal = {Journal of Econometrics},
  year    = {1996},
  volume  = {75},
  pages   = {369--395}
}

@article{ritz2008,
  author  = {Ritz, John and Demidenko, Eugene and Spiegelman, Donna},
  title   = {Multivariate Meta-Analysis for Data Consortia, Individual Patient Meta-Analysis, and Pooling Projects},
  journal = {Journal of Statistical Planning and Inference},
  year    = {2008},
  volume  = {138},
  pages   = {1919--1933}
}

@article{rubin1981,
  author  = {Rubin, Donald B.},
  title   = {Estimation in Parallel Randomized Experiments},
  journal = {Journal of Educational Statistics},
  year    = {1981},
  volume  = {6},
  pages   = {377--401}
}

@article{rosenblatt2016,
  author  = {Rosenblatt, Jonathan D. and Nadler, Boaz},
  title   = {On the Optimality of Averaging in Distributed Statistical Learning},
  journal = {Information and Inference: A Journal of the IMA},
  year    = {2016},
  volume  = {5},
  pages   = {379--404}
}

@article{swamy1970,
  author  = {Swamy, P. A. V. B.},
  title   = {Efficient Inference in a Random Regression Model},
  journal = {Econometrica},
  year    = {1970},
  volume  = {38},
  pages   = {311--323}
}

@article{tsitsiklis1986,
  author  = {Tsitsiklis, John N. and Bertsekas, Dimitri P. and Athans, Michael},
  title   = {Distributed Asynchronous Deterministic and Stochastic Gradient Optimization Algorithms},
  journal = {IEEE Transactions on Automatic Control},
  year    = {1986},
  volume  = {31},
  pages   = {803--812}
}

@article{varian2014,
  author  = {Varian, Hal R.},
  title   = {Big Data: New Tricks for Econometrics},
  journal = {Journal of Economic Perspectives},
  year    = {2014},
  volume  = {28},
  pages   = {3--28}
}

@article{vivalt2020,
  author  = {Vivalt, Eva},
  title   = {How Much Can We Generalize from Impact Evaluations?},
  journal = {Journal of the European Economic Association},
  year    = {2020},
  volume  = {18},
  pages   = {3045--3089}
}

@article{volgushev2019,
  author  = {Volgushev, Stanislav and Chao, Shih-Kang and Cheng, Guang},
  title   = {Distributed Inference for Quantile Regression Processes},
  journal = {The Annals of Statistics},
  year    = {2019},
  volume  = {47},
  pages   = {1634--1662}
}

@article{xue2020,
  author  = {Xue, Xiaofei and Reed, W. Robert and Menclova, Andrea},
  title   = {Social Capital and Health: A Meta-Analysis},
  journal = {Journal of Health Economics},
  year    = {2020},
  volume  = {72},
  pages   = {102317}
}

@article{yang2015,
  author  = {Yang, Zaichao},
  title   = {A General Method for Third-Order Bias and Variance Corrections on a Nonlinear Estimator},
  journal = {Journal of Econometrics},
  year    = {2015},
  volume  = {186},
  pages   = {178--200}
}

@misc{wang2025,
  author = {Wang, Yuting},
  title  = {Beliefs, Information Trust, and Air Pollution},
  year   = {2025},
  note   = {SSRN, \url{https://dx.doi.org/10.2139/ssrn.5648090}}
}

@article{wolfson2019,
  author  = {Wolfson, Paul and Belman, Dale},
  title   = {15 Years of Research on US Employment and the Minimum Wages},
  journal = {Labour},
  year    = {2019},
  volume  = {33},
  pages   = {488--506}
}

@article{zhang2013,
  author  = {Zhang, Yuchen and Duchi, John C. and Wainwright, Martin J.},
  title   = {Communication-Efficient Algorithms for Statistical Optimization},
  journal = {Journal of Machine Learning Research},
  year    = {2013},
  volume  = {14},
  pages   = {3221--3363}
}

@article{zeng2015,
  author  = {Zeng, Donglin and Lin, D. Y.},
  title   = {On Random-Effects Meta-Analysis},
  journal = {Biometrika},
  year    = {2015},
  volume  = {102},
  pages   = {281--294}
}

@article{zhang2015,
  author  = {Zhang, Yuchen and Duchi, John and Wainwright, Martin},
  title   = {Divide and Conquer Kernel Ridge Regression: A Distributed Algorithm with Minimax Optimal Rates},
  journal = {Journal of Machine Learning Research},
  year    = {2015},
  volume  = {16},
  pages   = {3299--3340}
}
\endgroup

\appendix
\clearpage

\refstepcounter{section}
\section*{\centering Appendix \thesection: Additional simulation results }
\label{app:additional-simulation-results}

\setcounter{subsection}{0}
\renewcommand{\thesubsection}{\thesection.\arabic{subsection}}
\renewcommand{\theHsubsection}{\thesection.\arabic{subsection}}
\setcounter{figure}{0}
\renewcommand{\thefigure}{A.\arabic{figure}}
\setcounter{equation}{0}
\renewcommand{\theequation}{A.\arabic{equation}}
\renewcommand{\theHequation}{A.\arabic{equation}}

\subsection{Additional simulation results for the discrete choice specification \eqref{DCM}}
\label{app:dcm-additional-results}

\begin{figure}[!htbp]
	\centering
	\includegraphics[width=0.4\textwidth]{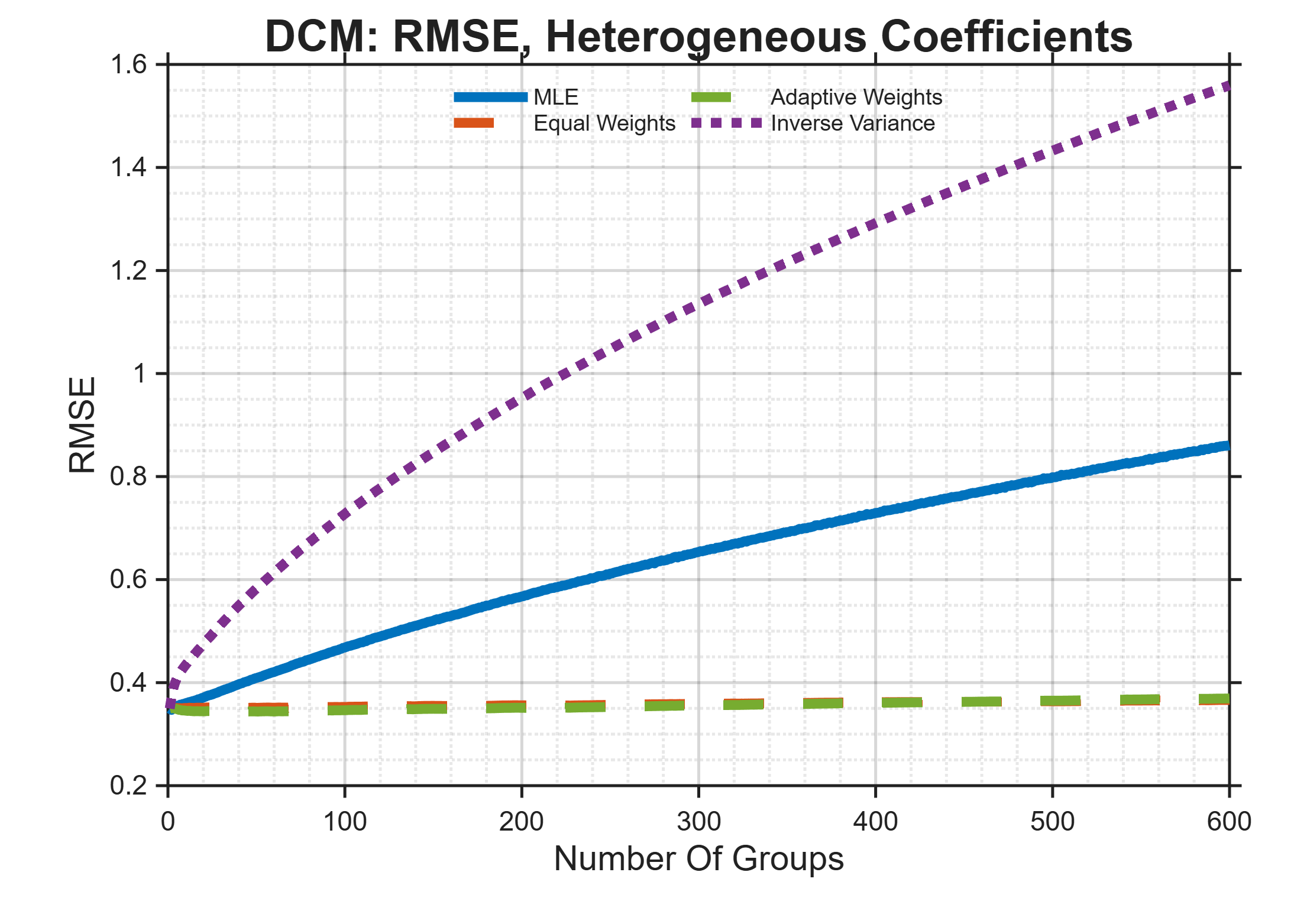}	
	\caption{RMSE for the DCM specification \eqref{DCM} under heterogeneity, $\theta_1=1$ for all weights.  RMSE on $y$-axis with number of groups $G_N$ on $x$-axis.}
	\label{fig:SuppRMSEDCM}
\end{figure}

\begin{figure}[!htbp]
	\centering
		\includegraphics[width=0.4\textwidth]{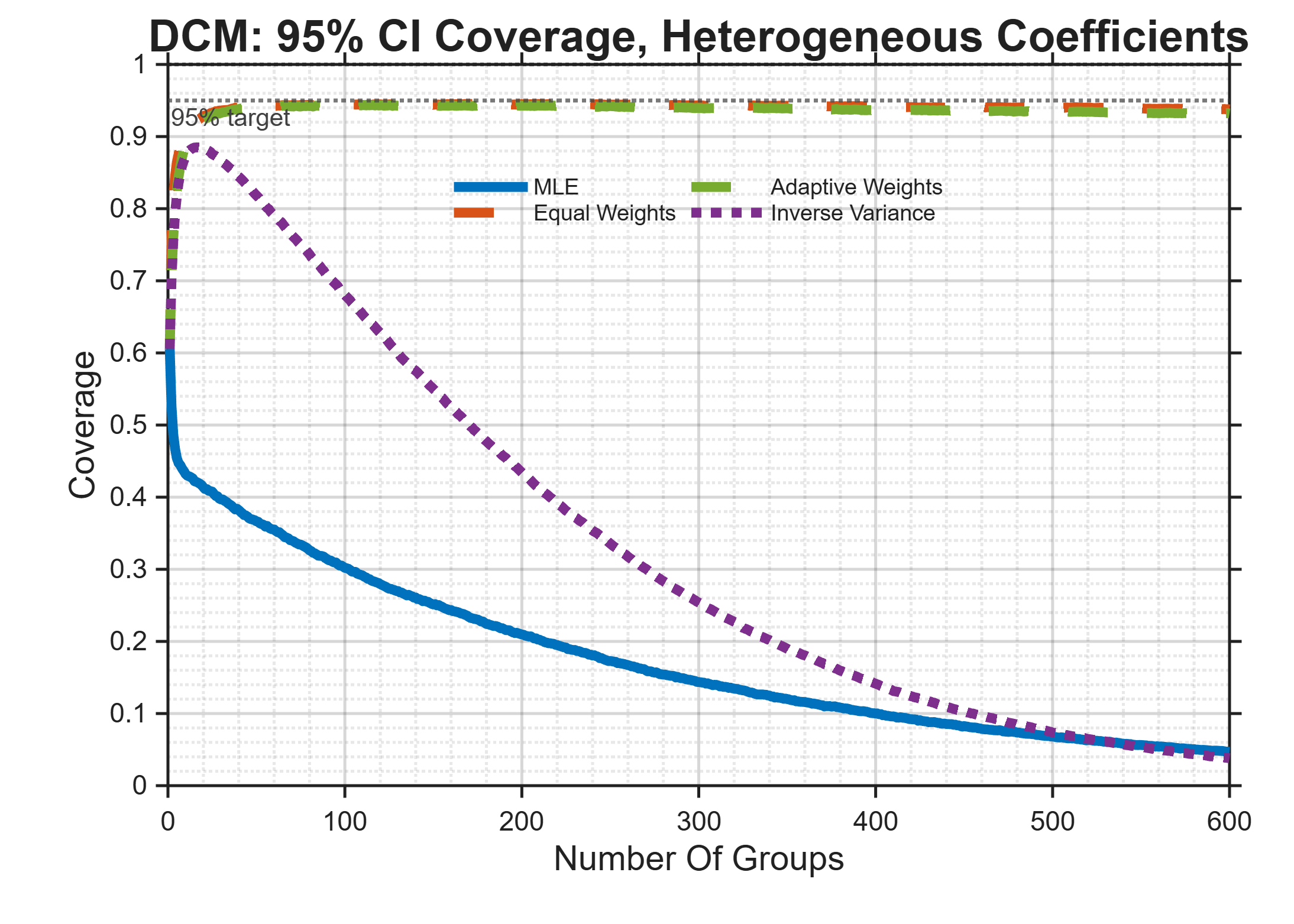}		
	\caption{Coverage of a $95\%$ confidence interval under heterogeneity for the DCM specification \eqref{DCM}, $\theta_1=1$ for all weights.  }
	\label{fig:CovheteroDCM}
\end{figure}

\subsection{Instrumental variable regression specification}
\label{app:iv-regression-specification}

In this specification, the dependent variable $y_i$, the endogenous covariate $x_{1i}$, the instrument $z_i$, and the exogenous covariates $x_{2i}$ and $x_{3i}$ are generated according to, for all $i$ in group $\mathcal{G}_g$,
\begin{align}
	\left\{
	\begin{array}{ll}
		y_i = \theta_{1g} x_{1i} + \theta_2 x_{2i} + \theta_3 x_{3i} + \sigma_g u_i + \psi v_i, & \quad \psi=.6,
		\\
		x_{1i} = \gamma z_i + v_i,  \quad \gamma=1,&  
		\left[z_i,x_{2i}, x_{3i},  u_i,v_i\right]^{\prime} \stackrel{i.i.d.}{\sim} \mathcal{N} \left(0, \mathrm{Id}\right),
		\\
		\theta_{1g}\sim \mathcal{N} \left(1,\sigma_{\delta}^2\right), \quad \theta_2= \theta_3=1, \quad \sigma_g^{2} \sim \chi^2 (2), &
	\end{array}
	\right.
	\label{IVmodel}
\end{align}
where $\theta_{1g}$, whose expectation is the parameter of interest, and $\sigma_g$ are both i.i.d. and mutually independent, and the draws of instrument, covariate, and error term are independent of these parameters as well as across groups.  The parameter $\psi$ causes endogeneity of $x_{1i}$. The variance term $\sigma_g^2$ ensures that the variance of $\widehat{\theta}_{1g}$ varies across groups. The slope standard deviation  $\sigma_{\delta}$ is fixed at $0.1$ under heterogeneity.

The slope parameters $\theta = \left[\theta_1,\theta_2,\theta_3 \right]^{\prime}$ are estimated using a just-identified GMM with $Z_i = \left[z_{i},x_{2i},x_{3i}\right]$, so that for $X_i = \left[x_{1i},x_{2i},x_{3i}\right]$, $\widehat{\theta}_g = \left(\sum_{i \in \mathcal{G}_g} Z_i^{\prime} X_i\right)^{-1}   \sum_{i \in \mathcal{G}_g} Z_i^{\prime} y_i$. Define $\widehat{res}_i \left(\widehat{\theta}_g\right)=y_i - X_i \widehat{\theta}_g$, $i$ in $\mathcal{G}_g$, and let
\begin{align*}
	\widehat{\Omega}_{g} \left(\widehat{\theta}_g\right)
	=
	\left[1 \quad 0 \quad 0 \right]
	\left(\frac{1}{n}\sum_{i \in \mathcal{G}_g} Z_i^{\prime} X_i\right)^{-1}
	\frac{1}{n}
	\sum_{i \in \mathcal{G}_g} \widehat{res}_i^{2} \left(\widehat{\theta}_g\right)Z_i^{\prime} Z_i
	\left(
	\frac{1}{n}\sum_{i \in \mathcal{G}_g} X_i^{\prime} Z_i\right)^{-1}
	\left[\begin{array}{c}
		1 \\ 0 \\ 0
	\end{array} \right]
	,
\end{align*} 
be the robust GMM estimator for the asymptotic variance of $\sqrt{n_g} \left(\widehat{\theta}_{1g} - \theta_{1g}\right)$.

\begin{figure}[!htbp]
	\centering
	\begin{subfigure}[t]{0.4\textwidth}
		\includegraphics[width=\textwidth]{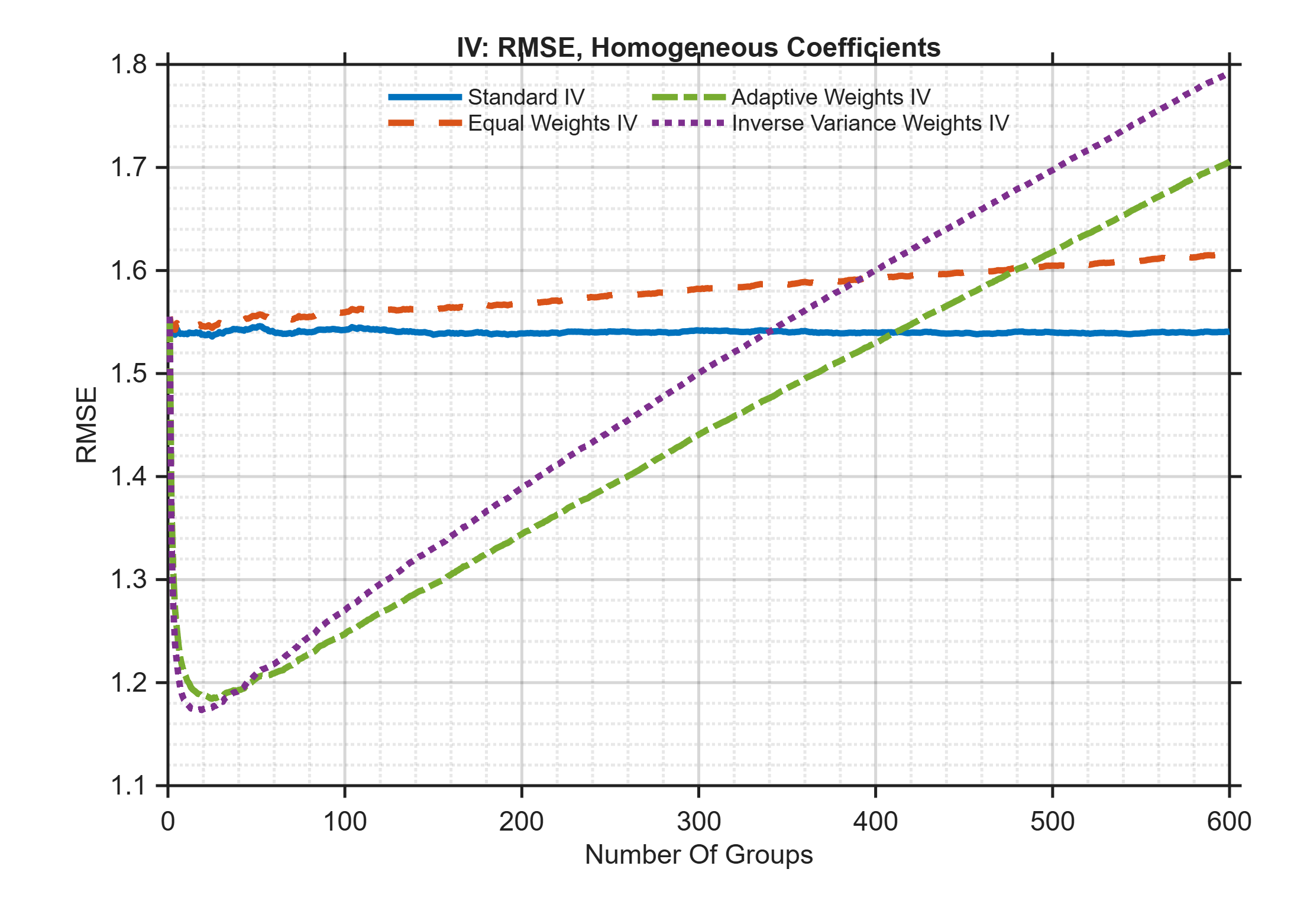}
		\caption{RMSE under homogeneity}
		\label{fig:RMSEhomoIV}
	\end{subfigure}
	~
	\begin{subfigure}[t]{0.4\textwidth}
		\includegraphics[width=\textwidth]{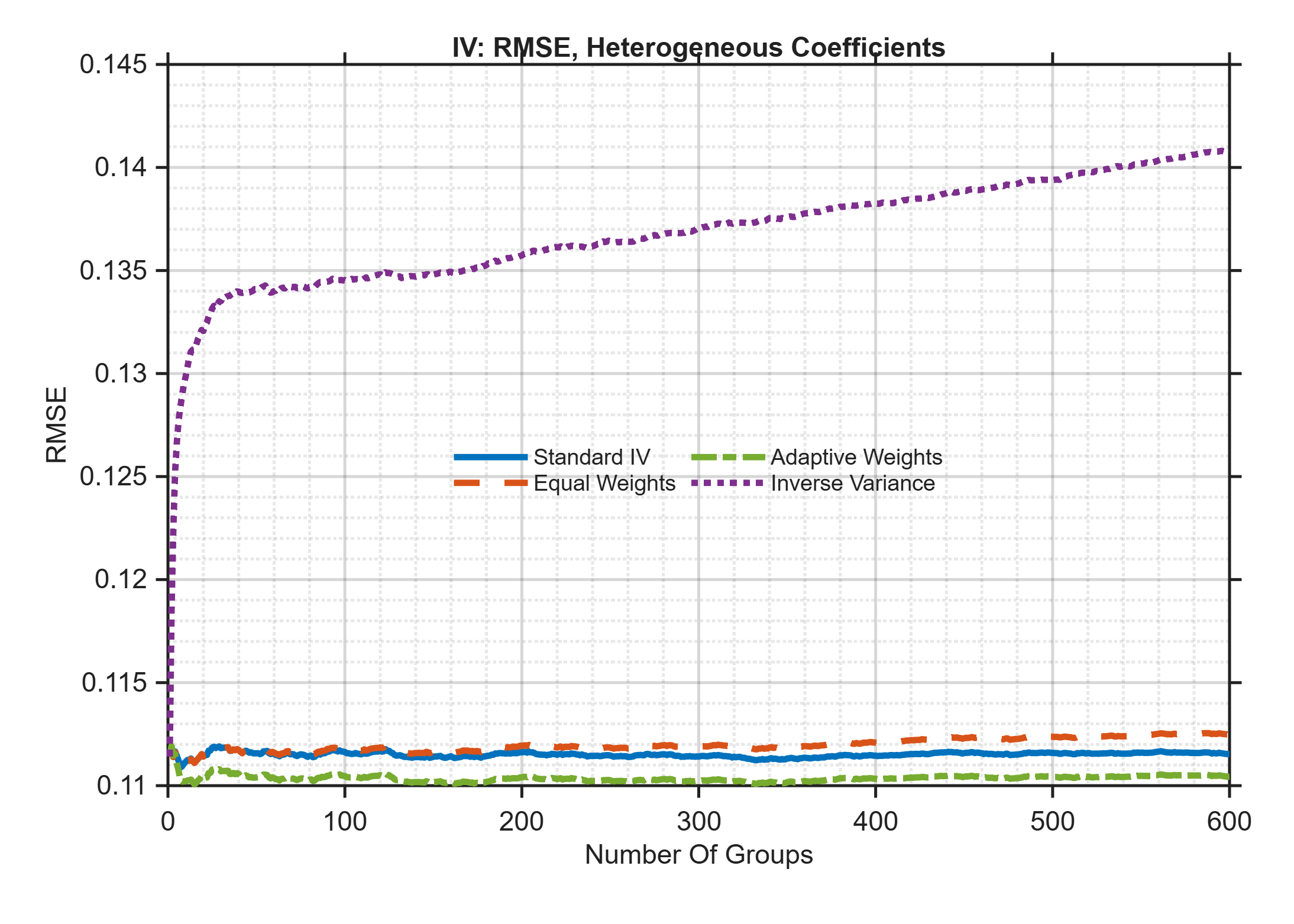}
		\caption{RMSE under heterogeneity}
		\label{fig:RMSEheteroIV}
	\end{subfigure}
	\caption{RMSE for the IV model \eqref{IVmodel}. The RMSE is standardized as in \eqref{RMSE}. Homogeneity in the left panel and heterogeneity in the right panel. RMSE on $y$-axis with number of groups $G_N$ on $x$-axis.}
	\label{fig:RMSEIV}
\end{figure}

\paragraph{RMSE.} Figure \ref{fig:RMSEIV} reports the RMSE. Under homogeneity, adaptive and inverse-variance weights yield RMSE improvements  of up to $25\%$ for around $20$ groups compared with equal-weight or even the standard IV estimator. However, increasing the number of groups $G_N$ reduces this improvement, with the standard and equal-weight IV dominating the other weighted estimators for $G_N$ around  $350$.  The adaptive-weight IV estimator appears to dominate under both homogeneity and heterogeneity. Under heterogeneity, the inverse-variance IV estimator has an RMSE about $18\%$ higher than the other estimators for most values of $G_N$, although its RMSE remains reasonable. As expected, the estimators appear less sensitive to $G_N$ under heterogeneity.
A comparison of Figures~\ref{fig:RMSEhomoIV} and~\ref{fig:RMSEhomoDCM} suggests that, under homogeneity, the performance of the inverse-variance and adaptive-weight estimators deteriorates more slowly in the discrete-choice model than in the IV specification.

\begin{figure}[!htbp]
	\centering
	\begin{subfigure}[t]{0.4\textwidth}
		\includegraphics[width=\textwidth]{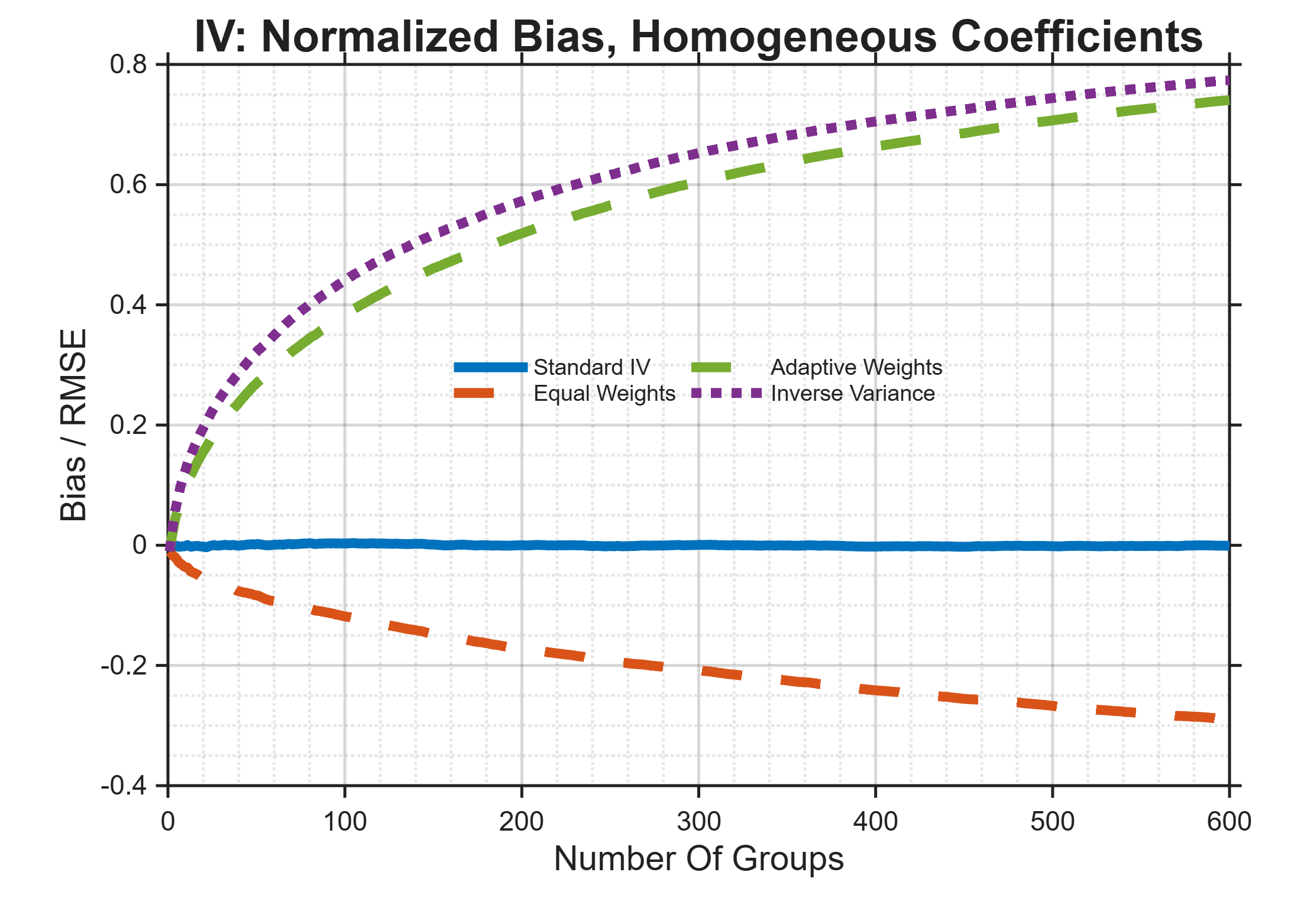}
		\caption{Bias/RMSE under homogeneity}
		\label{fig:BiashomoIV}
	\end{subfigure}
	~
	\begin{subfigure}[t]{0.4\textwidth}
		\includegraphics[width=\textwidth]{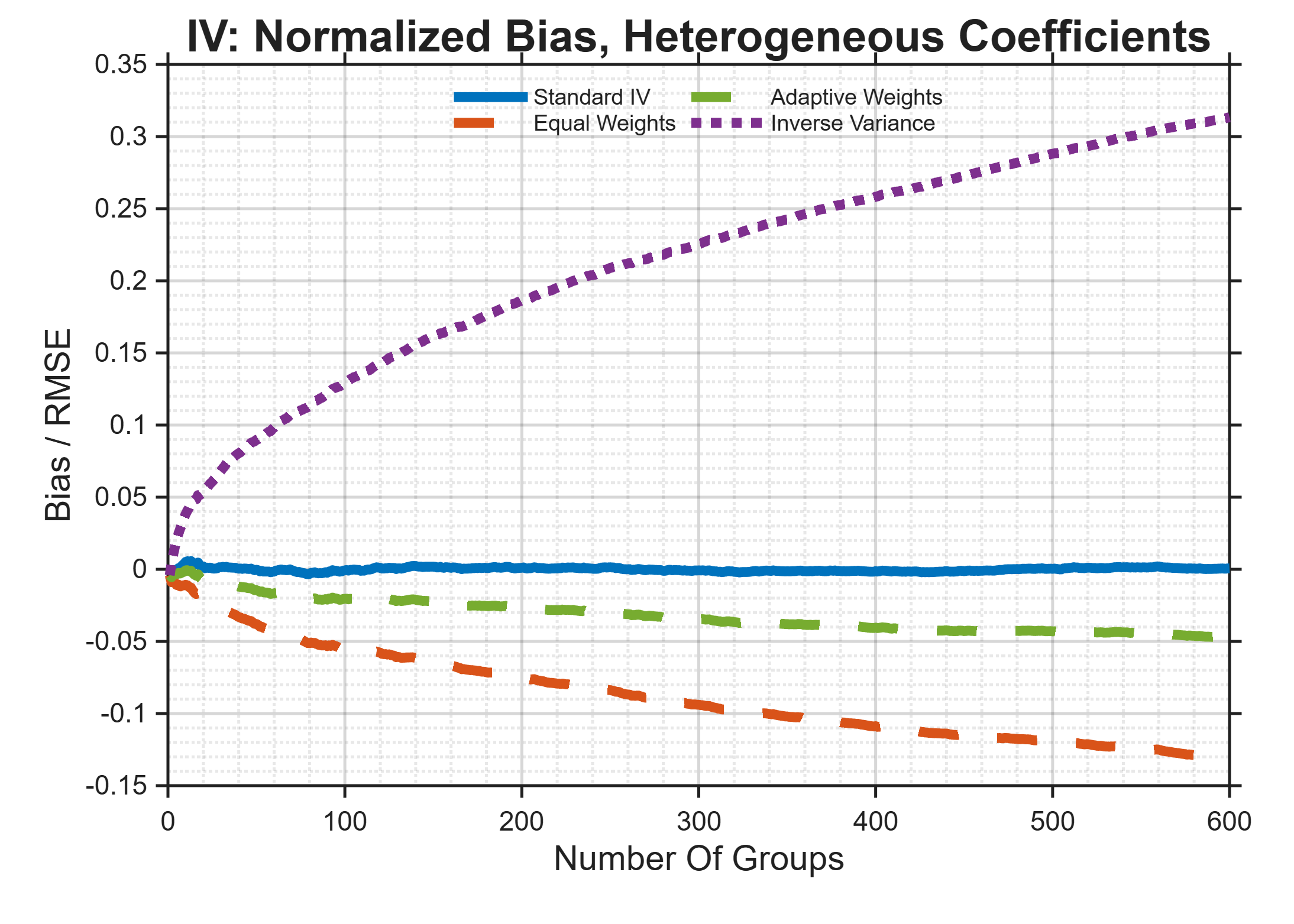}
		\caption{Bias/RMSE under heterogeneity}
		\label{fig:BiasheteroIV}
	\end{subfigure}
	\caption{Bias/RMSE for the IV model \eqref{IVmodel}. The bias is standardized as in \eqref{Biassim}. Homogeneity in the left panel and heterogeneity in the right panel. Bias/RMSE on $y$-axis with number of groups $G_N$ on $x$-axis.}
	\label{fig:BiasIV}
\end{figure}

\paragraph{Bias and confidence interval.} 
Figure \ref{fig:BiasIV} illustrates that random weights can increase bias relative to deterministic ones. Under homogeneity, the bias of the inverse-variance and adaptive-weight IV estimators can account for up to $80\%$ of the RMSE, especially for large $G_N$, whereas the bias of the equal-weight estimator represents at most $20\%$ of the RMSE. Hence, the poor RMSE performance of these estimators for large $G_N$ under homogeneity is largely driven by bias. 

By contrast, under heterogeneity, the bias of the inverse-variance weighted IV estimator accounts for at most $32\%$ of the RMSE. The bias of the adaptive-weight IV estimator switches from being potentially large under homogeneity to only about $5\%$ of the RMSE under heterogeneity, even less than for  equal weights. It also changes sign, mirroring the bias of the equal-weight estimator, as is also observed for the RMSE. 

\begin{figure}[!htbp]
	\centering
	\begin{subfigure}[t]{0.4\textwidth}
		\includegraphics[width=\textwidth]{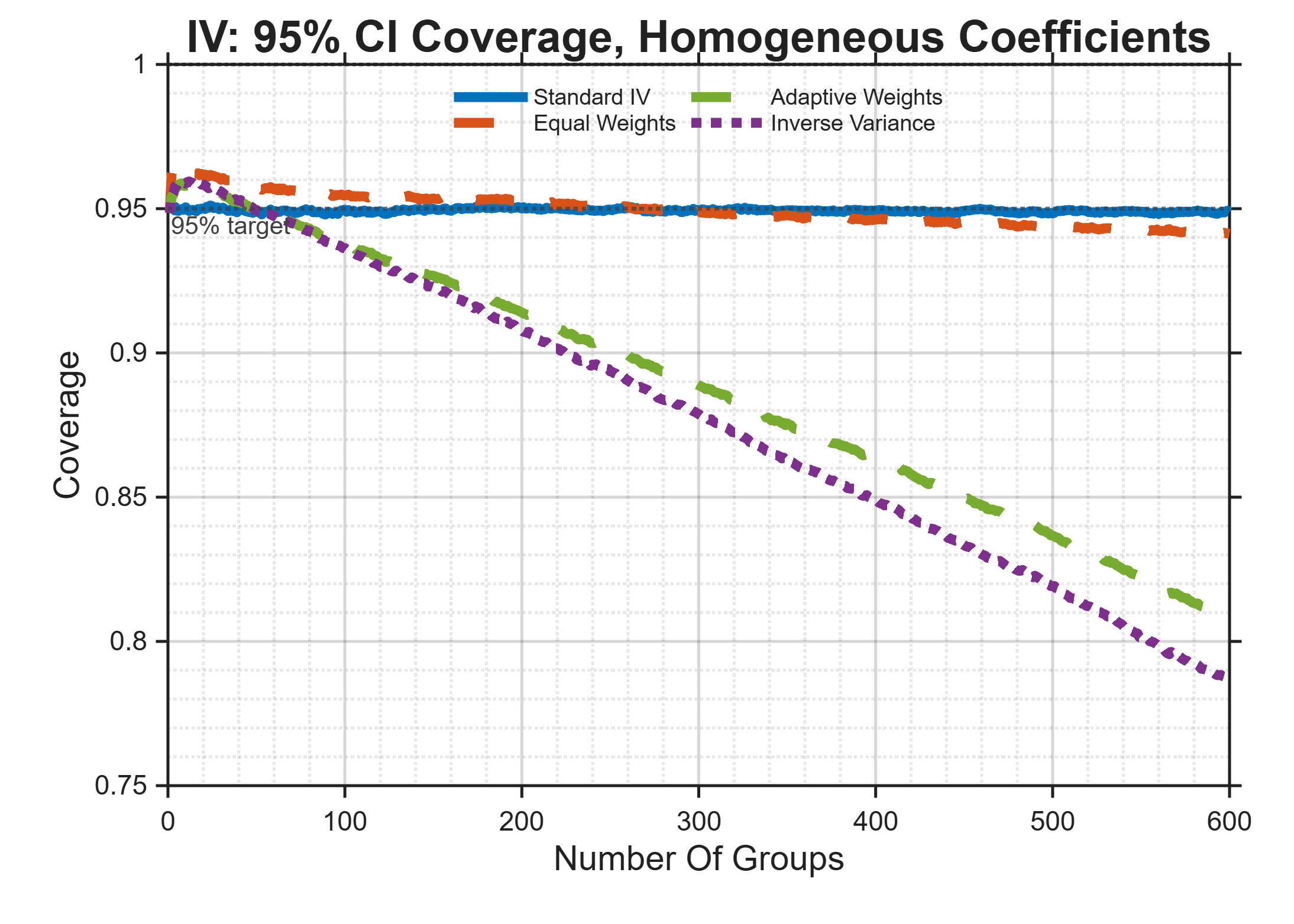}
		\caption{Coverage under homogeneity}
		\label{fig:CovhomoIV}
	\end{subfigure}
	~
	\begin{subfigure}[t]{0.4\textwidth}
		\includegraphics[width=\textwidth]{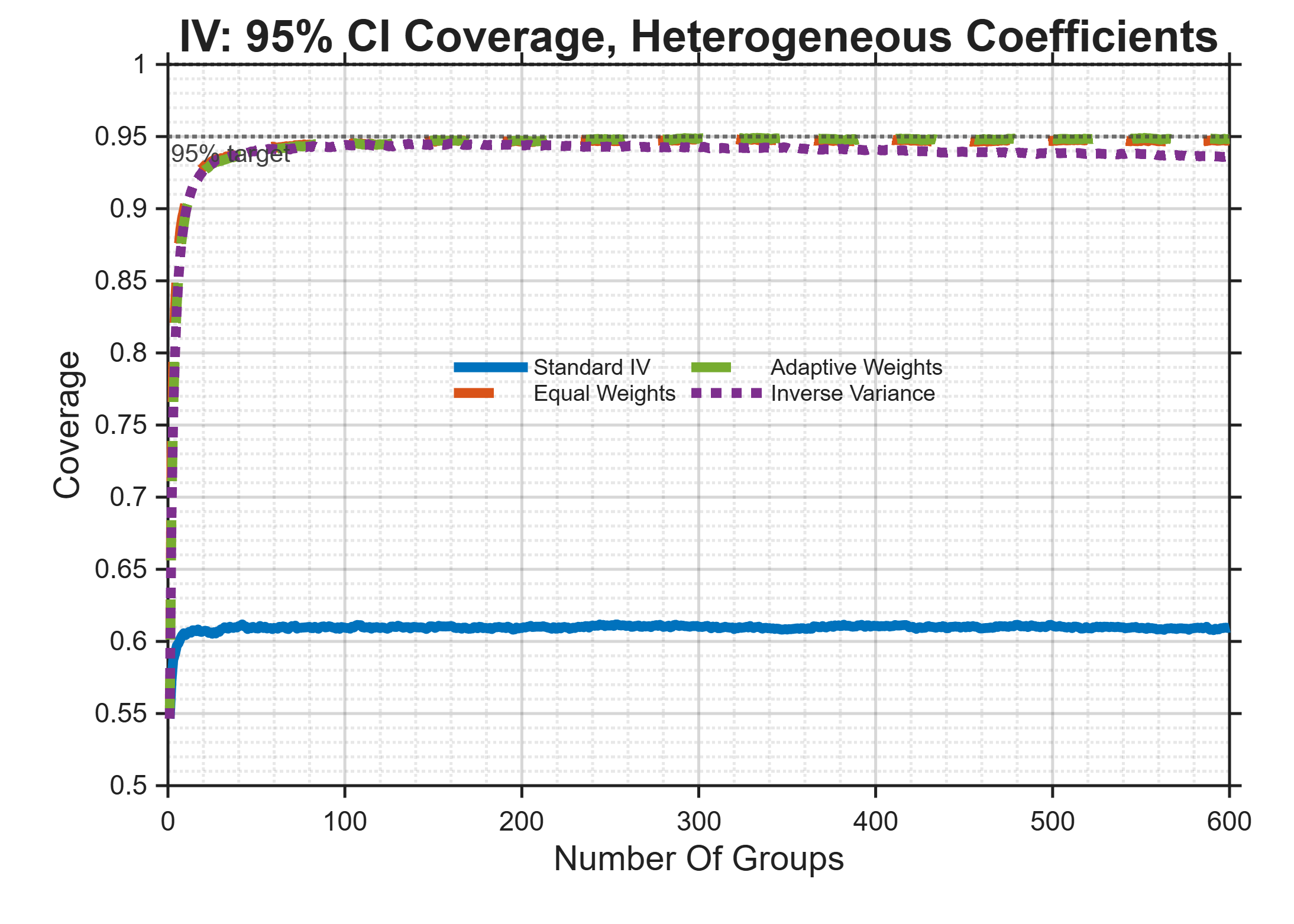}
		\caption{Coverage under heterogeneity}
		\label{fig:CovheteroIV}
	\end{subfigure}
	\caption{Coverage of a $95\%$ confidence interval for the IV model \eqref{IVmodel}.  Homogeneity in the left panel and heterogeneity in the right panel. Coverage on $y$-axis with number of groups $G_N$ on $x$-axis.}
	\label{fig:CovIV}
\end{figure}

\paragraph{Coverage.} Figure \ref{fig:CovIV} reports the coverage of $95\%$ confidence intervals based on the variance estimator $\widehat{\boldsymbol{V}}_{\Sigma}(\widehat{W})$ in \eqref{Varest}. Under homogeneity, the coverage of the confidence interval based on equal weighting is close to its nominal level for all values of $G_N$. Under heterogeneity, this remains the case for $G_N \geq 40$. The same pattern holds for the inverse-variance and adaptive weights. 

Under homogeneity, however, the larger bias of the inverse-variance and adaptive-weight IV estimators reduces coverage to below $0.92$ for $G_N \geq 150$. Nevertheless, this still leaves a wide range of group sizes for which inference based on these weighted IV estimators performs reasonably well.

	
	\paragraph{Homogeneity Q test.} The simulation experiment considers a nominal level of $5\%$, for $G_N=5,10,50,100$. The $y$-axis gives power as a function of the slope heterogeneity standard deviation $\sigma_{\delta}$ in \eqref{IVmodel}. The empirical level, achieved for $\sigma_{\delta}=0$, is close to the nominal level. As expected, the power increases with $G_N$. 
	\begin{figure}[!htbp]
		\centering
		\includegraphics[scale=.37]{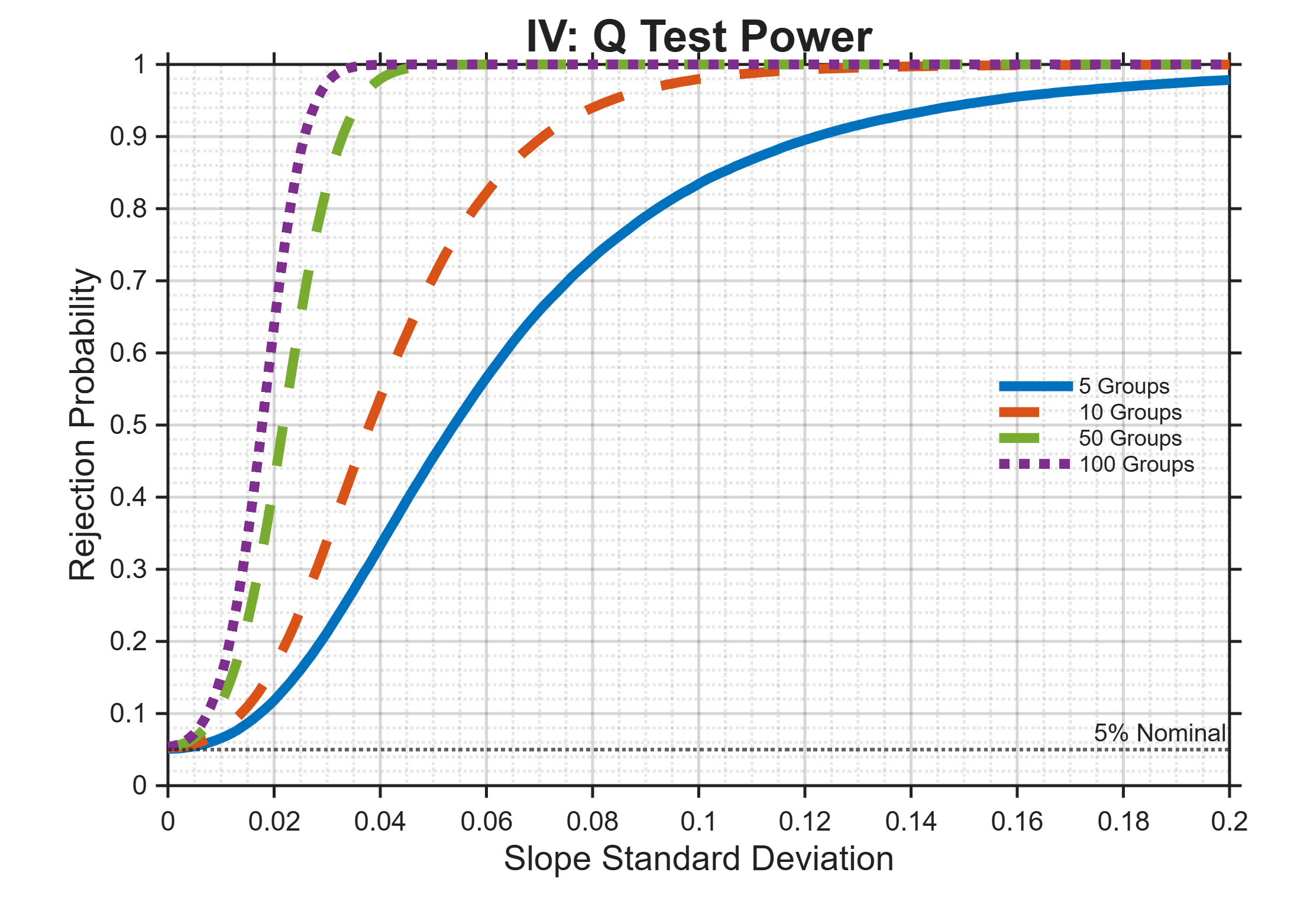}
		\caption{Homogeneity Q-test for the IV model \eqref{IVmodel}, nominal $5\%$ level.  Power on the $y$-axis, heterogeneity standard deviation $\sigma_{\delta}$ on the $x$-axis.}
		\label{fig:QIV}
	\end{figure}

\clearpage

\refstepcounter{section}
\section*{\centering Appendix \thesection: Proof section}
\label{app:proof-section}
\setcounter{subsection}{0}
\renewcommand{\thesubsection}{\thesection.\arabic{subsection}}
\renewcommand{\theHsubsection}{\thesection.\arabic{subsection}}
\renewcommand{\thetheorem}{B.\arabic{theorem}}
\setcounter{equation}{0}
\renewcommand{\theequation}{B.\arabic{equation}}
\renewcommand{\theHequation}{B.\arabic{equation}}

For the sake of brevity, abbreviate $\mathrm{Var} \left[ \mathrm{Vec} \left(\widehat{M} \right)\right]$ into $\mathrm{Var}  \left(\widehat{M} \right)$, and proceed similarly for higher-order multilinear form representations. For matrices $\widehat{M}=\widehat{M}_N$ or $M_N$ and a sequence $\{r_N, N \geq 1\}$ of real numbers or matrices, $\widehat{M} \asymp_{\mathbb{P}} r_N$ or $M_N \asymp r_N$ means that the eigenvalues of the concerned matrix have the same  exact order as the ones of $r_N$, in probability for $\widehat{M}$.
As in the main section, $C$ denotes a positive constant which can vary from line to line.

\subsection{Proof of Proposition \ref{Expansion}}
\label{app:proof-expansion}
Uniformly over $\mathcal{D}_{\kappa}$ holds by definition of $\mathcal{D}_{\kappa}$, because the proof of the Proposition only uses Markov-type inequalities.    It holds 
\begin{align*}
\widehat{\theta} \left(\widehat{W} \right)  = \theta +  \left[\sum_{g=1}^{G_N} \widehat{W}_g \left(\widehat{\theta}_g \right)\right]^{-1}\left(\widehat{E}_{He} + \widehat{E}_{\theta} + \widehat{B}\right) ,
\end{align*}
with:
\begin{align*}
\widehat{E}_{He} & =  \sum_{g=1}^{G_N} \widehat{W}_g \left(\widehat{\theta}_g \right) \delta_g ,
\\
\widehat{E}_{\theta}
&
 =  \sum_{g=1}^{G_N} \widehat{W}_g \left(\widehat{\theta}_g \right)
\widehat{\varepsilon}_g,
\quad
\widehat{\varepsilon}_g = \widehat{\theta}_g
-
\theta_g
-
\frac{B_{n_g,g} \left(\theta_g\right)}{n_g^b}
\\
\widehat{B}
&
=
\sum_{g=1}^{G_N} \widehat{W}_g \left(\widehat{\theta}_g \right)
\frac{B_{n_g,g}\left(\theta_g\right)}{n_g^b}
.
\end{align*}
We study each of these terms starting with the numerator $\sum_{g=1}^{G_N} \widehat{W}_g \left(\widehat{\theta}_g \right)$.

\paragraph{The numerator and bias term $\widehat{B}$.} A first-order Taylor expansion gives:
\begin{align*}
	\sum_{g=1}^{G_N} \widehat{W}_g \left(\widehat{\theta}_g \right)
	-
	\sum_{g=1}^{G_N} W_g  \left(\theta_g \right)
	& =
	\sum_{g=1}^{G_N} 
	\left[
	\widehat{W}_g \left(\theta_g \right)
	-
	W_g  \left(\theta_g \right)
	\right]
	+
	\sum_{g=1}^{G_N}
	\int_0^1
	\widehat{W}_g^{(1)} \left(\theta_g +t \left(\widehat{\theta}_g -\theta_g\right) \right) dt \left[\widehat{\theta}_g -\theta_g\right].
\end{align*}

It holds for the first  item, by Assumptions W and E,
\begin{align*}
	\mathbb{E}
	\left[
	\left\|
	\sum_{g=1}^{G_N} 
	\left[
	\widehat{W}_g \left(\theta_g \right)
	-
	W_g  \left(\theta_g \right)
	\right]
	\right\|
	\right]
	&
	\leq
	\sum_{g=1}^{G_N} 
	 \frac{1}{\left(n_g \|\Sigma\|+1\right)\sqrt{n_g}}
	\mathbb{E}^{\frac{1}{2}}
		\left[\left\|
	\left(n_g \|\Sigma\|+1\right) \sqrt{n_g}
	\left(
	\widehat{W}_g \left(\theta_g \right)
	-
	W_g  \left(\theta_g \right)
	\right)
	\right\|^2\right]
\\
& 
    =
    O
    \left[   
    \frac{\left(\frac{G_N}{N}\right)^{\frac{1}{2}}}{
    	\frac{N}{G_N} \|\Sigma\| +1
    }
    \sum_{g=1}^{G_N} 
    \left\| W_g \right\|
    \right],
\end{align*}
and for the last second term,
\begin{align*}
	&
	\sum_{g=1}^{G_N}
	\mathbb{E}
	\left[
	 \left\|
	\int_0^1
	\widehat{W}_g^{(1)} \left(\theta_g +t \left(\widehat{\theta}_g -\theta_g\right) \right) dt \left[\widehat{\theta}_g -\theta_g\right]
	\right\|
	\right]
	\\
	&
	\qquad
	\leq
	\sum_{g=1}^{G_N}
	\mathbb{E}^{\frac{1}{2}}
	\left[
	\sup_t
	\left\|
	\widehat{W}_g^{(1)} (t)
	\right\|^2
	\right]
	\mathbb{E}^{\frac{1}{2}}
	\left[
	\left\|
	\widehat{\theta}_g -\theta_g
	\right\|^2
	\right]
	=
	O
	\left[
	\frac{\left(\frac{G_N}{N}\right)^{\frac{1}{2}}}{
		\frac{N}{G_N} \|\Sigma\| +1
	}
	\sum_{g=1}^{G_N} 
	\left\| W_g \right\|
	\right].
\end{align*}
Hence, by the Markov  Inequality and since the population weights are well-conditioned, it holds uniformly over $\mathcal{D}_{\kappa}$,
\begin{align}
	\sum_{g=1}^{G_N} \widehat{W}_g \left(\widehat{\theta}_g \right)
	& =
	\left(
	1
	+
	O_{\mathbb{P}}
	\left[
	\left(\frac{G_N}{N}\right)^{\frac{1}{2}}
	\right]
	\right)
	\sum_{g=1}^{G_N} W_g  \left(\theta_g \right)
\label{Numerator}
\end{align}
Arguing similarly for the bias term gives, since $b>1/2$,
\begin{align}
	\widehat{B}
	=
	\sum_{g=1}^{G_N} 
	W_g (\theta_g)\frac{B_{n_g,g} (\theta_g)}{n_g^b}
	+
	O_{\mathbb{P}}
	\left[
	\frac{\left(\frac{G_N}{N}\right)^{\frac{1}{2}}}{
		\frac{N}{G_N} \|\Sigma\| +1
	}
	G_N \left(\frac{G_N}{N}\right)^b
	\right]
	=
	\sum_{g=1}^{G_N} 
	W_g (\theta_g)\frac{B_{n_g,g} (\theta_g)}{n_g^b}
	+
	o_{\mathbb{P}}
	\left(
	G_N r_{\theta}
	\right)
     .
\label{Bias}
\end{align}

\paragraph{The heterogeneity term $\widehat{E}_{He}$. } As above:
\begin{align*}
	\widehat{E}_{He}
	-
	\sum_{g=1}^{G_N} W_g  \left(\theta_g \right)\delta_g
	& =
	\sum_{g=1}^{G_N} 
	\left[
	\widehat{W}_g \left(\theta_g \right)
	-
	W_g  \left(\theta_g \right)
	\right]
	\delta_g
	\\
	&
	\quad
	+
	\sum_{g=1}^{G_N}
	\int_0^1
	\widehat{W}_g^{(1)} \left(\theta_g +t \left(\widehat{\theta}_g -\theta_g\right) \right) dt \left[\widehat{\theta}_g -\theta_g\right]\delta_g.
\end{align*}
Under Assumptions He and W, arguing as above,
\begin{align*}
	\mathbb{E}
	\left[
	\left\|
	\sum_{g=1}^{G_N} 
	\left[
	\widehat{W}_g \left(\theta_g \right)
	-
	W_g  \left(\theta_g \right)
	\right]
	\delta_g
	\right\|
	\right]
	&
	\leq
	\sum_{g=1}^{G_N} 
	\mathbb{E}^{\frac{1}{2}}
		\left[\left\|
		\widehat{W}_g \left(\theta_g \right)
		-
		W_g  \left(\theta_g \right)
		\right\|^2\right]
	\mathbb{E}^{\frac{1}{2}}
	\left[
	\left\|
	\delta_g
	\right\|^2
	\right]	
	\\
	& =
	O
	\left[
	\frac{\left(\frac{G_N}{N} \left\|\Sigma\right\|\right)^{\frac{1}{2}}}{
		\frac{N}{G_N} \|\Sigma\| +1
	}
	\sum_{g=1}^{G_N} 
	\left\| W_g \right\|
	\right],
\end{align*}
and
\begin{align*}
	&
	\sum_{g=1}^{G_N}
	\mathbb{E}
	\left[
	 \left\|
	\int_0^1
	\widehat{W}_g^{(1)} \left(\theta_g +t \left(\widehat{\theta}_g -\theta_g\right) \right) dt \left[\widehat{\theta}_g -\theta_g\right]
	\delta_g
	\right\|
	\right]
	\\
	&
	\qquad
	\leq
	\sum_{g=1}^{G_N}
\mathbb{E}^{\frac{1}{3}}	
\left[
\sup_t
\left\|
\widehat{W}^{(1)}_g (t)
\right\|^3
\right]
\mathbb{E}^{\frac{1}{3}}	
\left[
\left\|
\delta_g
\right\|^3
\right]
	\mathbb{E}^{\frac{1}{2}}
	\left[
	\left\|
	\widehat{\theta}_g -\theta_g
	\right\|^4
	\right]
	=
	O
	\left[
		\frac{\left(\frac{G_N}{N} \left\|\Sigma\right\|\right)^{\frac{1}{2}}}{
			\frac{N}{G_N} \|\Sigma\| +1
		}
	\sum_{g=1}^{G_N} 
	\left\| W_g \right\|
	\right].
\end{align*}
This gives, by the Markov Inequality, uniformly over $\mathcal{D}_{\kappa}$.
\begin{align}
\widehat{E}_{He}
=
\sum_{g=1}^{G_N} W_g  \left(\theta_g \right)\delta_g	
+
O_{\mathbb{P}}
\left[
\frac{\left(\frac{G_N}{N} \left\|\Sigma\right\|\right)^{\frac{1}{2}}}{
	\frac{N}{G_N} \|\Sigma\| +1
}
\sum_{g=1}^{G_N} 
\left\| W_g \right\|
\right].
\label{HatEhe}	
\end{align}

\paragraph{The estimation error $\widehat{E}_{\theta}$.}
Recall $\widehat{\varepsilon}_g = \widehat{\theta}_g
-
\theta_g
-
\frac{B_{n_g,g}}{n_g^b}$, which is such that
\begin{align*}
	\mathbb{E}^{\frac{1}{4}}
	\left[
	\left\|
	\widehat{\varepsilon}_g
	\right\|^4
	\right]
	\leq 
	C \cdot n_g^{-\frac{1}{2}},
	\text{ for all $g=1,\ldots,G_N$ and $N$,}
\end{align*}
under Assumption E.
 As above:
\begin{align*}
	\widehat{E}_{\theta}
	-
	\sum_{g=1}^{G_N} W_g  \left(\theta_g \right)\widehat{\varepsilon}_g
	& =
	\sum_{g=1}^{G_N} 
	\left[
	\widehat{W}_g \left(\theta_g \right)
	-
	W_g  \left(\theta_g \right)
	\right]
	\widehat{\varepsilon}_g
	\\
	&
	\quad
	+
	\sum_{g=1}^{G_N}
	\int_0^1
	\widehat{W}_g^{(1)} \left(\theta_g +t \left(\widehat{\theta}_g -\theta_g\right) \right) dt \left[\widehat{\theta}_g -\theta_g\right]\widehat{\varepsilon}_g.
\end{align*}
Under Assumptions E, He and W, arguing as above gives:
\begin{align*}
	\mathbb{E}
	\left[
	\left\|
	\sum_{g=1}^{G_N} 
	\left[
	\widehat{W}_g \left(\theta_g \right)
	-
	W_g  \left(\theta_g \right)
	\right]
	\widehat{\varepsilon}_g
	\right\|
	\right]
	&
	\leq
	\sum_{g=1}^{G_N} 
	\mathbb{E}^{\frac{1}{2}}
		\left[\left\|
		\widehat{W}_g \left(\theta_g \right)
		-
		W_g  \left(\theta_g \right)
		\right\|^2\right]
	\mathbb{E}^{\frac{1}{2}}
	\left[
	\left\|
	\widehat{\varepsilon}_g
	\right\|^2
	\right]	
	\\
	& =
	O
	\left[
\frac{\frac{G_N}{N}}{\frac{N}{G_N} \| \Sigma\| +1 }
	\sum_{g=1}^{G_N} 
	\left\| W_g \right\|
	\right],
\end{align*}
and
\begin{align*}
	&
	\sum_{g=1}^{G_N}
	\mathbb{E}
	\left[
	 \left\|
	\int_0^1
	\widehat{W}_g^{(1)} \left(\theta_g +t \left(\widehat{\theta}_g -\theta_g\right) \right) dt \left[\widehat{\theta}_g -\theta_g\right]
		\widehat{\varepsilon}_g
	\right\|
	\right]
	\\
	&
	\qquad
	\leq
	\sum_{g=1}^{G_N}
	\mathbb{E}^{\frac{1}{3}}	
	\left[
	\sup_t
	\left\|
	\widehat{W}^{(1)}_g (t)
	\right\|^3
	\right]
	\mathbb{E}^{\frac{1}{3}}	
	\left[
	\left\|
	\widehat{\varepsilon}_g
	\right\|^3
	\right]
	\mathbb{E}^{\frac{1}{3}}
	\left[
	\left\|
	\widehat{\theta}_g -\theta_g
	\right\|^3
	\right]
	=
	O
	\left[
	\frac{\frac{G_N}{N}}{\frac{N}{G_N} \| \Sigma\| +1 }
	\sum_{g=1}^{G_N} 
	\left\| W_g \right\|
	\right].
\end{align*}
This gives, by the Markov Inequality and uniformly over $\mathcal{D}_{\kappa}$,
\begin{align}
	\widehat{E}_{\theta}
	=
	\sum_{g=1}^{G_N} W_g  \left(\theta_g \right)\widehat{\varepsilon}_g	
	+
	O_{\mathbb{P}}
	\left[
\frac{\frac{G_N}{N}}{\frac{N}{G_N} \| \Sigma\| +1 }
	\sum_{g=1}^{G_N} 
	\left\| W_g \right\|
	\right].
	\label{HatEtheta}	
\end{align}

\paragraph{Conclusion.} Proposition \ref{Expansion} follows from \eqref{Numerator}, \eqref{Bias}, \eqref{HatEhe} and \eqref{HatEtheta}. \hfill $\Box$

\subsection{Proof of Proposition \ref{Randweightshet}}
\label{app:proof-randweightshet}

\paragraph{Expectation.} We first find a rate for
$\mathbb{E}
\left[
\sum_{g=1}^{G_N}
W_g \left(\theta_g\right) \left(\delta_g + \widehat{\varepsilon}_g\right)
\right]$. As
$\mathbb{E} \left[W_g \left(\theta_g\right) \widehat{\varepsilon}_g\right] = \mathbb{E} \left[W_g \left(\theta\right) \delta_g\right]=0$
by definition of $\widehat{\varepsilon}_g$, Assumptions W and He, it holds, 
\begin{align*}
	\left\|
	\mathbb{E}
	\left[
	\sum_{g=1}^{G_N}
	W_g \left(\theta_g\right) \left(\delta_g + \widehat{\varepsilon}_g\right)
	\right]
	\right\|
	& =
	\left\|
	\sum_{g=1}^{G_N}
	\mathbb{E}
	\left[W_g \left(\theta_g\right) \delta_g\right]
	\right\|
	\leq
	\sum_{g=1}^{G_N}
	\left\|
	\mathbb{E}
	\left[\left(W_g \left(\theta_g\right) -W_g \left(\theta\right) \right)\delta_g\right]
	\right\|
	.
\end{align*}
Assumption W then gives:
\begin{align*}
	\left\|
\mathbb{E}
\left[\left(W_g \left(\theta_g\right) -W_g \left(\theta\right) \right)\delta_g\right]
\right\|
& \leq
\frac{\mathbb{E}
\left[
\sup_{\tau \in \mathcal{T}}
\left\|\left(\frac{N}{G_N} \| \Sigma \| +1 \right)W_g^{(1)} \left(\tau\right) \right\|
\left\|\delta_g\right\|^2
\right]}{\frac{N}{G_N} \| \Sigma \| +1}
\leq \kappa
\frac{\| \Sigma\|}{\frac{N}{G_N} \| \Sigma\| +1},
\end{align*}
for all $1 \leq g \leq G_N$ and all $G_N$. Hence:
\begin{align*}
	\left\|
	\mathbb{E}
	\left[
	\sum_{g=1}^{G_N}
	W_g \left(\theta_g\right) \left(\delta_g + \widehat{\varepsilon}_g\right)
	\right]
	\right\|
	\leq 
	\kappa
	G_N
	\frac{\| \Sigma\|}{\frac{N}{G_N} \| \Sigma\| +1},
	\text{ uniformly over $\mathcal{D}_{\kappa}$.}
\end{align*}
which gives \eqref{CLT_esp}.

\paragraph{Variance.}
We now show that $\boldsymbol{V} \left(W\right)$ and $\widetilde{\boldsymbol{V}} \left(W\right)$ are close. Recall that, by Assumption E and W,
\begin{align*}
	\boldsymbol{V} \left(W\right)
	& =
	\mathrm{Var}
	\left[
	\sum_{g=1}^{G_N}
	W_g \left(\theta_g\right) \left(\delta_g + \widehat{\varepsilon}_g\right)
	\right]	
	=
	\sum_{g=1}^{G_N}
	\mathrm{Var}
	\left[W_g \left(\theta_g\right) \left(\delta_g + \widehat{\varepsilon}_g\right)\right]
	\\
	&
	=
	\boldsymbol{V}_{Het} \left(W\right) 
	- 
	\sum_{g=1}^{G_N}
	\mathbb{E}
	\left[W_g \left(\theta_g\right) \delta_g\right]
	\cdot
	\mathbb{E}^{\prime}
	\left[W_g \left(\theta_g\right) \delta_g\right]
	+ 
	\boldsymbol{V}_{Hom} \left(W\right)
	\text{ with: }
	\\
	&
	\quad
	\boldsymbol{V}_{Het} \left(W\right) 
	=
	\sum_{g=1}^{G_N}
	\mathbb{E}
	\left[W_g \left(\theta_g\right) \delta_g \delta_g^{\prime} W_g^{\prime} \left(\theta_g\right)\right]
	,
	\quad
	\boldsymbol{V}_{Hom} \left(W\right) 
	=
	\sum_{g=1}^{G_N}
	n_g^{-1}
	\mathbb{E}
	\left[W_g \left(\theta_g\right) \Omega_{n_g,g} \left(\theta_g \right) W_g^{\prime} \left(\theta_g\right)\right].
\end{align*} 
Arguing as for \eqref{CLT_Esp} gives, uniformly over $\mathcal{D}_{\kappa}$,
\begin{align*}
	\left\|
	\sum_{g=1}^{G_N}
	\mathbb{E}
	\left[W_g \left(\theta_g\right) \delta_g\right]
	\cdot
	\mathbb{E}^{\prime}
	\left[W_g \left(\theta_g\right) \delta_g\right]
	\right\|
	=
	O
	\left[
	\frac{G_N\left\|\Sigma\right\|^2}{\left(\frac{N}{G_N} \left\|\Sigma\right\|+1\right)^2}
	\right]
.
\end{align*}
For $\boldsymbol{V}_{Hom} \left(W\right)$, the Markov Inequality yields, uniformly over $\mathcal{D}_{\kappa}$ under Assumptions He, W and V,
\begin{align*}
	& \boldsymbol{V}_{Hom} \left(W\right) 
	=
	\sum_{g=1}^{G_N}
	n_g^{-1}
	W_g \left(\theta_g\right) \Omega_{n_g,g} \left(\theta_g \right) W_g^{\prime} \left(\theta_g\right)
	\\
	&\qquad
	+
	\sum_{g=1}^{G_N}
	n_g^{-1}
	\left(
	\mathbb{E}
	\left[W_g \left(\theta_g\right) \Omega_{n_g,g} \left(\theta_g \right) W_g^{\prime} \left(\theta_g\right)\right]
	-
	W_g \left(\theta_g\right) \Omega_{n_g,g} \left(\theta_g \right) W_g^{\prime} \left(\theta_g\right)
	\right)
	\\
	&
	\qquad
	=
	\sum_{g=1}^{G_N}
	n_g^{-1}
	W_g \left(\theta_g\right) \Omega_{n_g,g} \left(\theta_g \right) W_g^{\prime} \left(\theta_g\right)
	+
	O_{\mathbb{P}}
	\left(
	\frac{G_N \sqrt{G_N}}{N}
	\right)
	.
\end{align*}
For $\boldsymbol{V}_{Het} \left(W\right) $, note first that, uniformly in $g$ by Assumption W:
\begin{align*}
	\left\|
	\mathbb{E}
	\left[
	\left(W_g (\theta_g) - W_g (\theta)\right)
	\delta_g \delta_g^{\prime}
	W_g^{\prime} (\theta_g)
	\right]
	\right\|
	& \leq
	\kappa
	\sup_{\tau \in \mathcal{T}} 
	\left\|
	W_g^{(1)} (\tau)
	\right\|
	\mathbb{E}
	\left[
	\left\|\delta_g \right\|^3
	\right]
	\\
	&
	\leq
	C
	\kappa^2
	\frac{ \|\Sigma \|^{\frac{3}{2}}}{\frac{N}{G_N} \| \Sigma \| +1},
	\\
	\left\|
	\mathbb{E}
	\left[
	\left(W_g (\theta_g) - W_g (\theta)\right)
	\delta_g \delta_g^{\prime}
	\left(W_g (\theta_g) - W_g (\theta)\right)^{\prime} 
	\right]
	\right\|
	&\leq
	C
	\kappa^2
	\frac{ \|\Sigma \|^{2}}{\frac{N}{G_N} \| \Sigma \| +1}.
\end{align*}
As
\begin{align*}
	&\boldsymbol{V}_{Het} \left(W\right) =
	\sum_{g=1}^{G_N} W_g (\theta) \Sigma W_g^{\prime} (\theta)
	\\
	&
	\quad
	+
	 \sum_{g=1}^{G_N}
	 \left[
	 \mathbb{E}
	 \left[
	 \left(W_g (\theta_g) - W_g (\theta)\right)
	 \delta_g \delta_g^{\prime}
	 W_g^{\prime} (\theta_g)
	 \right]
	 +
	 \mathbb{E}
	 \left[
	 W_g (\theta_g)
	 \delta_g \delta_g^{\prime}
	 \left(W_g (\theta_g) - W_g (\theta)\right)^{\prime} 
	 \right]
	 \right]
	 \\
	 &
	 \quad
	 +
	 \sum_{g=1}^{G_N}
	 \mathbb{E}
	 \left[
	 \left(W_g (\theta_g) - W_g (\theta)\right)
	 \delta_g \delta_g^{\prime}
	 \left(W_g (\theta_g) - W_g (\theta)\right)^{\prime} 
	 \right],
\end{align*}
it holds, uniformly over $\mathcal{D}_{\kappa}$,
\begin{align*}
\boldsymbol{V}_{Het} \left(W\right) =
\sum_{g=1}^{G_N} W_g (\theta) \Sigma W_g^{\prime} (\theta)	
+
G_N
O
\left(
\frac{ \|\Sigma \|^{\frac{3}{2}}+ \|\Sigma \|^{2}}{\frac{N}{G_N} \| \Sigma \| +1}
\right).
\end{align*}
Replacing the $W_g (\theta)$ by $W_g (\theta_g)$ and arguing as above with the Markov inequality yields:
\begin{align*}
	\boldsymbol{V}_{Het} \left(W\right) =
	\sum_{g=1}^{G_N} W_g (\theta_g) \Sigma W_g^{\prime} (\theta_g)	
	+
	G_N
	O_{\mathbb{P}}
	\left(
	\frac{ \|\Sigma \|^{\frac{3}{2}}+ \|\Sigma \|^{2}}{\frac{N}{G_N} \| \Sigma \| +1}
	\right).
\end{align*}
Hence, uniformly over $\mathcal{D}_{\kappa}$,
\begin{align*}
	\boldsymbol{V} \left(W\right) =
	\widetilde{\boldsymbol{V}} \left(W\right) 
	+
	G_N
	O_{\mathbb{P}}
	\left(
	\frac{\sqrt{G_N}}{N}
	+
	\frac{ \|\Sigma \|^{\frac{3}{2}}\left(1+ \|\Sigma \|^{\frac{1}{2}}\right)}{\frac{N}{G_N} \| \Sigma \| +1}
	\right).
\end{align*}
Recall also that, under Assumption W,
\begin{align*}
	\widetilde{\boldsymbol{V}}\left(W\right) 
	\asymp 
	G_N \left(\left\| \Sigma \right\| + \frac{G_N}{N} \right) 
	=
	\frac{G_N^2}{N}
	\left(\frac{N}{G_N} \left\| \Sigma \right\| + 1 \right).
\end{align*}
Combining these last results gives, uniformly over $\mathcal{D}_{\kappa}$,
\begin{align*}
	\widetilde{\boldsymbol{V}}\left(W\right)^{-\frac{1}{2}}
	\left(
	\boldsymbol{V}\left(W\right) 
	-
	\widetilde{\boldsymbol{V}}\left(W\right) 
	\right)
	\widetilde{\boldsymbol{V}}\left(W\right)^{-\frac{1}{2}}
	& =
	O_{\mathbb{P}}
	\left(\frac{1}{\sqrt{G_N}}\right)
	+
	\frac{N}{G_N}
	O_{\mathbb{P}}
	\left(
	\left(\frac{G_N}{N}\right)^{\frac{3}{2}}
	\frac{\left(\frac{N}{G_N} \|\Sigma \|\right)^{\frac{3}{2}}\left(1+ \frac{N}{G_N}\|\Sigma \|\right)^{\frac{1}{2}}}{\left(\frac{N}{G_N} \| \Sigma \| +1\right)^2}
	\right)
	\\
	& =
	O_{\mathbb{P}}
	\left(\frac{1}{\sqrt{G_N}}
	+
	\left(\frac{G_N}{N}\right)^{\frac{1}{2}}
	\right),
\end{align*}
which is \eqref{Vareq}.

It then follows that:
\begin{align*}
	\boldsymbol{V}\left(W\right) 
	\asymp
	\frac{G_N^2}{N}
	\left(\frac{N}{G_N} \left\| \Sigma \right\| + 1 \right),
\end{align*}
so that \eqref{CLT_esp} yields:
\begin{align*}
	\boldsymbol{V}^{-\frac{1}{2}}\left(W\right) 
	\mathbb{E}
	\left[
	\sum_{g=1}^{G_N}
	W_g \left(\theta_g\right) \left(\delta_g + \widehat{\varepsilon}_g\right)
	\right]
	=
	O_{\mathbb{P}}
\left[	
	\sqrt{N}
	\frac{\| \Sigma\|}{\left(\frac{N}{G_N} \| \Sigma\| +1\right)^{\frac{3}{2}}}\right]
	=
	O_{\mathbb{P}}
	\left[	
	\frac{\sqrt{\frac{G_N^2}{N}}}{\left(\frac{N}{G_N} \| \Sigma\| +1\right)^{\frac{1}{2}}}\right],
\end{align*}
which is \eqref{CLT_Esp}. \hfill $\Box$
\subsection{Proof of Proposition \ref{Varconsistency}}
\label{app:proof-varconsistency}

Again uniformity follows from the systematic use of Markov-type inequalities, so that all orders stated in the proof hold uniformly over $\mathcal{D}_{\kappa}$.  
We start with \eqref{Checksigwrtng} and then study  $\widetilde{\boldsymbol{V}}^{-1/2} \left(W\right)\widehat{\boldsymbol{V}}_{\Sigma} \left(\widehat{W}\right)\widetilde{\boldsymbol{V}}^{-1/2} \left(W\right)$.

\subparagraph{The variance $\widehat{\Sigma}$.} 
Let  $\underline{v} = \min \left(v,\frac{1}{2}\right)$. 
Note  that:
\begin{align*}
	\overline{\widehat{\theta}}
	=
	\theta
	+
	O_{\mathbb{P}}
	\left[
	\frac{\left\|\Sigma\right\|^{\frac{1}{2}}}{\sqrt{G_N}}
	+
	\frac{1}{\sqrt{N}}
	+
	\left( \frac{G_N}{N}\right)^b\right].
\end{align*}

Since $\theta_g-\theta = \delta_g$,  it holds, recalling $\widehat{\varepsilon}_g = \widehat{\theta}_g - \theta_g - B_{n_g,g}/n_g^b$,
\begin{align*}
	\widehat{\Sigma} 
	&
	=
	\widehat{\Sigma}_{He} 
	+
	\widehat{\Sigma}_{Ho}
	+
	\widehat{\Sigma}_{B,He} 
	+
	\widehat{\Sigma}_{B,He} ^{\prime}
	\\
	&
	\quad
	+
	\widehat{\Sigma}_{He,1}+ \widehat{\Sigma}_{He,1}^{\prime}
	-
	\widehat{\Sigma}_{He,2}- \widehat{\Sigma}_{He,2}^{\prime}
	-
	\widehat{\Sigma}_{3}-\widehat{\Sigma}_{3}^{\prime}
	+\widehat{\Sigma}_{4}
	+
	\widehat{\Sigma}_{5}+\widehat{\Sigma}_{5}^{\prime}
	+\widehat{\Sigma}_{6}
	\text{ where,}
	\\
	&
	\qquad
	\widehat{\Sigma}_{He} 
	=
	\frac{1}{G_N}
	\sum_{g=1}^{G_N}
	\delta_g \delta_g^{\prime},
	\quad
	\widehat{\Sigma}_{Ho} 
	=
	\frac{1}{G_N}
	\sum_{g=1}^{G_N}
	\widehat{\varepsilon}_g \widehat{\varepsilon}_g^{\prime},
	\quad
	\widehat{\Sigma}_{B,He}
	=
	\frac{1}{G_N}
	\sum_{g=1}^{G_N}
	\frac{B_{n_g,g}  \left(\theta_g\right)\delta_g^{\prime}}{n_g^b},
	\\
	&	
	\qquad
	\widehat{\Sigma}_{He,1}
	=
	\frac{1}{G_N}
	\sum_{g=1}^{G_N}
	\delta_g \widehat{\varepsilon}_g^{\prime},
	\quad
	\widehat{\Sigma}_{He,2}
	=
	\left[\frac{1}{G_N}
	\sum_{g=1}^{G_N}
	\delta_g\right] 
	\left(\overline{\widehat{\theta}}-\theta\right)^{\prime},
	\\
	&
	\qquad
	\widehat{\Sigma}_{3}
	=
	\left[\frac{1}{G_N}
	\sum_{g=1}^{G_N}
	\left(\widehat{\theta}_g-\theta_g\right)\right]
	\left(\overline{\widehat{\theta}}-\theta\right)^{\prime},
	\quad
	\widehat{\Sigma}_{4} =\left(\overline{\widehat{\theta}}-\theta\right) \left(\overline{\widehat{\theta}}-\theta\right)^{\prime},
	\\
	&
	\qquad
	\widehat{\Sigma}_{5}
	=
	\frac{1}{G_N}
	\sum_{g=1}^{G_N}
	\frac{B_{n_g,g}}{n_g^b}
	\widehat{\varepsilon}_g^{\prime},
	\quad
	\widehat{\Sigma}_{6}
	=
	\frac{1}{G_N}
	\sum_{g=1}^{G_N}
	\frac{B_{n_g,g}B_{n_g,g}^{\prime}}{n_g^{2b}}.
\end{align*}

We now aim for the order of the remainder terms $\widehat{\Sigma}_{He,k}$, $k=1,2$ and $\widehat{\Sigma}_{k}$, $k=3,4$.  Note that, under Assumptions He and  E, and for any conformable $u$, $v$ with unit norms,
\begin{align*}
	& 
	\mathbb{E}
	\left[
	\delta_g \widehat{\varepsilon}_g^{\prime}
	\right]
	=
	\mathbb{E}
	\left[
	\delta_g \mathbb{E}\left[\left.\widehat{\varepsilon}_g^{\prime}\right| \theta_g \right]
	\right]=0
	,
	\\
	& 
	\sup_g
	\left\|\mathrm{Var} \left(u' \delta_g \widehat{\varepsilon}_g^{\prime} v\right)\right\|
	\leq
	\left\|\Sigma u \right\| \| v \|
	\sup_g
	\left\{\mathbb{E}^{\frac{1}{2}}
	\left[
	\left\| \Sigma^{-1/2 }\delta_g \right\|^4
	\right]
	\mathbb{E}^{\frac{1}{2}}
	\left[
	\left\|n_{g}^{\frac{1}{2}}\widehat{\varepsilon}_g\right\|^4
	\right]\right\}
	\leq
	C \| \Sigma\|
	.
\end{align*}
Hence, applying the Chebyshev Inequality to each entry of the matrix below gives:
\begin{align*}
	\widehat{\Sigma}_{He,1}
	=
	\frac{1}{G_N}
	\sum_{g=1}^{G_N}
	\delta_g \widehat{\varepsilon}_g^{\prime}
	= 
	\left\|\Sigma \right\|^{\frac{1}{2}}
	O_{\mathbb{P}} 
	\left[
	\left(
	\frac{1}{G_N^2}
	\sum_{g=1}^{G_N}
	\frac{1}{n_g}
	\right)^{\frac{1}{2}}
	\right]
	=
 O_{\mathbb{P}} \left(\frac{	\left\|\Sigma \right\|^{\frac{1}{2}}}{\sqrt{N}}\right).
\end{align*} 
Hence, under Assumption E and He, and using $2ab \leq a^2+b^2$,\footnote{Since $a^2+b^2-2ab = (a-b)^2 \geq 0$},
 $(a+b)^2 \leq 2(a^2+b^2)$,
\begin{align*}
	& 
	\widehat{\Sigma}_{He,2}
	=
	O_{\mathbb{P}}
	\left[
	\frac{\left\|\Sigma \right\|^{\frac{1}{2}}}{\sqrt{G_N}}
	\left(
	\frac{\left\|\Sigma \right\|^{\frac{1}{2}}}{\sqrt{G_N}}
	+
	\frac{1}{\sqrt{N}}
	+
	\left(\frac{G_N}{N}\right)^b
	\right)
	\right]
	=
	O_{\mathbb{P}}
	\left( 
	\frac{\left\|\Sigma \right\|}{G_N} + \frac{1}{N} + \left(\frac{G_N}{N}\right)^{2b}
	\right)
	,
	\\
	& 
	\widehat{\Sigma}_{3}
	=
	O_{\mathbb{P}}
	\left[
	\left(
	\frac{1}{\sqrt{N}}
	+
	\left(\frac{G_N}{N}\right)^b
	\right)
	\left(
	\frac{\left\|\Sigma\right\|^{\frac{1}{2}}}{\sqrt{G_N}}
	+
	\frac{1}{\sqrt{N}}
	+
	\left(\frac{G_N}{N}\right)^b
	\right)
	\right]
	=
	O_{\mathbb{P}}
	\left( 
	\frac{\left\|\Sigma \right\|}{G_N} + \frac{1}{N} + \left(\frac{G_N}{N}\right)^{2b}
	\right)
	,
	\\
	&
	\widehat{\Sigma}_{4}
	=
	O_{\mathbb{P}}
	\left[
	\left(
	\frac{\left\|\Sigma\right\|^{\frac{1}{2}}}{\sqrt{G_N}}
	+
	\frac{1}{\sqrt{N}}
	+
	\left(\frac{G_N}{N}\right)^b
	\right)^2
	\right]
	=
	O_{\mathbb{P}}
	\left( 
	\frac{\left\|\Sigma \right\|}{G_N} + \frac{1}{N} + \left(\frac{G_N}{N}\right)^{2b}
	\right)
	,
	\\
	& 
	\widehat{\Sigma}_{5}
	=
	O_{\mathbb{P}}
	\left[
	\left(
	\frac{1}{G_N^2}
	\sum_{g=1}^{G_N}
	\left(\frac{G_N}{N}\right)^{2b+1}
	\right)^{\frac{1}{2}}
	\right]
	=
	O_{\mathbb{P}}
	\left(\frac{G_N^{b}}{N^{b+\frac{1}{2}}}\right)
	,
	\quad
	\widehat{\Sigma}_{6}
	=
	O_{\mathbb{P}}
	\left[
	\left(\frac{G_N}{N}\right)^{2b}
	\right].
\end{align*} 
Hence, since $2 \frac{G_N^{b}}{N^{b+\frac{1}{2}}} \leq \frac{1}{N} + \left(\frac{G_N}{N}\right)^{2b}$,
\begin{align*}
	\widehat{\Sigma}
	& =
	\frac{1}{G_N}
	\sum_{g=1}^{G_N}
	\delta_g \delta_g^{\prime}
	+
	\frac{1}{G_N}
	\sum_{g=1}^{G_N}
	\widehat{\varepsilon}_g \widehat{\varepsilon}_g^{\prime}
	+
	\frac{1}{G_N}
	\sum_{g=1}^{G_N}
	\frac{B_{n_g,g}  \left(\theta_g\right)\delta_g^{\prime}}{n_g^b}
	\\
	& \quad
	+
	O_{\mathbb{P}}
	\left(
	\frac{	\left\|\Sigma \right\|^{\frac{1}{2}}}{\sqrt{N}}
	+
	\frac{	\left\|\Sigma \right\|}{G_N}
	+
	\frac{1}{N}
	+
	\left(\frac{G_N}{N}\right)^{2b}
\right)
	.
\end{align*}

\subparagraph{The expansion  \eqref{Checksigwrtng}  of  $\check{\Sigma}$.} Recall first $\widehat{\Sigma}_{w}= G_N^{-1}  \sum_{g=1}^{G_N} n_g^{-1} \cdot \widehat{\Omega}_g \left(\widehat{\theta}_g\right)$, $\Sigma_w= G_N^{-1}  \sum_{g=1}^{G_N} n_g^{-1} \cdot \Omega_g$, so that
$\check{\Sigma} = \widehat{\Sigma} - \widehat{\Sigma}_{w}$. Assumption V  and the Taylor formula with integral remainder yield:
\begin{align*}
	\widehat{\Sigma}_{w}
	& =
	\Sigma_{w}+ \widehat{\omega}_1 + \widehat{\omega}_2 \text{ with }
	\\
	&
	\quad
	\widehat{\omega}_1 
	=
	\frac{1}{G_N}\sum_{g=1}^{G_N} n_g^{-1} \cdot\left[\widehat{\Omega}_g \left(\theta_g\right) - \Omega_g  \left(\theta_g\right)\right],
	\\
	&
	\quad
	\widehat{\omega}_2
	=
	\frac{1}{G_N}\sum_{g=1}^{G_N} n_g^{-1} \cdot 
	\int_0^1
	\widehat{\Omega}_g^{(1)} 
	\left(
	\theta_g 
	+
	t
	\left(\widehat{\theta}_g -\theta_g \right)
	\right)dt
	\left[\widehat{\theta}_g -\theta_g\right].
\end{align*}
The Markov Inequality, Assumptions V and E then give:
\begin{align*}
	\widehat{\omega}_1
	&= 
	\mathbb{E}\left[ \widehat{\omega}_1 \right]
	+
	O_{\mathbb{P}}
	\left(
	\mathrm{Var}^{1/2} \left( \widehat{\omega}_1 \right)
	\right)
	=
	O \left(\frac{1}{G_N} \sum_{g=1}^{G_N} \frac{1}{n_g^{1+\frac{1}{2}}}\right)
	+
	O_{\mathbb{P}}
	\left[
	\left(
	\frac{1}{G_N^2}
	\sum_{g=1}^{G_N} \frac{1}{n_g^3}
	\right)^{\frac{1}{2}}
	\right]
	\\
	&
	=
	O_{\mathbb{P}}
	\left[
	\left(\frac{G_N}{N}\right)^{\frac{3}{2}}
	\right],
	\\
	\left\| \widehat{\omega}_2 \right\|
	&
	\leq
	\frac{1}{G_N}
	\sum_{g=1}^{G_N}
	\frac{\sup_t
		\left\|
		\widehat{\Omega}^{(1)}
		(t)
		\right\|}{n_g}
	\left\|
	\widehat{\theta}_g-\theta_g
	\right\|
	=
	O_{\mathbb{P}}
	\left(
	\frac{1}{G_N}
	\sum_{g=1}^{G_N}\frac{1}{n_g}
	\mathbb{E}^{\frac{1}{2}}
	\left[
	\left(
     \sup_t
		\left\|
		\widehat{\Omega}^{(1)}_{g}
		(t)
		\right\|
	\right)^2
	\right]
	\mathbb{E}^{\frac{1}{2}}
	\left[
	\left\|
	\widehat{\theta}_g-\theta_g
	\right\|^2
	\right]
	\right)
	\\
	&
	=
	O_{\mathbb{P}}
	\left[\left(\frac{G_N}{N} \right)^{\frac{3}{2}}\right]
	.
\end{align*}
Hence:
\begin{align*}
	\widehat{\Sigma}_{w}
	=
	\Sigma_{w}+
	O_{\mathbb{P}}
	\left[
	\left(\frac{G_N}{N}\right)^{\frac{3}{2}}
	\right].
\end{align*}
Combining this with the last approximation for $\widehat{\Sigma}$ of the previous paragraph then gives, 
\begin{align}
	\check{\Sigma}
	& =
	\frac{1}{G_N}
	\sum_{g=1}^{G_N}
	\delta_g \delta_g^{\prime}
	+
	\frac{1}{G_N}
	\sum_{g=1}^{G_N}
	\frac{B_{n_g,g}  \left(\theta_g\right)\delta_g^{\prime}}{n_g^b}
	+
	\frac{1}{G_N}
	\sum_{g=1}^{G_N}
	\left(\widehat{\varepsilon}_g \widehat{\varepsilon}_g^{\prime}-\frac{\Omega_{n_g,g}}{n_g}\right)
	+
	\frac{1}{G_N}
	\sum_{g=1}^{G_N}
	\frac{\Omega_{n_g,g}-\Omega_{g}}{n_g}
	\nonumber
	\\
	&
	\qquad
	+
	O_{\mathbb{P}}
	\left(
	\frac{	\left\|\Sigma \right\|^{\frac{1}{2}}}{\sqrt{N}}
	+
	\frac{	\left\|\Sigma \right\|}{G_N}
	+
	\frac{1}{N}
	+
	\left(\frac{G_N}{N}\right)^{2b}
	+
	\left(
	\frac{G_N}{N}
	\right)^\frac{3}{2}
	\right)
.
	\label{Checksig2heho}
\end{align}
Now Assumption E and the Chebyshev Inequality yield that:
\begin{align*}
	\frac{1}{G_N}
	\sum_{g=1}^{G_N}
	\left(\widehat{\varepsilon}_g \widehat{\varepsilon}_g^{\prime}-\frac{\Omega_{n_g,g}}{n_g}\right)
	& =
	O_{\mathbb{P}}
	\left[
	\left(
	\frac{1}{G_N^2}
	\sum_{g=1}^{G_N}
	\frac{1}{n_g^2}
	\right)^{\frac{1}{2}}
	\right]
	=
	O_{\mathbb{P}}
	\left(
	\frac{G_N^{\frac{1}{2}}}{N}
	\right),
	\\
	\frac{1}{G_N}
	\sum_{g=1}^{G_N}
	\frac{\Omega_{n_g,g}-\Omega_{g}}{n_g}
	& = O \left[\left( \frac{G_N}{N} \right)^{1+v}\right],
	\\
	\frac{1}{G_N}
	\sum_{g=1}^{G_N}
	\frac{B_{n_g,g}  \left(\theta_g\right)\delta_g^{\prime}}{n_g^b}
	&
	=
	\frac{1}{G_N}
	\sum_{g=1}^{G_N}
	\frac{\mathbb{E}\left[B_{n_g,g}  \left(\theta_g\right)\delta_g^{\prime}\right]}{n_g^b}
	+
	O_{\mathbb{P}}
	\left[\frac{\left\|\Sigma\right\|^{\frac{1}{2}}}{\sqrt{G_N}}
	\left(\frac{G_N}{N}\right)^{b}
	\right]
	\\
	& =
	O_{\mathbb{P}}
	\left[
	\left\|\Sigma\right\|^{\frac{1}{2}}
	\left(\frac{G_N}{N}\right)^{b}
	\right]
	.
\end{align*}
Substituting in \eqref{Checksig2heho} gives, since $\underline{v} = \min (v,1/2)$,
\begin{align*}
	\check{\Sigma}
	& =
	\Sigma
	\left(
	1
	+
	O_{\mathbb{P}}
	\left(
	\frac{1}{\sqrt{G_N}}
	\right)
	\right)
    +
    \left\|\Sigma\right\|^{\frac{1}{2}}
    O_{\mathbb{P}} 
    \left[
    \frac{1}{\sqrt{N}}
    +
    \left(\frac{G_N}{N}\right)^{b}
    \right]
	+
	O_{\mathbb{P}} \left[\left( \frac{G_N}{N} \right)^{1+\underline{v}}+ 
	\left(\frac{G_N}{N}\right)^{2b}
	+
	\frac{\sqrt{G_N}}{N} 
	\right]
	,
\end{align*}
which implies \eqref{Checksigwrtng}.

\subparagraph{Notations for $\widehat{\boldsymbol{V}}_{\Sigma}\left(\widehat{W}\right)$.}
Define
\begin{align*}
	V_{He} = \sum_{g=1}^{G_N} W_g \left(\theta_g\right) \Sigma W_g^{\prime}\left(\theta_g\right),
	\qquad
	V_{Ho}
	=
	\sum_{g=1}^{G_N} n_g^{-1} \cdot W_g \left(\theta_g\right) \Omega_{n_g,g} \left(\theta_g \right)W_g^{\prime}\left(\theta_g\right),
\end{align*}
noting that $\widetilde{\boldsymbol{V}} (W)$ in \eqref{Var} satisfies $\widetilde{\boldsymbol{V}} (W) = V_{He}+V_{Ho}$.
Under Assumptions W, E, $V_{He} \asymp  G_N$ under heterogeneity while, for balanced designs and $G_N = o (N)$,
\begin{align*}
	V_{Ho} \asymp \frac{G_N}{N}  \sum_{g=1}^{G_N}  W_g \left(\theta_g\right) \Omega_{n_g,g}\left(\theta_g\right) W_g^{\prime}\left(\theta_g\right) \asymp \frac{G_N^2}{N}=o(G_N).
\end{align*}

\subparagraph{Decomposition of $\widehat{\boldsymbol{V}}_{\Sigma} \left(\widehat{W}\right)$.} From \eqref{Varest},
$\widehat{\boldsymbol{V}}_{\Sigma} \left(\widehat{W}\right) =\widehat{V}^{\Sigma}_{He} + \widehat{V}^{\Sigma}_{Ho}$ with
\begin{align*}
	\widehat{V}^{\Sigma}_{He} = \sum_{g=1}^{G_N} \widehat{W}_g \left(\widehat{\theta}_g\right) \check{\Sigma} \widehat{W}_g^{\prime}\left(\widehat{\theta}_g\right),
	\qquad
	\widehat{V}^{\Sigma}_{Ho}
	=
	\sum_{g=1}^{G_N} n_g^{-1} \cdot \widehat{W}_g \left(\widehat{\theta}_g\right)  \widehat{\Omega}_g \left(\widehat{\theta}_g\right)\widehat{W}_g^{\prime}\left(\widehat{\theta}_g\right).
\end{align*} 
It holds, for $\widehat{V}^{\Sigma}_{He}$, using
$\max_{1 \leq g \leq G_N} \left\| W_g(\theta_g) \right\| = O(1)$
and the H\"older inequality:
\begin{align*}
	\widehat{V}^{\Sigma}_{He}
	& = 
	V_{He}
	+
	\widehat{V}^{\Sigma}_{He,1}+\widehat{V}^{\Sigma\prime}_{He,1}
	+
	\widehat{V}^{\Sigma}_{He,2}+\widehat{V}^{\Sigma\prime}_{He,2}
	+
	\widehat{V}^{\Sigma}_{He,3}+\widehat{V}^{\Sigma\prime}_{He,3}
	+
	\widehat{V}^{\Sigma}_{He,4}
	+
	\widehat{V}^{\Sigma}_{He,5},
	\\
	\text{with }
	&
	\widehat{V}^{\Sigma}_{He,1}
	=
	\sum_{g=1}^{G_N} W_g \left(\theta_g\right) \Sigma 
	\left[
	\widehat{W}_g\left(\widehat{\theta}_g\right)
	-
	W_g\left(\theta_g\right)
	\right]^{\prime},
	=
	O\left( \left\|\Sigma\right\|\right)
	G_N^{\frac{2}{3}}
	\left[
	\sum_{g=1}^{G_N}
	\left\|
	\widehat{W}_g\left(\widehat{\theta}_g\right)
	-
	W_g\left(\theta_g\right)
	\right\|^3
	\right]^{\frac{1}{3}},
	\\
	&
	\widehat{V}^{\Sigma}_{He,2}
	=
	\sum_{g=1}^{G_N} W_g \left(\theta_g\right)
	\left[\check{\Sigma}-\Sigma\right] 
	W_g^{\prime}\left(\theta_g\right)
	=
	O \left(G_N\right)
	\left\|\check{\Sigma}-\Sigma\right\|
	\\
	&
	\widehat{V}^{\Sigma}_{He,3}
	=
	\sum_{g=1}^{G_N} W_g \left(\theta_g\right)
	\left[\check{\Sigma}-\Sigma\right] 
	\left[
	\widehat{W}_g\left(\widehat{\theta}_g\right)
	-
	W_g\left(\theta_g\right)
	\right]^{\prime}
	\\
	&
	\qquad \qquad
	=
	O \left(1\right)
	G_N^{\frac{2}{3}}
	\left\|\check{\Sigma}-\Sigma\right\|
	\left[
	\sum_{g=1}^{G_N}
	\left\|
	\widehat{W}_g\left(\widehat{\theta}_g\right)
	-
	W_g\left(\theta_g\right)
	\right\|^3
	\right]^{\frac{1}{3}},
	\\
	& 
	\widehat{V}^{\Sigma}_{He,4}
	=
	\sum_{g=1}^{G_N} 
	\left[
	\widehat{W}_g\left(\widehat{\theta}_g\right)
	-
	W_g\left(\theta_g\right)
	\right] 
	\Sigma 
	\left[
	\widehat{W}_g\left(\widehat{\theta}_g\right)
	-
	W_g\left(\theta_g\right)
	\right]^{\prime}
	\\
	&
	\qquad \qquad
	=
	O\left( \left\|\Sigma\right\|\right)
	G_N^{\frac{1}{3}}
	\left[
	\sum_{g=1}^{G_N}
	\left\|
	\widehat{W}_g\left(\widehat{\theta}_g\right)
	-
	W_g\left(\theta_g\right)
	\right\|^3
	\right]^{\frac{2}{3}},
	\\
	&
	\widehat{V}^{\Sigma}_{He,5}
	=
	\sum_{g=1}^{G_N} 
	\left[
	\widehat{W}_g\left(\widehat{\theta}_g\right)
	-
	W_g\left(\theta_g\right)
	\right] 
	\left[\check{\Sigma}-\Sigma\right] 
	\left[
	\widehat{W}_g\left(\widehat{\theta}_g\right)
	-
	W_g\left(\theta_g\right)
	\right]^{\prime},
	\\
	&
	\qquad\qquad
      =
	O(1)
	\left\|\check{\Sigma}-\Sigma\right\|
	G_N^{\frac{1}{3}}
	\left[
	\sum_{g=1}^{G_N}
	\left\|
	\widehat{W}_g\left(\widehat{\theta}_g\right)
	-
	W_g\left(\theta_g\right)
	\right\|^3
	\right]^{\frac{2}{3}}.
\end{align*}
As, by the Markov Inequality,
\begin{align*}
	& \sum_{g=1}^{G_N} \left\|\widehat{W}_g \left(\widehat{\theta}_g\right)-W_g \left(\theta_g\right)\right\|^3 
	\\
	& \quad \leq
	4
	\sum_{g=1}^{G_N} \left\|\widehat{W}_g \left(\theta_g\right)-W_g \left(\theta_g\right)\right\|^3 
	+
	\left(\sum_{g=1}^{G_N}\sup_{\tau} \left\|\widehat{W}_g^{(1)} \left(\tau \right)\right\|^6\right)^{\frac{1}{2}}
	\left(\sum_{g=1}^{G_N} \left\|\widehat{\theta}_g - \theta_g\right\|^6\right)^{\frac{1}{2}}
	\\
	&
	\quad
	=
	O_{\mathbb{P}}
	\left(
	\sum_{g=1}^{G_N} 
	\mathbb{E}
	\left[\left\|\widehat{W}_g \left(\theta_g\right)-W_g \left(\theta_g\right)\right\|^3 \right]
	+
	\left(\sum_{g=1}^{G_N} \mathbb{E} \left[\sup_{\tau} \left\|\widehat{W}_g^{(1)} \left(\tau \right)\right\|^6\right]\right)^{\frac{1}{2}}
	\left(\sum_{g=1}^{G_N} \mathbb{E}\left[\left\|\widehat{\theta}_g - \theta_g\right\|^6\right]\right)^{\frac{1}{2}}
	\right),
\end{align*}
Assumptions E and W give, since the weights are asymptotically bounded,
\begin{align}
	\sum_{g=1}^{G_N} \left\|\widehat{W}_g \left(\widehat{\theta}_g\right)-W_g \left(\theta_g\right)\right\|^3
	=
	G_N
	O_{\mathbb{P}}
	\left( 
	\left(\frac{G_N}{N}\right)^{\frac{3}{2}}
	\right).
	\label{DW3}
\end{align}
It then follows from the bounds of the items of the decomposition of $\widehat{V}_{He}^{\Sigma}$ and \eqref{Checksigwrtng}:
\begin{align}
	\widehat{V}_{He}^{\Sigma}
	-
	V_{He}
	&=
	G_N
	\left\|\Sigma\right\|
	O_{\mathbb{P}}
	\left[
	\left(\frac{G_N}{N}\right)^{\frac{1}{2}}
	+
	\frac{1}{\sqrt{G_N}}
	\right]
	+
	G_N
	\left\|\Sigma\right\|^{\frac{1}{2}}
	O_{\mathbb{P}}
	\left[
	\frac{1}{\sqrt{N}}
	+
	\left(\frac{G_N}{N}\right)^{b}
	\right]
	\nonumber \\
	&
	\quad
	+
	G_N
	O_{\mathbb{P}} 
	\left[
	\frac{\sqrt{G_N}}{N} 
	+
	\left( \frac{G_N}{N} \right)^{1+\underline{v}}
	+ 
	\left(\frac{G_N}{N}\right)^{2b}
	\right].
	\label{Hatvhesig}
\end{align}

For $\widehat{V}^{\Sigma}_{Ho}$, observe first that, arguing as for \eqref{DW3} under Assumption V gives, for $\underline{v} = \min \left(v,\frac{1}{2}\right)$:
\begin{align*}
	\sum_{g=1}^{G_N} 
	\left\|
	\widehat{\Omega}_g \left(\widehat{\theta}_g\right)
	-
	\Omega_{n_g,g} \left(\theta_g \right)
	\right\|^3
	=
	G_N
	O_{\mathbb{P}}
	\left[
	\left(\frac{G_N}{N}\right)^{3 \underline{v}}
	\right].
\end{align*}
Now, $\widehat{V}^{\Sigma}_{Ho}$ decomposes as,
\begin{align*}
	\widehat{V}^{\Sigma}_{Ho}
	& = 
	V_{Ho}
	+
	\widehat{V}^{\Sigma}_{Ho,1}+\widehat{V}^{\Sigma\prime}_{Ho,1}
	+
	\widehat{V}^{\Sigma}_{Ho,2}+\widehat{V}^{\Sigma\prime}_{Ho,2}
	+
	\widehat{V}^{\Sigma}_{Ho,3}+\widehat{V}^{\Sigma\prime}_{Ho,3}
	+
	\widehat{V}^{\Sigma}_{Ho,4}
	+
	\widehat{V}^{\Sigma}_{Ho,5},
\end{align*}
with, by the H\"older Inequality,
\begin{align*}
	&
	\widehat{V}^{\Sigma}_{Ho,1}
	=
	\sum_{g=1}^{G_N} 
	n_g^{-1}
	W_g \left(\theta_g\right) 
	\Omega_{n_g,g} \left(\theta_g \right)
	\left[
	\widehat{W}_g\left(\widehat{\theta}_g\right)
	-
	W_g\left(\theta_g\right)
	\right]^{\prime}
	\\
	&
	\qquad\qquad
	=
	O\left[\frac{G_N}{N} G_N^{\frac{2}{3}}\left(\sum_{g=1}^{G_N} \left\|\widehat{W}_g \left(\widehat{\theta}_g\right)-W_g \left(\theta_g\right)\right\|^3\right)^{\frac{1}{3}}\right]
	=
	o_{\mathbb{P}}
	\left( \frac{G_N^2}{N} \right),
\end{align*}
\begin{align*}
	&
	\widehat{V}^{\Sigma}_{Ho,2}
	=
	\sum_{g=1}^{G_N} 
	n_g^{-1}
	W_g \left(\theta_g\right)
	\left[\widehat{\Omega}_g \left(\widehat{\theta}_g\right)-\Omega_{n_g,g} \left(\theta_g \right)\right] 
	W_g^{\prime}\left(\theta_g\right)
	\\
	&
	\qquad\qquad
	=
	O
	\left[
	\frac{G_N}{N} 
	G_N^{\frac{2}{3}}
	\left(
	\sum_{g=1}^{G_N} 
	\left\|
	\widehat{\Omega}_g \left(\widehat{\theta}_g\right)
	-
	\Omega_{n_g,g} \left(\theta_g \right)
	\right\|^3
	\right)^{\frac{1}{3}}
	\right]
	=
	o_{\mathbb{P}}
	\left( \frac{G_N^2}{N} \right),
\end{align*}
\begin{align*}
	&
	\widehat{V}^{\Sigma}_{Ho,3}
	=
	\sum_{g=1}^{G_N} 
	n_g^{-1}
	W_g \left(\theta_g\right)
	\left[\widehat{\Omega}_g \left(\widehat{\theta}_g\right)-\Omega_{n_g,g} \left(\theta_g \right)\right] 
	\left[
	\widehat{W}_g\left(\widehat{\theta}_g\right)
	-
	W_g\left(\theta_g\right)
	\right]^{\prime}
		\\
	&
	\qquad\qquad
	=
	O_{\mathbb{P}}
	\left[
	\frac{G_N}{N} 
	G_N^{\frac{2}{3}}
	\left(
	\sum_{g=1}^{G_N} 
	\left\|
	\widehat{\Omega}_g \left(\widehat{\theta}_g\right)
	-
	\Omega_{n_g,g} \left(\theta_g \right)
	\right\|^3
	\right)^{\frac{1}{3}}
	\right]=
	o_{\mathbb{P}}
	\left( \frac{G_N^2}{N} \right),
\end{align*}
\begin{align*}
	& 
	\widehat{V}^{\Sigma}_{Ho,4}
	=
	\sum_{g=1}^{G_N} 
	n_g^{-1}
	\left[
	\widehat{W}_g\left(\widehat{\theta}_g\right)
	-
	W_g\left(\theta_g\right)
	\right] 
	\Omega_{n_g,g} \left(\theta_g \right) 
	\left[
	\widehat{W}_g\left(\widehat{\theta}_g\right)
	-
	W_g\left(\theta_g\right)
	\right]^{\prime}
	=
	o_{\mathbb{P}}
	\left( \frac{G_N^2}{N} \right),
\end{align*}
\begin{align*}
	&
	\widehat{V}^{\Sigma}_{Ho,5}
	=
	\sum_{g=1}^{G_N} 
	n_g^{-1}
	\left[
	\widehat{W}_g\left(\widehat{\theta}_g\right)
	-
	W_g\left(\theta_g\right)
	\right] 
	\left[\widehat{\Omega}_g \left(\widehat{\theta}_g\right)-\Omega_{n_g,g} \left(\theta_g \right)\right] 
	\left[
	\widehat{W}_g\left(\widehat{\theta}_g\right)
	-
	W_g\left(\theta_g\right)
	\right]^{\prime}
	\\
	&
	\qquad \qquad
	=
	O \left(\frac{G_N}{N}\right)
	\left(
	\sum_{g=1}^{G_N} 
	\left\|
	\widehat{W}_g \left(\widehat{\theta}_g\right)
	-
	W_g \left(\theta_g \right)
	\right\|^3
	\right)^{\frac{2}{3}}
	\left(
	\sum_{g=1}^{G_N} 
	\left\|
	\widehat{\Omega}_g \left(\widehat{\theta}_g\right)
	-
	\Omega_{n_g,g} \left(\theta_g \right)
	\right\|^3
	\right)^{\frac{1}{3}}
	=
	o_{\mathbb{P}}
	\left( \frac{G_N^2}{N} \right).
\end{align*}
Since $V_{Ho} \asymp \frac{G_N^2}{N}$, it then follows:
\begin{align*}
	\widehat{V}_{Ho}^{\Sigma} = V_{Ho} + o_{\mathbb{P}} \left(\frac{G_N^2}{N}\right) = V_{Ho}  \left(1+o_{\mathbb{P}} (1)\right).
\end{align*}

Now, the equality above, \eqref{Hatvhesig} with $b>1/2$, $\underline{v}>0$, and $V_{He} \asymp G_N \Sigma$, $V_{Ho} \asymp \frac{G_N^2}{N}$ then gives
\begin{align*}
	&\left\|
\widetilde{\boldsymbol{V}}^{-1/2} \left(W\right)\widehat{\boldsymbol{V}}_{\Sigma} \left(\widehat{W}\right)\widetilde{\boldsymbol{V}}^{-1/2} \left(W\right) - \mathrm{Id} 
	\right\|
	=
	\frac{
	\left\|\Sigma\right\| \left| O_{\mathbb{P}} \left(\frac{1}{\sqrt{G_N}}\right)\right|
	+
	\left\|\Sigma\right\|^{\frac{1}{2}}
	\left| O_{\mathbb{P}} \left(\frac{1}{\sqrt{N}} + \left(\frac{G_N}{N}\right)^{b}\right)\right|
	+ o_{\mathbb{P}} \left( \frac{G_N}{N}\right)
	}{\left\|\Sigma\right\|+\frac{G_N}{N}}
\\
&
\qquad
=
\frac{
	\frac{N \left\|\Sigma\right\|}{G_N} \left| O_{\mathbb{P}} \left(\frac{1}{\sqrt{G_N}}\right)\right|
	+
	\left(\frac{N \left\|\Sigma\right\|}{G_N} \right)^{\frac{1}{2}}
	\left| O_{\mathbb{P}} \left(\frac{1}{\sqrt{G_N}} + \left(\frac{G_N}{N}\right)^{b-\frac{1}{2}}\right)\right|
	+ \left|o_{\mathbb{P}} \left( 1 \right)\right|
}{\frac{N \left\|\Sigma\right\|}{G_N}+1}
\\
&
\qquad
=
\frac{
	\left(\left(\frac{N \left\|\Sigma\right\|}{G_N}+1\right) 
	+
	\left(\frac{N \left\|\Sigma\right\|}{G_N} +1\right)^{\frac{1}{2}}
	+1\right)
	 \left|o_{\mathbb{P}} \left( 1 \right)\right|
}{\frac{N \left\|\Sigma\right\|}{G_N}+1}
\\
&
\qquad
=
\left[1+\left(\frac{N \left\|\Sigma\right\|}{G_N} +1\right)^{-\frac{1}{2}}+\left(\frac{N \left\|\Sigma\right\|}{G_N} +1\right)^{-1}\right]
o_{\mathbb{P}} \left( 1 \right)
=
o_{\mathbb{P}} \left( 1 \right),
\end{align*}
uniformly in $\Sigma$, since
$\frac{N \left\|\Sigma\right\|}{G_N}+1$ is bounded away from $0$. \hfill $\Box$

\subsection{Proof of Proposition \ref {Inference}}
\label{app:proof-inference}

\paragraph{Effects of bias.}
We show here that the estimation bias $\widetilde{B}$ and the remainder terms of order $r_{\theta}$ and $r_{He}$ from Proposition \ref{Expansion} and the random weight bias \eqref{CLT_Esp} can be neglected under the considered rate for $G_N$.

Proposition \ref{Varconsistency} shows that it is sufficient to study the asymptotic normality of 
\begin{align*}
	\boldsymbol{V}^{-\frac{1}{2}} \left(W\right)
	\left[
	\sum_{g=1}^{G_N} \widehat{W}_g  \left(\widehat{\theta}_g\right)
	\right]  \left( \widehat{\theta}\left(\widehat{W}\right)-\theta\right)
\end{align*}
where $\boldsymbol{V}\left(W\right)$ is as in \eqref{Var} and satisfies by Proposition \ref{Randweightshet},
\begin{align*}
	\boldsymbol{V}\left(W\right) 
	\asymp 
	\frac{G_N^2}{N}
	\left(\frac{N}{G_N} \left\| \Sigma \right\| + 1 \right).
\end{align*}
Recall $\theta_g-\theta = \delta_g$ and  $\widehat{\varepsilon}_g = \widehat{\theta}_g - \theta_g - B_{n_g,g}/n_g^b$.
 \eqref{CLT_Esp} in Proposition \ref{Randweightshet}, Proposition \ref{Expansion} and its proof,  and Assumptions E, W give, uniformly over $\mathcal{D}_{\kappa}$,
\begin{align*}
  &	\boldsymbol{V}^{-\frac{1}{2}} \left(W\right)
	\left[
	\sum_{g=1}^{G_N} \widehat{W}_g  \left(\widehat{\theta}_g\right)
	\right]  \left( \widehat{\theta}\left(\widehat{W}\right)-\theta\right)
	\\
	&
	\quad 
	=
	\boldsymbol{V}^{-\frac{1}{2}} \left(W\right)
	\sum_{g=1}^{G_N}
	W_g \left(\theta_g\right)
	\left(
	\delta_g  	
	+
	\widehat{\varepsilon}_g
	\right)
	+
	O_{\mathbb{P}}
	\left(
	\frac{\sqrt{\frac{G_N^2}{N}}}{\left(\frac{N}{G_N} \left\| \Sigma \right\| + 1 \right)^{\frac{1}{2}}}
	\right)
	+
	O_{\mathbb{P}}
	\left[
	\frac{\sqrt{N}}{G_N}
	\frac{
		G_N
		\left(
		\left(\frac{G_N}{N}\right)^{b}
		+
		r_{\theta}
		+
		r_{He}
		\right)
	}{\left(\frac{N}{G_N} \left\| \Sigma \right\| + 1 \right)^{\frac{1}{2}}}
	\right].
\end{align*}
Observe that, by Proposition \ref{Expansion} and uniformly in $\Sigma$,
\begin{align*}
	\frac{\sqrt{N}}{G_N}
	\frac{
		G_N
		\left(
		\left(\frac{G_N}{N}\right)^{b}
		+
		r_{\theta}
		+
		r_{He}
		\right)
	}{\left(\frac{N}{G_N} \left\| \Sigma \right\| + 1 \right)^{\frac{1}{2}}}
& 
=
\frac{\frac{G_N^b}{N^{b-\frac{1}{2}}}}{\left(\frac{N}{G_N} \left\| \Sigma \right\| + 1 \right)^{\frac{1}{2}}}
+
\frac{\sqrt{\frac{G_N^2}{N}}}{\left(\frac{N}{G_N} \left\| \Sigma \right\| + 1 \right)^{\frac{3}{2}}}
+
\frac{\sqrt{G_N \|\Sigma\|}}{\left(\frac{N}{G_N} \left\| \Sigma \right\| + 1 \right)^{\frac{3}{2}}},
\end{align*}
with, since $\frac{\sqrt{G_N \|\Sigma\|}}{\left(\frac{N}{G_N} \left\| \Sigma \right\| + 1 \right)^{\frac{3}{2}}}
=
\sqrt{\frac{G_N^2}{N}} \frac{\sqrt{\frac{N}{G_N} \|\Sigma\|}}{\left(\frac{N}{G_N} \left\| \Sigma \right\| + 1 \right)^{\frac{3}{2}}}
\leq \sqrt{\frac{G_N^2}{N}} \frac{1}{\frac{N}{G_N} \left\| \Sigma \right\| + 1 }$:
\begin{align*}
	&\frac{\frac{G_N^b}{N^{b-\frac{1}{2}}}}{\left(\frac{N}{G_N} \left\| \Sigma \right\| + 1 \right)^{\frac{1}{2}}}
	=
	\left\{
	\begin{array}{ll}
	O
	\left(
	\frac{G_N^b}{N^{b-\frac{1}{2}}}
	\right)
	=
	o(1)
	&
	\text{uniformly in $\| \Sigma\| \in [0,\infty)$ for $G_N = o\left(N^{1-\frac{1}{2 b}}\right)$},
	\\
	O
	\left(
	\frac{G_N^{b+\frac{1}{2}}}{N^{b}}
	\right)
	=
	o(1)
	&
	\text{uniformly in $\| \Sigma\| \in [1/\kappa,\infty)$ for $G_N = o\left(N^{1-\frac{1}{2b+1}}\right)$},
	\end{array}
	\right.
	\\
	&\frac{\sqrt{\frac{G_N^2}{N}}}{\left(\frac{N}{G_N} \left\| \Sigma \right\| + 1 \right)^{\frac{3}{2}}}
	=
	\left\{
	\begin{array}{ll}
		O
		\left(
		\frac{G_N}{N^{\frac{1}{2}}}
		\right)
		=
		o(1)
		&
		\text{uniformly in $\| \Sigma\| \in [0,\infty)$ for $G_N =o\left(N^{\frac{1}{2}}\right)$},
		\\
		O
		\left(
		\frac{G_N^{\frac{5}{2}}}{N^{2}}
		\right)
		=
		o(1)
		&
		\text{uniformly in $\| \Sigma\| \in [1/\kappa,\infty)$ for $G_N = o\left(N^{\frac{4}{5}}\right)$},
	\end{array}
	\right.
		\\
	&\frac{\sqrt{G_N \|\Sigma\|}}{\left(\frac{N}{G_N} \left\| \Sigma \right\| + 1 \right)^{\frac{3}{2}}}
	=
	\left\{
	\begin{array}{ll}
		O
		\left(
		\frac{G_N}{N^{\frac{1}{2}}}
		\right)
		=
		o(1)
		&
		\text{uniformly in $\| \Sigma\| \in [0,\infty)$ for $G_N =o\left(N^{\frac{1}{2}}\right)$},
		\\
		O
		\left(
		\frac{G_N^{2}}{N^{\frac{3}{2}}}
		\right)
		=
		o(1)
		&
		\text{uniformly in $\| \Sigma\| \in [1/\kappa,\infty)$ for $G_N = o\left(N^{\frac{3}{4}}\right)$},
	\end{array}
	\right.
		\\
	&\frac{\sqrt{\frac{G_N^2}{N}}}{\left(\frac{N}{G_N} \left\| \Sigma \right\| + 1 \right)^{\frac{1}{2}}}
	=
	\left\{
	\begin{array}{ll}
		O
		\left(
		\frac{G_N}{N^{\frac{1}{2}}}
		\right)
		=
		o(1)
		&
		\text{uniformly in $\| \Sigma\| \in [0,\infty)$ for $G_N =o\left(N^{\frac{1}{2}}\right)$},
		\\
		O
		\left(
		\frac{G_N^{\frac{3}{2}}}{N}
		\right)
		=
		o(1)
		&
		\text{uniformly in $\| \Sigma\| \in [1/\kappa,\infty)$ for $G_N = o\left(N^{\frac{2}{3}}\right)$}.
	\end{array}
	\right.
\end{align*}
Hence, since $\underline{b} = \min (b,1)$,
\begin{align*}
	&	\boldsymbol{V}^{-\frac{1}{2}} \left(W\right)
	\left[
	\sum_{g=1}^{G_N} \widehat{W}_g  \left(\widehat{\theta}_g\right)
	\right]  \left( \widehat{\theta}\left(\widehat{W}\right)-\theta\right)
	\\
	&
	\quad 
	=
	\boldsymbol{V}^{-\frac{1}{2}} \left(W\right)
	\sum_{g=1}^{G_N}
	W_g \left(\theta_g\right)
	\left(
	\delta_g  	
	+
	\widehat{\varepsilon}_g
	\right)
	+
	o_{\mathbb{P}}
	\left(
	1
	\right),
	\quad
	\left\{
	\begin{array}{l}
	\text{uniformly over $\mathcal{D}_{\kappa}$ for $G_N = o\left(N^{1-\frac{1}{2 \underline{b}}}\right)$,}
		\\
	\text{uniformly over $\mathcal{D}_{\kappa}\cap \{\|\Sigma\| \geq \frac{1}{\kappa}\}$ for $G_N = o\left(N^{1-\frac{1}{2\underline{b}+1}}\right)$.}
	\end{array}
	\right.
\end{align*}

\paragraph{Central Limit Theorem.}
Hence the uniform CLTs of the Proposition hold if 
\begin{align}
	\widehat{S}
	=
	\boldsymbol{V}^{-\frac{1}{2}} \left(W\right)
	\sum_{g=1}^{G_N}
	W_g \left(\theta_g\right)
	\left(
	\delta_g  	
	+
	\widehat{\varepsilon}_g
	\right)
	\stackrel{d}{\longrightarrow}
	\mathcal{N} \left(0,\mathrm{Id}\right),
	\label{CLTS}
\end{align}
uniformly over $\mathcal{D}_{\kappa}$.

We check here that \eqref{CLTS} holds pointwise, based on Theorem 7.19 in \citet[p. 179]{pollard2002}. 
\eqref{CLT_Esp} yields that, uniformly over $\mathcal{D}_{\kappa}$,
\begin{align*}
	\mathbb{E} \left[\widehat{S}\right] = O\left( \sqrt{\frac{G_N^2}{N}} \frac{1}{\left(\frac{N}{G_N} \left\|\Sigma\right\|+1\right)^{\frac{1}{2}}}\right)=o(1)
	\text{ while } \mathrm{Var} \left( \widehat{S}\right) = \mathrm{Id},
\end{align*}
by definition of $\boldsymbol{V} \left(W\right)$.

Assumptions He, E and W also yield:
\begin{align*}
	\sum_{g=1}^{G_N}
	\mathbb{E}
	\left[
	\left\|
	\boldsymbol{V}^{-\frac{1}{2}} \left(W\right)
	W_g \left(\theta_g\right)
	\left(
	\delta_g  	
	+
	\widehat{\varepsilon}_g
	\right)
	\right\|^{3}
	\right]
	&
	\leq
	O
	\left\{
	G_N 	
	\left[
	G_N
   \left( \left\|
		\Sigma
		\right\|
		+
		\frac{G_N}{N}\right)
	\right]^{-\frac{3}{2}}
	\left[
	\left\|
	\Sigma
	\right\|
	+
	\frac{G_N}{N}
	\right]^{\frac{3}{2}}
	\right\}
	\\
	&
	\leq
	O
	\left(
	\frac{1}{\sqrt{G_N}}
	\right)
,
\end{align*}
uniformly over $\mathcal{D}_{\kappa}$. 
Hence  the limit distribution of $\widehat{S}$ is a standard normal by Theorem 7.19 in \citet{pollard2002}.

\paragraph{Uniformity in the CLT.} This follows from 
\begin{align*}
	\sup_{\mathcal{D}_{\kappa}}
	\sum_{g=1}^{G_N}
	\mathbb{E}
	\left[
	\left\|
	\boldsymbol{V}^{-\frac{1}{2}} \left(W\right)
	W_g \left(\theta_g\right)
	\left(
	\delta_g  	
	+
	\widehat{\varepsilon}_g
	\right)
	\right\|^{3}
	\right]
	=
	o(1),
\end{align*}
as established above, see Theorem 15, p. 369, in \citet{ibragimov1981}.
Proposition \ref{Inference} follows from the CLT above and Proposition \ref{Varconsistency} by the Slutsky Proposition. \hfill $\Box$

\subsection{Proof of Theorem \ref{Opt}}
\label{app:proof-opt}

Recall first that the adaptive weights $\widehat{W}^{\star}_g$ of \eqref{Hatwstar} satisfy
\begin{align*}
\widehat{W}^{\star}_g = \left(\check{\Sigma} + \frac{\widehat{\Omega}_{g} \left(\widehat{\theta}_g\right)}{n_g}\right)^{-1}
=
n_g 
\left(n_g \check{\Sigma} + \widehat{\Omega}_{g} \left(\widehat{\theta}_g\right)\right)^{-1},	
\end{align*}
and, under balanced designs, are equivalent to
\begin{align*}
	\widehat{W}_g^{\ast} \left(\widehat{\theta}_g, \check{\Sigma}\right) 
	= \left(\frac{N \|\Sigma\|}{G_N}+1\right) \frac{G_N}{N}\widehat{W}_g^{\star} \left(\widehat{\theta}_g\right) 
	=
	\nu_g (\Sigma)
		\left(n_g \check{\Sigma} + \widehat{\Omega}_{g} \left(\widehat{\theta}_g\right) \right)^{-1},
		\quad
		\nu_g (\Sigma) = \frac{G_N n_g}{N} \left(\frac{N \|\Sigma\|}{G_N}+1\right) ,
\end{align*}
that is $\widehat{\theta} \left(\widehat{W}_g^{\star}\right)=\widehat{\theta} \left(\widehat{W}_g^{\ast}\right)$. Although infeasible, these new weights
 are bounded away from 0 and infinity uniformly in $g$ and $\Sigma$. They are therefore well behaved for mathematical purposes. Let their population counterparts be:
 \begin{align*}
W_g^{\ast} \left(\theta_g,\Sigma\right) 
 =
 \nu_g (\Sigma)
 \left(n_g \Sigma + \Omega_{n_g,g} \left(\theta_g\right) \right)^{-1}.
 \end{align*}
 
\subsubsection{Regularized estimated variance and associated weights}
\label{app:regularized-estimated-variance}

Although the population weights $W_g^{\ast} \left(\theta_g,\Sigma\right)$ are bounded, this is not the case of their sample counterparts $\widehat{W}_g^{\ast} \left(\widehat{\theta}_g, \check{\Sigma}\right)$, so that some regularization is needed here.

Let $\underline{\lambda} (M)$ and $\overline{\lambda} (M)$ be respectively the minimal and maximal eigenvalues of the matrix M, and set
\begin{align*}
	\underline{\lambda}_{\Omega}
	=
	\inf_{N}\min_{1\leq g \leq G_N} \inf_{\tau \in \mathcal{T}} \underline{\lambda} \left(\Omega_{n_g,g} (\tau)\right)>0.
\end{align*}
Let $\Lambda_{\Sigma} (\cdot)$ and $\Lambda_{\Omega} (\cdot)$  be infinitely differentiable functions valued in $[0,1]$, with bounded derivatives and such that, for $\underline{v}$ as in Proposition \ref{Varconsistency} , $\epsilon>0$, and for $N$ large enough:
\begin{align*}
	&\Lambda_{\Sigma} (M) = 1
	 \text{ if } \frac{2}{3} \left[\underline{\lambda} \left( \frac{N}{G_N}\Sigma\right) -\frac{N}{G_N}  C_N\right] \leq \underline{\lambda} (M) \leq \overline{\lambda} (M)  \leq \frac{3}{2} \left[\overline{\lambda} \left( \frac{N}{G_N}\Sigma\right) +\frac{N}{G_N}  C_N\right] ,
	 \\
	 &\Lambda_{\Sigma} (M) = 0
	 \text{ if } \frac{1}{3}  \left[\underline{\lambda} \left( \frac{N}{G_N}\Sigma\right) -\epsilon \underline{\lambda}_{\Omega}\right]\geq \underline{\lambda} (M)  \text{ or } \overline{\lambda} (M)  \geq 2 \left[\overline{\lambda} \left( \frac{N}{G_N}\Sigma\right) +\frac{N}{G_N} \epsilon\right],
	 \\
	 & \text { where }
	 C_N 
	 =
	 c_N
	 \left[
	 \frac{\sqrt{G_N}}{N} + \left(\frac{G_N}{N}\right)^{1+\underline{v}} + \left(\frac{G_N}{N}\right)^{2b} 
	 \right],
	 \text{ $c_N \uparrow \infty$ with $C_N = o \left(\frac{G_N}{N}\right)$,}
	 \\
	 &\Lambda_{\Omega} (M) = 1
	 \text{ if } \frac{2}{3} \inf_g \inf_{\tau \in \mathcal{T}} \underline{\lambda} \left(\Omega_{n_g,g} (\tau)\right) 
	 \leq 
	 \underline{\lambda} (M) \leq \overline{\lambda} (M)  
	 \leq 
	 \frac{3}{2}
	 \sup_g \sup_{\tau \in \mathcal{T}} 
	  \overline{\lambda} \left(\Omega_{n_g,g} (\tau)\right) ,
	 \\
	 & \Lambda_{\Omega} (M) = 0
	 \text{ if } \frac{1}{3} \inf_g \inf_{\tau \in \mathcal{T}} \underline{\lambda} \left(\Omega_{n_g,g} (\tau)\right) \geq \underline{\lambda} (M)  
	 \text{ or } \overline{\lambda} (M)  \geq 2\sup_g \sup_{\tau \in \mathcal{T}} 
	 \overline{\lambda} \left(\Omega_{n_g,g} (\tau)\right), 
\end{align*}

The associated regularized variance estimators are:
\begin{align*}
	\widetilde{\Sigma} = \frac{G_N}{N}\left[\frac{N}{G_N}\check{\Sigma} \Lambda_{\Sigma} \left( \frac{N}{G_N}\check{\Sigma}\right) + \frac{N}{G_N}\Sigma \left(1-\Lambda_{\Sigma} \left( \frac{N}{G_N}\check{\Sigma}\right) \right)\right],
	\quad
	\widetilde{\Omega}_g = \widehat{\Omega}_g \Lambda_{\Omega} \left( \widehat{\Omega}_g\right) + \Omega_g (\theta_g) \left(1-\Lambda_{\Omega} \left( \widehat{\Omega}_g\right) \right),
\end{align*}
with associated weights
\begin{align*}
	\widetilde{W}_g
	=
	\widetilde{W}_g \left(\widehat{\theta}_g, \widetilde{\Sigma} \right)
	=
	\nu_g (\Sigma)
	\left(n_g \widetilde{\Sigma} + \widetilde{\Omega}_{g}  \right)^{-1},
	\quad
	W_g \left(\theta_g, \Sigma \right)=
	W_g^{\ast} \left(\theta_g,\Sigma\right) 
	=
	\nu_g (\Sigma)
	\left(n_g \Sigma + \Omega_{n_g,g} \left(\theta_g\right) \right)^{-1}.
\end{align*}
As the design is balanced, 
\begin{align*}
	\min_{1 \leq g \leq G_N} \underline{\lambda} \left(n_g \widetilde{\Sigma} + \widetilde{\Omega}_{g}  \right) \geq \frac{\underline{\lambda}_{\Omega}}{3} \left(1 - C \epsilon\right),
\end{align*}
which is strictly positive for $\epsilon>0$, ensuring in particular that all the $n_g \widetilde{\Sigma} + \widetilde{\Omega}_{g}$ are definite positive, uniformly over $\mathcal{D}_{\kappa}$.

Consider the event
\begin{align*}
	\mathcal{E}_N
	=
	\left\{
	 \widetilde{\Sigma} = \check{\Sigma},
	 \text{ and }
	 \widetilde{\Omega}_g = \widehat{\Omega}_g
	 \text{ for all } 1 \leq g \leq G_N
	\right\}
	\subset
	\left\{
	\widehat{\theta} \left(\widehat{W}^{\star}\right)
	=
	\widehat{\theta} \left(\widetilde{W}\right)
	\right\}.
\end{align*}
Note that, for some $\epsilon>0$ small enough, and by Assumption V,
\begin{align*}
	& \mathbb{P}
	\left(
	\widetilde{\Omega}_g \neq \widehat{\Omega}_g
	\text{ for some } 1 \leq g \leq G_N
	\right)
	\leq
	\sum_{g=1}^{G_N}
	\mathbb{P}
	\left(
	\left\|\widehat{\Omega}_g - \Omega_{n_g,g} (\theta_g)\right\| \geq \epsilon
	\right)
	\\
	&
	\qquad
	\leq
	\frac{1}{\epsilon^8}
	\sum_{g=1}^{G_N} 
	\mathbb{E}
	\left[
	\left\|\widehat{\Omega}_g - \Omega_{n_g,g} (\theta_g)\right\| ^8
	\right]
	=
	O \left(G_N \left(\frac{G_N}{N}\right)^4\right)
	=o(1)
	\text{ uniformly over } \mathcal{D}_{\kappa},
\end{align*}
since $G_N = o \left(N^{4/5}\right)$. It then follows by Proposition \ref{Varconsistency} that:
	\begin{align}
		\sup_{\mathbb{P} \in \mathcal{D}_{\kappa}}
		\left|\mathbb{P} \left(\mathcal{E}_N \right)-1\right|
		=
		o(1)
		\label{Event}
	\end{align}
It is therefore sufficient to establish the CLT of the Theorem for the mathematically convenient $\widehat{\theta} \left(\widetilde{W}\right)$. 

\subsubsection{Variance intermediary results}
\label{app:variance-intermediary-results}
Following \eqref{Var}, define:
\begin{align*}
\boldsymbol{V} \left(\widetilde{W}\right)
& =
\mathrm{Var}
\left[
\sum_{g=1}^{G_N}
\nu_g (\Sigma)
\left(n_g \Sigma + \Omega_{n_{g},g} (\theta_g) \right)^{-1}
\left(\delta_g + \widehat{\varepsilon}_g\right)
\right]=\mathrm{Var}
\left[
\sum_{g=1}^{G_N}
\widetilde{W}_g (\theta_g,\Sigma)
\left(\delta_g + \widehat{\varepsilon}_g\right)
\right],
\\
\widetilde{\boldsymbol{V}} \left(\widetilde{W}\right)
& 
=
\sum_{g=1}^{G_N}
\frac{\nu_g^2 (\Sigma)}{n_g} 
\left(n_g \Sigma + \Omega_{n_{g},g} (\theta_g) \right)^{-1}
=
\sum_{g=1}^{G_N}
\widetilde{W}_g (\theta_g,\Sigma)
\left(\Sigma + \frac{\Omega_{n_{g},g} (\theta_g)}{n_g}\right)
\text{$\displaystyle \widetilde{W}_g^{\prime}(\theta_g,\Sigma)$},
\\
\widetilde{\boldsymbol{V}}_{\Sigma} \left(\widetilde{W}\right)
& 
=
\sum_{g=1}^{G_N}
\frac{\nu_g^2 (\Sigma)}{n_g} 
\left(n_g \widetilde{\Sigma} + \widetilde{\Omega}_g \right)^{-1}.
\end{align*}

By Assumptions He, V and the definition of $\nu_g (\Sigma)$, it holds uniformly in $g$, $\tau$ and $\Sigma$, $\left\|W_g (\tau,\Sigma)\right\| \leq C $, $\left\|W_g^{-1} (\tau,\Sigma)\right\| \leq C$ and:
\begin{align*}
\left\|
\left(
\sum_{g=1}^{G_N}
W_g (\tau,\Sigma)
\right)^{-1}
\right\|
&
\leq 
\left\|
\left(\sum_{g=1}^{G_N}
\left\| W_g^{-1} (\tau,\Sigma) \right\|^{-1}
\mathrm{Id} \right)^{-1}
\right\|
\leq
\left(\sum_{g=1}^{G_N}
\left\| W_g^{-1} (\tau,\Sigma) \right\|^{-1} \right)^{-1}
\\
&
\leq 
C
\left(\sum_{g=1}^{G_N}
\left\| W_g (\tau,\Sigma) \right\| \right)^{-1},
\\
\left\|
\frac{\partial}{\partial \theta} W_g \left(\tau,\Sigma\right)
\right\|
&
\leq
\left(n_g \| \Sigma \|+1\right)
\left\|
\left(
n_g \Sigma + \Omega_{n_{g},g} (\tau)
\right)^{-1}
\right\|^2
\left\|
\Omega_{n_{g},g}^{(1)} (\tau)
\right\|
\leq\frac{ C \kappa}{n_g \| \Sigma \|+1},
\end{align*}
implying that the population weights $W_g (\cdot,\Sigma)$ satisfy the corresponding part of Assumption W, up to $\kappa$ being multiplied by a constant. 
Arguing as for \eqref{Vareq} then gives:
\begin{align}
	\widetilde{\boldsymbol{V}}^{-\frac{1}{2}} \left(\widetilde{W}\right)\boldsymbol{V} \left(\widetilde{W}\right)\widetilde{\boldsymbol{V}}^{-\frac{1}{2}} \left(\widetilde{W}\right)
	-
	\mathrm{Id}
	=
	O_{\mathbb{P}}
	\left(
	\sqrt{\frac{G_N}{N}}
	+
	\frac{1}{\sqrt{G_N}}
	\right)
	=
	o_{\mathbb{P}} (1),
	\text{ uniformly over $\mathcal{D}_{\kappa}$.}
	\label{Vareqopt}	
\end{align}

We now turn to variance estimation. It holds, by definition of $\widetilde{\Sigma}$ and $\widetilde{\Omega}_g$ and the Taylor Inequality:
\begin{align*}
	&
	\left\|
	\widetilde{\boldsymbol{V}}^{-\frac{1}{2}} \left(\widetilde{W}\right)
	\widetilde{\boldsymbol{V}}_{\Sigma} \left(\widetilde{W}\right)
	\widetilde{\boldsymbol{V}}^{-\frac{1}{2}} \left(\widetilde{W}\right)
	-
	\mathrm{Id}
	\right\|
	=
	\left\|
	\widetilde{\boldsymbol{V}}^{-\frac{1}{2}} \left(\widetilde{W}\right)
	\left[
	\widetilde{\boldsymbol{V}}_{\Sigma} \left(\widetilde{W}\right)
	-
	\widetilde{\boldsymbol{V}} \left(\widetilde{W}\right)
	\right]
	\widetilde{\boldsymbol{V}}^{-\frac{1}{2}} \left(\widetilde{W}\right)
	\right\|
	\\
	& \quad
	\leq
	C
	\left[
	\sup_{M} \left\|\Lambda_{\Sigma}^{(1)}(M) \right\|
	+
	\sup_{M} \left\|\Lambda_{\Omega}^{(1)}(M)\right\|
	\right]
	\frac{\frac{N}{G_N^2}}{\frac{N}{G_N}\|\Sigma\|+1}
	\frac{G_N}{N}
	\sum_{g=1}^{G_N}
	\left(
	\frac{N}{G_N}
	\left\|\check{\Sigma} - \Sigma\right\|
	+
	\left\|\widehat{\Omega}_g \left(\widehat{\theta}_g\right) - \Omega_{n_g,g} \left(\theta_g\right)\right\|
	\right),
\end{align*}
uniformly over $\mathcal{D}_{\kappa}$. Hence Assumptions V and E, Proposition \ref{Varconsistency} yield, by the Markov Inequality:
\begin{align*}
	&
	\left\|
	\widetilde{\boldsymbol{V}}^{-\frac{1}{2}} \left(\widetilde{W}\right)
	\widetilde{\boldsymbol{V}}_{\Sigma} \left(\widetilde{W}\right)
	\widetilde{\boldsymbol{V}}^{-\frac{1}{2}} \left(\widetilde{W}\right)
	-
	\mathrm{Id}
	\right\|
	\\
	&
	\quad
	=
	\frac{\frac{N}{G_N^2}}{\frac{N}{G_N}\|\Sigma\|+1}
	\frac{G_N^2}{N}
	O_{\mathbb{P}}
	\left[
	\frac{N}{G_N}
	\left(
	\frac{\|\Sigma\|}{\sqrt{G_N}}
	+
	\|\Sigma\|^{\frac{1}{2}}
	\left(
	\frac{1}{\sqrt{N}}
	+
	\left(\frac{G_N}{N}\right)^b
	\right)
	+
	o
	\left(\frac{G_N}{N}\right)	
	\right)
	+
	\sqrt{\frac{G_N}{N}}
	\right]
	\\
	&
	\quad
	=
	\frac{1}{\frac{N}{G_N}\|\Sigma\|+1}
	O_{\mathbb{P}}
	\left[
	\frac{\frac{N}{G_N}\|\Sigma\|+1}{\sqrt{G_N}}
	+
	\left(\frac{N}{G_N}\|\Sigma\|+1\right)^{\frac{1}{2}}
	\sqrt{\frac{N}{G_N}}
	\left(
	\frac{1}{\sqrt{N}}
	+
	\left(\frac{G_N}{N}\right)^b
	\right)
	+
	o(1)	
	+
	\sqrt{\frac{G_N}{N}}
	\right]
	,
\end{align*}
so that, uniformly over $\mathcal{D}_{\kappa}$,
\begin{align}
	\widetilde{\boldsymbol{V}}^{-\frac{1}{2}} \left(\widetilde{W}\right)
	\widetilde{\boldsymbol{V}}_{\Sigma} \left(\widetilde{W}\right)
	\widetilde{\boldsymbol{V}}^{-\frac{1}{2}} \left(\widetilde{W}\right)
	=
	\mathrm{Id}
	+
	o_{\mathbb{P}} (1).
	\label{Varestiopt}
\end{align}

\subsubsection{Effect of the estimated heterogeneity variance $\check{\Sigma}$}
\label{app:effect-estimated-heterogeneity-variance}
Define:
\begin{align*}
	&
	\nu \left(\Sigma\right)
	=
	\frac{N}{G_N} \|\Sigma \| +1,
	\text{ so that uniformly, by \eqref{Vareqopt}, Assumptions He and W, } \boldsymbol{V} \left(\widetilde{W}\right) \asymp \frac{G_N^2}{N} \nu \left(\Sigma\right),
	\\
	& \widetilde{w}^{\Sigma}_g 
    =
    \nu \left(\Sigma\right)
    \nu_g \left(\Sigma\right)
    \frac{G_N n_g}{N}
    \left(n_g \Sigma + \widetilde{\Omega}_{g} \left(\widehat{\theta}_g\right) \right)^{-1}
    \frac{N}{G_N}
    \left(
    \check{\Sigma}-\Sigma
    \right)
    \left(n_g \Sigma + \widetilde{\Omega}_{g} \left(\widehat{\theta}_g\right) \right)^{-1}.
\end{align*}
The second-order Taylor Inequality gives, on $\mathcal{E}
_N$ and by definition of the regularized matrices:
\begin{align*}
	\left\|
	\sum_{g=1}^{G_N}
	\left[
	\widetilde{W}_g \left(\widehat{\theta}_g, \widetilde{\Sigma} \right)
	-
	\widetilde{W}_g \left(\widehat{\theta}_g, \Sigma \right)	
	-
	\frac{\widetilde{w}^{\Sigma}_g}{\nu (\Sigma)}
	\right]\left(\delta_g + \widehat{\varepsilon}_g \right)
	\right\|
	\leq
	C
	\frac{\left\|\frac{N}{G_N}
		\left(
		\check{\Sigma}-\Sigma
		\right)\right\|^2}{\nu^2 \left(\Sigma\right)}
    \sum_{g=1}^{G_N}
    \left\|\delta_g + \widehat{\varepsilon}_g\right\|^2,
\end{align*}
with, by Proposition \ref{Varconsistency}  and the Markov Inequality for $\sum_{g=1}^{G_N}
\left\|\delta_g + \widehat{\varepsilon}_g\right\|^2$,
\begin{align*}
	&
	\frac{1}{\left\|\boldsymbol{V}^{\frac{1}{2}} \left(\widetilde{W}\right)\right\|}
	\frac{\left\|\frac{N}{G_N}
		\left(
		\check{\Sigma}-\Sigma
		\right)\right\|^2}{\nu^2 \left(\Sigma\right)}
	\sum_{g=1}^{G_N}
	\left\|\delta_g + \widehat{\varepsilon}_g\right\|^2
	\\
	&
	\qquad
	=
	O_{\mathbb{P}}
	\left[
	\frac{\sqrt{N }}{G_N\nu^{\frac{1}{2}} (\Sigma)}
	\left(
	\frac{N}{G_N}
	\left(
	\frac{\|\Sigma\|}{\sqrt{G_N}}
	+
	\|\Sigma\|^{\frac{1}{2}}\left(\frac{1}{\sqrt{N}}+\left(\frac{G_N}{N}\right)^{b}\right)
	+
	o
	\left(
	\frac{G_N}{N}\right)
	\right)
	\right)^2
	\frac{\frac{G_N^2}{N}\nu (\Sigma)}{\nu^2 (\Sigma)}
	\right]
	\\
	&
	\qquad
	=
	O_{\mathbb{P}}
	\left[
	\frac{G_N}{\sqrt{N }\left(\frac{N}{G_N} \|\Sigma\|+1\right)^{\frac{3}{2}}}
	\left(
	\frac{\frac{N}{G_N} \|\Sigma\|+1}{\sqrt{G_N}}
	+
	\left(\frac{N}{G_N}\|\Sigma\|+1\right)^{\frac{1}{2}}
	\left(
	\sqrt{\frac{N}{G_N}}
	\frac{1}{\sqrt{N}}
	+
	\left(\frac{G_N}{N}\right)^{b-\frac{1}{2}}
	\right)
	+
	o(1)
	\right)^2
	\right]
	\\
	&
	\qquad
	=
	O_{\mathbb{P}}
	\left[
	\frac{G_N}{\sqrt{N}}
	\left(
	\frac{1}{\sqrt{G_N}}
	+
	\left(\frac{G_N}{N}\right)^{b-\frac{1}{2}}
	+
	o(1)
	\right)
	\right]
	= o_{\mathbb{P}}(1), \text{ as $G_N = o(\sqrt{N})$ and $b>\frac{1}{2}$,}
\end{align*}
uniformly over $\mathcal{D}_{\kappa}$.

We now show that $\boldsymbol{V}^{-\frac{1}{2}}\left(\widetilde{W}\right) \sum_{g=1}^{G_N} \widetilde{w}^{\Sigma}_g \left(\delta_g+\widehat{\varepsilon}_g\right)/\nu (\Sigma)$ can be neglected. Factoring out of the sum the contribution of $\check{\Sigma}-\Sigma$ in each entry of the vector above shows that it is sufficient to do the proof for the case of dimension 1, for which it holds:
\begin{align*}
\widetilde{w}^{\Sigma}_g 
=
\frac{N}{G_N}
\left(
\check{\Sigma}-\Sigma
\right)
\cdot
\frac{\nu \left(\Sigma\right)
\nu_g \left(\Sigma\right)
\frac{G_N n_g}{N}}{
\left(n_g \Sigma + \widetilde{\Omega}_{g} \left(\widehat{\theta}_g\right) \right)^{2}
}
=
\frac{N}{G_N}
\left(
\check{\Sigma}-\Sigma
\right)
\widetilde{U}^{\Sigma}_g 
\text{ where }
\widetilde{U}^{\Sigma}_g
=
\frac{\nu \left(\Sigma\right)
	\nu_g \left(\Sigma\right)
	\frac{G_N n_g}{N}}{
	\left(n_g \Sigma + \widetilde{\Omega}_{g} \left(\widehat{\theta}_g\right) \right)^{2}
}.
\end{align*}
Define:
\begin{align*}
U^{\Sigma}_g (\theta_g)
=
\frac{\nu \left(\Sigma\right)
	\nu_g \left(\Sigma\right)
	\frac{G_N n_g}{N}}{
	\left(n_g \Sigma + \Omega_{n_g,g} \left(\theta_g\right) \right)^{2}
}.
\end{align*}
The Taylor Inequality under regularization and Assumption W ensure that:
\begin{align*}
&
\mathbb{E} 
\left[
\left\|
\sqrt{n_g}
\left(
\widetilde{U}^{\Sigma}_g (\theta_g)
-
U^{\Sigma}_g (\theta_g)
\right)
\right\|^8
\right]
\leq
C
\sup_{M}
\left\|
\Lambda_{\Omega}^{(1)} (M)
\right\|
\mathbb{E} 
\left[
\left\|
\sqrt{n_g}
\left(
\widehat{\Omega}_g (\theta_g)
-
\Omega_g (\theta_g)
\right)
\right\|^8
\right]
\leq C \kappa,
\\
&
\mathbb{E} 
\left[
\sup_{\tau}
\left\|
\widetilde{U}^{\Sigma (1)}_g (\tau)
\right\|^8
\right]
\leq
C \kappa,
\\
&
\sup_{\tau}
\left\|
\left(n_g \left\|\Sigma\right\|+1\right)
U^{\Sigma (1)}_g (\tau)
\right\|
\leq C \kappa,
\end{align*}
uniformly in $g$ and over $\mathcal{D}_{\kappa}$. Then Proposition \ref{Inference} yields, uniformly over $\mathcal{D}_{\kappa}$:
\begin{align*}
\left\|
\boldsymbol{V}^{-\frac{1}{2}} \left(\widetilde{W}\right)
\sum_{g=1}^{G_N}
U^{\Sigma }_g 
\left(\delta_g+\widehat{\varepsilon}_g\right)
\right\|
=
O_{\mathbb{P}} (1).
\end{align*}
It then follows, by Proposition \ref{Varconsistency} and since $G_N = o \left(\sqrt{N}\right)$ and uniformly over $\mathcal{D}_{\kappa}$:
\begin{align*}
	&\left\|
	\boldsymbol{V}^{-\frac{1}{2}} \left(\widetilde{W} \right)
	\frac{\sum_{g=1}^{G_N}
	\widetilde{w}^{\Sigma }_g 
	\left(\delta_g+\widehat{\varepsilon}_g\right)}{\nu (\Sigma)}
	\right\|
	=
	O_{\mathbb{P}}
	\left[
	\frac{N}{G_N \nu (\Sigma)}
	\left\|\check{\Sigma} - \Sigma\right\|
	\right]
	\\
	& 
	\qquad
	=
	O_{\mathbb{P}}
	\left[
	\frac{1}{ \frac{N}{G_N} \| \Sigma\| +1}
	\frac{N}{G_N}
	\left(
	\frac{\|\Sigma\|}{\sqrt{G_N}}
	+
	\|\Sigma\|^{\frac{1}{2}}\left(\frac{1}{\sqrt{N}}+\left(\frac{G_N}{N}\right)^{b}\right)
	+
	o\left(\frac{G_N}{N}\right)
	\right)
	\right]
	\\
	& 
	\qquad
	=
	O_{\mathbb{P}}
	\left[
	\frac{1}{ \frac{N}{G_N} \| \Sigma\| +1}
	\left(
	\frac{\frac{N}{G_N}\|\Sigma\|+1}{\sqrt{G_N}}
	+
	\left(\frac{N}{G_N}\|\Sigma\|+1\right)^{\frac{1}{2}}
	\left(\sqrt{\frac{N}{G_N}}\frac{1}{\sqrt{N}}+\left(\frac{G_N}{N}\right)^{b-\frac{1}{2}}\right)
	+
	o(1)
	\right)
	\right]
	\\
	& 
	\qquad
	=
	O_{\mathbb{P}}
	\left[
	\frac{1}{\sqrt{G_N}}
	+
	\left(\frac{G_N}{N}\right)^{b-\frac{1}{2}}
	+
	o(1)
	\right]
	=
	o_{\mathbb{P}}(1)
	\text{ as $b>\frac{1}{2}$.} 
\end{align*}

To conclude this section, it follows from the Taylor expansion above that:
\begin{align*}
	\boldsymbol{V}^{-\frac{1}{2}} \left(\widetilde{W} \right)
	\sum_{g=1}^{G_N}
	\widetilde{W}_g \left(\widehat{\theta}_g, \widetilde{\Sigma} \right)
    \left(\delta_g + \widehat{\varepsilon}_g \right)
	=
	\boldsymbol{V}^{-\frac{1}{2}} \left(\widetilde{W} \right)
	\sum_{g=1}^{G_N}
	\widetilde{W}_g \left(\widehat{\theta}_g, \Sigma \right)	
	\left(\delta_g + \widehat{\varepsilon}_g \right)
	+
	o_{\mathbb{P}} (1),
	\text{ uniformly over $\mathcal{D}_{\kappa}$.}
\end{align*}

\subsubsection{Conclusion for the CLT}
\label{app:conclusion-clt}

As 
$G_N = o \left(N^{1-\frac{1}{2b}}\right)$, it holds,
\begin{align*}
	\boldsymbol{V}^{-\frac{1}{2}} \left(\widetilde{W} \right)
	\sum_{g=1}^{G_N}
	\widetilde{W}_g \left(\widehat{\theta}_g, \widetilde{\Sigma} \right)
	\frac{B_{n_g,g} (\theta_g)}{n_g^{b}}
	=
	O_{\mathbb{P}}
	\left(
	\frac{G_N \left(\frac{G_N}{N}\right)^{b}}{\frac{G_N}{\sqrt{N}}
	\left(\frac{N}{G_N} \|\Sigma\|+1\right)^{\frac{1}{2}}}
	\right)
	=
	O_{\mathbb{P}}
	\left(
	\left(\frac{G_N}{N^{1-\frac{1}{2b}}}\right)^{b}
	\right)
	=
	o_{\mathbb{P}} (1),
\end{align*}
uniformly over $\mathcal{D}_{\kappa}$. It is shown here that
$\boldsymbol{V}^{-\frac{1}{2}} \left(\widetilde{W} \right) 
\sum_{g=1}^{G_N}
\widetilde{W}_g \left(\widehat{\theta}_g, \Sigma \right)	
\left(\delta_g + \widehat{\varepsilon}_g \right)$ satisfies a CLT uniformly over $\mathcal{D}_{\kappa}$ using Proposition \ref{Inference} and checking Assumption W.
Recall that the population counterpart $W_g (\cdot,\Sigma)$ satisfies their part of Assumption W as shown above. As well, Assumptions V and He together with the definition of $\Lambda_{\Omega} (\cdot)$ imply, uniformly over $\mathcal{D}_{\kappa}$ (and then also in $\Sigma$):
\begin{align*}
	&
	\max_{1 \leq g \leq G_N} 
	\mathbb{E}
	\left[
	\left\| \sqrt{n_g}
	\left(	\widetilde{W}_g \left(\theta_g,\Sigma\right) -W_g \left(\theta_g,\Sigma\right) \right)\right\|^8
	\right]
	\leq
	\sup_{M}\left\|\Lambda_{\Omega}^{(1)} (M)\right\|
	\max_{1 \leq g \leq G_N}
	\mathbb{E}
	\left[
	\left\| \frac{\sqrt{n_g}\left(	\widehat{\Omega}_g \left(\theta_g\right) -\Omega_g \left(\theta_g\right)\right)}{n_g \|\Sigma\|+1}
	 \right\|^8
	\right]
	\leq C \kappa,
	\\
	&
	\max_{1 \leq g \leq G_N} 
	\mathbb{E}
	\left[
	\sup_{\tau \in \mathcal{T}}
	\left\|   \widetilde{W}_g^{(1)} (\tau,\Sigma)\right\|^8 \right]
	\leq
	\max_{1 \leq g \leq G_N} 
	\sup_{M}\left\|\Lambda_{\Omega}^{(1)} (M)\right\|
	\mathbb{E}
	\left[
	\sup_{\tau \in \mathcal{T}}
	\left\|   \frac{\widehat{\Omega}_g^{(1)} (\tau)}{n_g \|\Sigma\| +1}\right\|^8 \right] 
	\leq
	C \kappa.
\end{align*}
Hence Assumption W holds, up to $\kappa$ which is multiplied by a constant, which is sufficient for Proposition \ref{Inference} to hold. Hence
$\boldsymbol{V}^{-\frac{1}{2}} \left(\widetilde{W} \right) 
\sum_{g=1}^{G_N}
\widetilde{W}_g \left(\widehat{\theta}_g, \Sigma \right)	
\left(\delta_g + \widehat{\varepsilon}_g \right)$ asymptotically has a multivariate standard normal distribution uniformly over $\mathcal{D}_{\kappa}$.
It then follows:
\begin{align*}
	\boldsymbol{V}^{-\frac{1}{2}} \left(\widetilde{W} \right)
	\left[\sum_{g=1}^{G_N} \widetilde{W}_g \left(\widehat{\theta}_g,\check{\Sigma}\right) \right]
	\left(
	\widehat{\theta} \left(\widetilde{W}\right) - \theta
	\right)
	\stackrel{d}{\longrightarrow}
	\mathcal{N}
	\left(0,\mathrm{Id}\right)
	\text{ uniformly over $\mathcal{D}_{\kappa}$}.
\end{align*}
As \eqref{Varestiopt} and \eqref{Vareqopt} yield, on $\mathcal{E}_N$ and uniformly on $\mathcal{D}_{\kappa}$:
\begin{align*}
	\boldsymbol{V}^{-\frac{1}{2}} \left(\widetilde{W} \right)
	\left[\sum_{g=1}^{G_N} \widetilde{W}_g \left(\widehat{\theta}_g,\check{\Sigma}\right) \right]
	\left(
	\widehat{\theta} \left(\widetilde{W}\right) - \theta
	\right)
	&
	=
	(1+o_{\mathbb{P}}(1))
\left[\sum_{g=1}^{G_N} W_g^{\star} \left(\theta_g,\Sigma\right)\right]^{\frac{1}{2}}
\left(
\widehat{\theta} \left(\widehat{W}_g^{\star}\right) - \theta
\right),
\\
\boldsymbol{V}^{-\frac{1}{2}} \left(\widetilde{W} \right)
\left[\sum_{g=1}^{G_N} \widetilde{W}_g \left(\widehat{\theta}_g,\check{\Sigma}\right) \right]
\left(
\widehat{\theta} \left(\widetilde{W}\right) - \theta
\right)
&
=
(1+o_{\mathbb{P}}(1))
\left[\sum_{g=1}^{G_N} \widehat{W}_g^{\star} \left(\widehat{\theta}_g ,\check{\Sigma}\right)\right]^{\frac{1}{2}}
\left(
\widehat{\theta} \left(\widehat{W}_g^{\star}\right) - \theta
\right),
\end{align*}
since $\widehat{W}^{\ast}=\widetilde{W}$, $\widehat{\theta} \left(\widehat{W}^{\star}\right)
=
\widehat{\theta} \left(\widehat{W}^{\ast}\right)
=
\widehat{\theta} \left(\widetilde{W}\right)$, the two CLT of the Theorem follow from the one above and from \eqref{Event}.

\subsubsection{The efficiency statement}
\label{app:efficiency-statement}
The efficiency part follows by considering the theoretical GMM estimator $\widehat{\tau} \left(W\right)$ which minimizes
\begin{align*}
	\left[
	\tau - \widetilde{\tau}_1, \ldots, \tau - \widetilde{\tau}_{G_N}
	\right]
	\mathrm{Diag} \left[W_1,\ldots,W_{G_N}\right]
	\left[\begin{array}{c}
		\tau - \widetilde{\tau}_1 \\
		\vdots \\
		\tau - \widetilde{\tau}_{G_N}
	\end{array}\right],
\end{align*}
where the $\widetilde{\tau}_g \sim \mathcal{N} \left(\theta,\Sigma + n_g^{-1} \cdot \Omega_{n_g,g} \right)$,
$\Omega_{n_{g},g}=\Omega_{n_{g},g}\left(\theta_g\right)$, are independent. It is easily seen that  the variance of $\widetilde{\tau} \left(W\right) = \left(\sum_{g=1}^{G_N} W_g\right)^{-1} \sum_{g=1}^{G_N} W_g \widehat{\tau}_g$ is 
\[\left(\sum_{g=1}^{G_N} W_g \right)^{-1} \left[\sum_{g=1}^{G_N} W_g \left( \Sigma + n_g^{-1} \Omega_{n_g,g} \right)W_g^{\prime}\right]\left(\sum_{g=1}^{G_N} W_g^{\prime} \right)^{-1} \] 
which is greater than or equal to $\left(\sum_{g=1}^{G_N} \left(\Sigma + n_g^{-1} \Omega_{n_g,g}\right)^{-1}\right)^{-1}$, see e.g. \citet[Theorem 5.2]{newey1994} and its proof, among others. This ends the proof of Theorem \ref{Opt}. \hfill $\Box$

\subsection{Proof of Theorem \ref{Homtest}}
\label{app:proof-homtest}
Assume all along the proof that $\theta=0$ without loss of generality.

\paragraph{Preliminary expansions.}
Define:
\begin{align*}
	\widetilde{\theta}_{IV}
	=
	\left[
	\sum_{g=1}^{G_N}
	n_g
	\Omega_{n_g,g}^{-1} \left(\theta_g\right)
	\right]^{-1}
	\sum_{g=1}^{G_N}
	n_g
	\Omega_{n_g,g}^{-1} \left(\theta_g\right)
	\widehat{\theta}_g,
	\\
	\theta_{IV}
	=
	\left[
	\sum_{g=1}^{G_N}
	n_g
	\mathbb{E}
	\left[
	\Omega_{n_g,g}^{-1} (\theta_g)
	\right]
	\right]^{-1}
	\sum_{g=1}^{G_N}
	n_g
	\mathbb{E}
	\left[
	\Omega_{n_g,g}^{-1} (\theta_g)\theta_g
	\right].
\end{align*}
Arguing as for Proposition \ref{Expansion} ensures that, under Assumptions E and V,
\begin{align}
	\widehat{\theta}_{IV}
	&=
	\widetilde{\theta}_{IV}
	+
	O_{\mathbb{P}}
	\left[
	\left(
	\frac{G_N}{N}
	\right)^{b+v}
	\right] 
	+
	\left\|\Sigma\right\|
	O_{\mathbb{P}}
	\left[
	\left(\frac{G_N}{N}\right)^{\frac{1}{2}}
	+
	\left(\frac{G_N}{N}\right)^{v}
	\right] 
	\nonumber
	\\
	&
	\quad
	+
	O_{\mathbb{P}}
	\left[
	\left(\frac{G_N}{N}\right)
	+
	\frac{1}{\sqrt{G_N}}
	\left(\frac{G_N}{N}\right)^{\frac{1}{2}+v}
	\right] 
	\nonumber \\
	& =
	\theta_{IV}+
	\left\|\Sigma\right\| O_{\mathbb{P}} \left[\frac{1}{\sqrt{G_N}}+\left(\frac{G_N}{N}\right)^{\underline{v}} \right]
	\label{Hatthetaom}
	\\
	&
	\quad
	+
	O_{\mathbb{P}} \left[\frac{1}{\sqrt{N}} + \left(\frac{G_N}{N}\right)^{\underline{b}} +\frac{1}{\sqrt{G_N}}
	\left(\frac{G_N}{N}\right)^{\frac{1}{2}+v}  \right]
	,
	\quad
	\left\{
	\begin{array}{l}
		\underline{b}=\min \left(b,1\right),
		\\
		\underline{v}
		=
		\min \left(v,\frac{1}{2}\right).
	\end{array}
	\right.		
 \nonumber
\end{align}

Let:
\begin{align*}
	&
	\ddot{\xi}_{g}
	=
	n_g
	\left(
	\widehat{\theta}_g - \widetilde{\theta}_{IV}
	\right)^{\prime}
	\widehat{\Omega}_{g}^{-1} \left(\widehat{\theta}_g\right)
	\left(
	\widehat{\theta}_g -\widetilde{\theta}_{IV}
	\right)
	-
	D,
	\\
	&
	\xi_{g}
	=
	n_g
	\left(
	\widehat{\theta}_g - \widetilde{\theta}_{IV}
	\right)^{\prime}
	\Omega_{n_g,g}^{-1} \left(\theta_g\right)
	\left(
	\widehat{\theta}_g -\widetilde{\theta}_{IV}
	\right)
	-
	D,
	\\
	&
	\widetilde{s}^2
	=
	\frac{1}{G_N}
	\sum_{g=1}^{G_N}
	\left(
	\xi_g - \overline{\xi}
	\right)^2,
	\quad
	\widetilde{\ddot{s}}^2
	=
	\frac{1}{G_N}
	\sum_{g=1}^{G_N}
	\left(
	\ddot{\xi}_g - \overline{\ddot{\xi}}
	\right)^2.
\end{align*}
Note that, under Assumptions V and E,
\begin{align*}
	\left|
	\overline{\widehat{\xi}} -\overline{ \ddot{\xi}}
	\right|
	&\leq
	C
	\left|
	\left(
	\widehat{\theta}_{IV}- \widetilde{\theta}_{IV}
	\right)^{\prime}
	\left[\frac{1}{G_N}
	\sum_{g=1}^{G_N}
	n_g
	\widehat{\Omega}_{g}^{-1} \left(\widehat{\theta}_g\right)\right]
	\left(
	\widehat{\theta}_{IV}-\widetilde{\theta}_{IV}
	\right)
	\right|
	\\
	&
	=
	O_{\mathbb{P}}
	\left[
	\left\|
	\left(\frac{N}{G_N}\right)^{\frac{1}{2}}
	\left(\widehat{\theta}_{IV}- \widetilde{\theta}_{IV}\right)
	\right\|^2
	\right]
	.
\end{align*}

Assumptions V and E give:
\begin{align*}
	&
	\left|
	\overline{\ddot{\xi}} - \overline{\xi}
	\right|
	\leq
	\left[
	\frac{1}{G_N}
	\sum_{g=1}^{G_N}
	\left\|
	\widehat{\Omega}_{g}^{-1} \left(\widehat{\theta}_g\right) - \Omega_{n_g,g}^{-1} \left(\theta_g\right)
	\right\|^2
	\right]^{1/2}
	\left[
	\frac{1}{G_N}
	\sum_{g=1}^{G_N}
	\left\| 
	\sqrt{n_g}\left(\widehat{\theta}_g - \widetilde{\theta}_{IV}\right)
	\right\|^4
	\right]^{1/2}
	\\
	&
	\quad
	\leq
	\Bigg\{
	3
	\left[\frac{1}{G_N}
	\sum_{g=1}^{G_N}
	\sup_{\tau}
	\left\|
	\left[\widehat{\Omega}_{g}^{-1}\right]^{(1)} \left(\tau\right) 
	\right\|^2\right]
	\left[
	\frac{1}{G_N}
	\sum_{g=1}^{G_N}
	\left\| 
	\widehat{\theta}_g - \theta_g
	\right\|^2
	\right]
	+
	\frac{3}{G_N}
	\sum_{g=1}^{G_N}
	\left\|
	\widehat{\Omega}_{g}^{-1} \left(\theta_g\right) - \Omega_{g}^{-1} \left(\theta_g\right)
	\right\|^2
	\\
	&
	\qquad
	\qquad
	+
	\frac{3}{G_N}
	\sum_{g=1}^{G_N}
	\left\|
	\widehat{\Omega}_{g}^{-1} \left(\theta_g\right) - \Omega_{n_g,g}^{-1} \left(\theta_g\right)
	\right\|^2
	\Bigg\}^{1/2}
	\times
	\left[
	\frac{1}{G_N}
	\sum_{g=1}^{G_N}
	\left\| 
	\sqrt{n_g}\left(\widehat{\theta}_g - \widetilde{\theta}_{IV}\right)
	\right\|^4
	\right]^{1/2}
	\\
	& 
	\quad
	=
	O_{\mathbb{P}} \left[\left(\frac{G_N}{N}\right)^{\frac{1}{2}}+\left(\frac{G_N}{N}\right)^{v}\right]
	\left[
	\frac{1}{G_N}
	\sum_{g=1}^{G_N}
	\left\| 
	\sqrt{n_g}\left(\widehat{\theta}_g - \widetilde{\theta}_{IV}\right)
	\right\|^4
	\right]^{1/2}.
\end{align*}

For $\widehat{s} - \widetilde{s}$, observe first:
\begin{align*}
	& \left|
	  \widehat{s} - \widetilde{\ddot{s}}
	  \right|
	\leq
	\left[
	\frac{1}{G_N}
	\sum_{g=1}^{G_N}
	\left(
	\widehat{\xi}_g
	-
	\ddot{\xi}_g
	\right)^2
	\right]^{\frac{1}{2}}
	+
	\left|
	\overline{\widehat{\xi}} - \overline{\ddot{\xi}}
	\right|,
\end{align*}
with:
\begin{align*}
	&
	\frac{1}{G_N}
	\sum_{g=1}^{G_N}
	\left(
	\widehat{\xi}_g
	-
	\ddot{\xi}_g
	\right)^2
	\\
	&
	\quad
	\leq
	\frac{C}{G_N}
	\sum_{g=1}^{G_N}
	\left\|
	\left(\frac{N}{G_N}\right)^{\frac{1}{2}}
	\left(\widehat{\theta}_{IV}- \widetilde{\theta}_{IV}\right)
	\right\|
	\left\|\widehat{\Omega}_g^{-1} \left(\widehat{\theta}_g\right)\right\|
	\left[
	\left\|
	\left(\frac{N}{G_N}\right)^{\frac{1}{2}}
	\left(\widehat{\theta}_{IV}- \widetilde{\theta}_{IV}\right)
	\right\|
	+
	\left\|
	\sqrt{n_g}\left(\widehat{\theta}_g - \widetilde{\theta}_{IV}\right)
	\right\|
	\right]
	\\
	&
	\quad
	\leq
	C
	\left\|
	\left(\frac{N}{G_N}\right)^{\frac{1}{2}}
	\left(\widehat{\theta}_{IV}- \widetilde{\theta}_{IV}\right)
	\right\|
	\times
	\left[
	\frac{1}{G_N}
	\sum_{g=1}^{G_N}
		\left\|\widehat{\Omega}_g^{-1} \left(\widehat{\theta}_g\right)\right\|^2
	\right]^{\frac{1}{2}}
	\\
	&
	\qquad \qquad
	\times
	\left(
	\left\|
	\left(\frac{N}{G_N}\right)^{\frac{1}{2}}
	\left(\widehat{\theta}_{IV}- \widetilde{\theta}_{IV}\right)
	\right\|
	+
	\left[
	\frac{1}{G_N}
	\sum_{g=1}^{G_N}
	\left\|
	\sqrt{n_g}\left(\widehat{\theta}_g - \widetilde{\theta}_{IV}\right)
	\right\|^2
	\right]^{\frac{1}{2}}
	\right)
	\\
	&
	\quad
	=
	O_{\mathbb{P}}
	\left[
	\left\|
	\left(\frac{N}{G_N}\right)^{\frac{1}{2}}
	\left(\widehat{\theta}_{IV}- \widetilde{\theta}_{IV}\right)
	\right\|^2
	\right]
	\\
	&
	\qquad
	+
	O_{\mathbb{P}}
	\left[
	\left\|
	\left(\frac{N}{G_N}\right)^{\frac{1}{2}}
	\left(\widehat{\theta}_{IV}- \widetilde{\theta}_{IV}\right)
	\right\|
	\right]
	\left[
	\frac{1}{G_N}
	\sum_{g=1}^{G_N}
	\left\|
	\sqrt{n_g}\left(\widehat{\theta}_g - \widetilde{\theta}_{IV}\right)
	\right\|^2
	\right]^{\frac{1}{2}}
.
\end{align*}
It holds, for $\widetilde{\ddot{s}} - \widetilde{s}$:
\begin{align*}
  & 	\left|
	\widetilde{\ddot{s}} - \widetilde{s}
	\right|
	\leq
	\left[
	\frac{1}{G_N}
	\sum_{g=1}^{G_N}
	\left(
	\ddot{\xi}_g
	-
	\xi_g
	\right)^2
	\right]^{\frac{1}{2}}
	+
		\left|
	\overline{\ddot{\xi}} - \overline{\xi}
	\right|,
\end{align*}
with, by the Minkowski Inequality:
\begin{align*}
	&
	\left[
	\frac{1}{G_N}
	\sum_{g=1}^{G_N}
	\left(
	\ddot{\xi}_g
	-
	\xi_g
	\right)^2
	\right]^{\frac{1}{2}}
	\leq
	\left[
	\frac{1}{G_N}
	\sum_{g=1}^{G_N}
	\left\| 
	\sqrt{n_g}\left(\widehat{\theta}_g - \widetilde{\theta}_{IV}\right)
	\right\|^4
	\left\|
	\widehat{\Omega}_{g}^{-1} \left(\widehat{\theta}_g\right) - \Omega_{n_g,g}^{-1} \left(\theta_g\right)
	\right\|^2
	\right]^{\frac{1}{2}}
	\\
	&
	\leq
	\left[
	\frac{2}{G_N}
	\sum_{g=1}^{G_N}
	\left\| 
	\sqrt{n_g}\left(\widehat{\theta}_g - \widehat{\theta}_{IV}\right)
	\right\|^4
	\left(
	\left\|
	\widehat{\Omega}_{g}^{-1} \left(\theta_g\right) - \Omega_{n_g,g}^{-1} \left(\theta_g\right)
	\right\|^2
	+
	\sup_{\tau}
	\left\|
	\left[\widehat{\Omega}_{g}^{-1}\right]^{(1)} \left(\tau\right) 
	\right\|^2
	\left\| 
	\left(\widehat{\theta}_g - \theta_g\right)
	\right\|^2
	\right)
	\right]^{\frac{1}{2}}
	\\
	&
	\leq
	\left[
	\frac{2}{G_N}
	\sum_{g=1}^{G_N}
	\left\| 
	\sqrt{n_g}\left(\widehat{\theta}_g - \widehat{\theta}_{IV}\right)
	\right\|^6
	\right]^{\frac{1}{3}}
	\left[
	\frac{1}{G_N}
	\sum_{g=1}^{G_N}
	\left\| 
	\widehat{\Omega}_{g}^{-1} \left(\theta_g\right) - \Omega_{n_g,g}^{-1} \left(\theta_g\right)
	\right\|^6
	\right]^{\frac{1}{6}}
	\\
	&
	\quad
	+
	\left[
	\frac{2}{G_N}
	\sum_{g=1}^{G_N}
	\left\| 
	\sqrt{n_g}\left(\widehat{\theta}_g - \widehat{\theta}_{IV}\right)
	\right\|^8
	\right]^{\frac{1}{4}}
	\left[
	\frac{1}{G_N}
	\sum_{g=1}^{G_N}
	\left\|
	\left[\widehat{\Omega}_{g}^{-1}\right]^{(1)} \left(\tau\right) 
	\right\|^8
	\right]^{\frac{1}{8}}
	\left[
	\frac{1}{G_N}
	\sum_{g=1}^{G_N}
	\left\|
	\widehat{\theta}_g - \theta_g
	\right\|^8
	\right]^{\frac{1}{8}},
\end{align*}
so that, by Assumption V and since the subsamples are balanced,
\begin{align}
	\left\{
	\begin{array}{lll}
		\left|
		\overline{\widehat{\xi}} - \overline{\xi}
		\right|
		& =
		&
		O_{\mathbb{P}} \left[\left(\frac{G_N}{N}\right)^{\underline{v}}\right]
		\left[
		\frac{1}{G_N}
		\sum_{g=1}^{G_N}
		\left\| 
		\sqrt{n_g}\left(\widehat{\theta}_g - \widetilde{\theta}_{IV}\right)
		\right\|^8
		\right]^{1/4}
		\\
		&&
		+
		O_{\mathbb{P}}
		\left[
		\left\|
		\left(\frac{N}{G_N}\right)^{\frac{1}{2}}
		\left(\widehat{\theta}_{IV}- \widetilde{\theta}_{IV}\right)
		\right\|^2
		\right],
		\\
		\left|
		\widehat{s} - \widetilde{s}
		\right|
		&
		=
		&
		O_{\mathbb{P}} \left[
		\left(\frac{G_N}{N}\right)^{\underline{v}}
		+
		\left\|
		\left(\frac{N}{G_N}\right)^{\frac{1}{2}}
		\left(
		\widehat{\theta}_{IV}- \widetilde{\theta}_{IV}
		\right)
		\right\|
		\right]
		\\
		&&
		\quad
		\times
		\left[
		\frac{1}{G_N}
		\sum_{g=1}^{G_N}
		\left\| 
		\sqrt{n_g}
		\left(
		\widehat{\theta}_g -  \widetilde{\theta}_{IV}
		\right)
		\right\|^8
		\right]^{\frac{1}{4}}
		\\
		&&
		+
		O_{\mathbb{P}}
		\left[
		\left\|
		\left(\frac{N}{G_N}\right)^{\frac{1}{2}}
		\left(\widehat{\theta}_{IV}- \widetilde{\theta}_{IV}\right)
		\right\|^2
		\right].
	\end{array}
	\right.
	\label{Prelims}
\end{align}

\paragraph{The homogeneity null.}
We first derive the order of the items in \eqref{Prelims}, recalling $\theta_g=\theta$ and $\Sigma=0$ under homogeneity. It holds by \eqref{Hatthetaom} and Assumption E:
\begin{align*}
\left\|
	\left(\frac{N}{G_N}\right)^{\frac{1}{2}}
	\left(\widehat{\theta}_{IV}- \widetilde{\theta}_{IV}\right)
\right\|
& = 
O_{\mathbb{P}}
\left[
\left(\frac{G_N}{N}\right)^{b-\frac{1}{2}+\underline{v}}
+
\left(\frac{G_N}{N}\right)^{\frac{1}{2}}
+
\frac{1}{\sqrt{G_N}}
\left(\frac{G_N}{N}\right)^v
\right]
=
O_{\mathbb{P}}
\left[
\left(\frac{G_N}{N}\right)^{\underline{v}}
\right],
\\
\left[
\frac{1}{G_N}
\sum_{g=1}^{G_N}
\left\| 
\sqrt{n_g}\left(\widehat{\theta}_g - \widetilde{\theta}_{IV}\right)
\right\|^8
\right]^{1/8}
& \leq
\left[
\frac{1}{G_N}
\sum_{g=1}^{G_N}
\left\| 
\sqrt{n_g}\left(\widehat{\theta}_g - \theta_g\right)
\right\|^8
\right]^{1/8}
+
\left\|\widetilde{\theta}_{IV}-\theta\right\|
=
O_{\mathbb{P}} (1).
\end{align*}

\subparagraph{A Central-limit Theorem.}
Consider first the mean, noting that:
\begin{align*}
	\sqrt{G_N} 
	\left|
	\overline{\widehat{\xi}} - \overline{\xi}
	\right|
	=
	O_{\mathbb{P}} 
	\left(
	\frac{G_{N}^{\frac{1}{2}+\underline{v}}}{N^{\underline{v}}}
	\right)
	=
	O_{\mathbb{P}} 
	\left[
	\left(
	\frac{G_{N}^{\frac{2\underline{v}+1}{2\underline{v}}}}{N}
	\right)^{\underline{v}}
	\right]
	=
	o_{\mathbb{P}}  (1) \text{ since }
	G_N =o \left(N^{1-\frac{1}{2\underline{v}+1}}\right)
	=
	o \left(N^{\frac{2\underline{v}}{2\underline{v}+1}}\right)
	.
\end{align*}
Note that
\begin{align*}
	\sqrt{G_N}\overline{\xi}
	=
	\frac{1}{\sqrt{G_N}}
	\sum_{g=1}^{G_N}
	\left(
	n_g \widehat{\theta}_g^{\prime} \Omega_{n_{g},g}^{-1} \widehat{\theta}_g
	-D
	\right)
	-
	\sqrt{G_N}
	\widetilde{\theta}_{IV}^{\prime}
	\left[\frac{1}{G_N} \sum_{g=1}^{G_N} \Omega_{n_g,g}^{-1} \right]
	\widetilde{\theta}_{IV},
\end{align*}
with, by \eqref{Hatthetaom} and because $\underline{b} >\frac{1}{2}$,
\begin{align*}
	\sqrt{G_N}
	\widetilde{\theta}_{IV}^{\prime}
	\left[\frac{1}{G_N} \sum_{g=1}^{G_N} \Omega_{n_g,g}^{-1} \right]
	\widetilde{\theta}_{IV}
	=
	\sqrt{G_N}
	O_{\mathbb{P}}
	\left[
	\frac{1}{N}
	+
	\frac{G_N}{N}
	+
	\frac{o(1)}{G_N}
	\right]
	=
	o_{\mathbb{P}} (1).
\end{align*}
Since
$\widehat{\theta}_g = \widehat{\varepsilon}_g + B_{n_g,g} \left(\theta\right)/n_g^{b}$ under homogeneity,
it also holds:
\begin{align*}
	\frac{1}{\sqrt{G_N}}
	\sum_{g=1}^{G_N}
	n_g \widehat{\theta}_g^{\prime} \Omega_{n_{g},g}^{-1} \widehat{\theta}_g
	& =
	\frac{1}{\sqrt{G_N}}
	\sum_{g=1}^{G_N}
	n_g \widehat{\varepsilon}_g^{\prime} \Omega_{n_{g},g}^{-1} \widehat{\varepsilon}_g
	+
	2
	\frac{1}{\sqrt{G_N}}
	\sum_{g=1}^{G_N}
	\sqrt{n_g} \widehat{\varepsilon}_g^{\prime} \Omega_{n_{g},g}^{-1} \frac{B_{n_g,g}}{n_{g}^{b-\frac{1}{2}}}
	\\
	&
	\qquad
	+
	\frac{1}{\sqrt{G_N}}
	\sum_{g=1}^{G_N}
	\frac{ B_{n_g,g}^{\prime} \Omega_{n_{g},g}^{-1} B_{n_g,g}}{n_g^{2b-1}}
	\\
	&
	=
	\frac{1}{\sqrt{G_N}}
	\sum_{g=1}^{G_N}
	n_g \widehat{\varepsilon}_g^{\prime} \Omega_{n_{g},g}^{-1} \widehat{\varepsilon}_g
	+
	\overbrace{O_{\mathbb{P}} \left[\left(\frac{G_N}{N}\right)^{b-\frac{1}{2}}\right]
	+
	O_{\mathbb{P}} \left[\frac{G_N^{2b-\frac{1}{2}}}{N^{2b-1}}\right]}^{=o_{\mathbb{P}}(1) \text{ since $b > \frac{1}{2}$ and $G_N=o\left(N^{1-\frac{1}{4b-1}}\right) $}}.
\end{align*}
Hence:
\begin{align*}
	\sqrt{G_N}\overline{\xi}
	=
	\frac{1}{\sqrt{G_N}}
	\sum_{g=1}^{G_N}
	\left(n_g \widehat{\varepsilon}_g^{\prime} \Omega_{n_{g},g}^{-1} \widehat{\varepsilon}_g-D\right)
	+
	o_{\mathbb{P}}
	(1).
\end{align*}
As $\min_{1 \leq g \leq G_N} \mathrm{Var} \left(n_g \widehat{\varepsilon}_g^{\prime }\Omega^{-1}_{n_g,g} (\theta_g) \widehat{\varepsilon}_g\right) \geq C(1+o(1))$, it also holds:
\begin{align*}
	\sum_{g=1}^{G_N}
	\mathbb{E}
	\left[
	\left|\frac{n_g \widehat{\varepsilon}_g^{\prime} \Omega_{n_{g},g}^{-1} \widehat{\varepsilon}_g-D}{\sqrt{G_N} s_N}\right|^{3}
	\right]
	=
	O\left(G_N^{-1/2}\right) = o(1),
	\quad
	s_N^2=
	\frac{1}{G_N} \sum_{g=1}^{G_N}  \mathrm{Var} \left(n_g \widehat{\varepsilon}_g^{\prime} \text{$\Omega_{n_{g},g}^{-1}$} \widehat{\varepsilon}_g-D\right).
\end{align*}
Hence Theorem 7.19 in \citet{pollard2002} yields that:
\begin{align*}
	\sqrt{G_N}
	\frac{\overline{\xi}}{s_N}
	\stackrel{d}{\longrightarrow}
	\mathcal{N}(0,1).
\end{align*}

\subparagraph{Variance Consistency.} From the orders of the items in \eqref{Prelims}, we get first:
\begin{align*}
	\widehat{s}
	& =
	\widetilde{s} + O_{\mathbb{P}} \left[ \left(\frac{G_N}{N}\right)^{\underline{v}}\right] \text{ with }
	\widetilde{s}^2
	= \frac{1}{G_N} \sum_{g=1}^{G_N}
	\xi_g^2
	-
	\overline{\xi}^2
	=
	\frac{1}{G_N} \sum_{g=1}^{G_N}
	\xi_g^2
	+
	O_{\mathbb{P}}
	\left(\frac{1}{G_N}\right).
\end{align*}
Let 
\begin{align*}
	\eta_g
	=
	n_g \widehat{\varepsilon}_g^{\prime} \Omega_{n_{g},g}^{-1} \widehat{\varepsilon}_g -D,
\end{align*}
so that, by Assumptions V and E:
\begin{align*}
	&\left|
	\left[
	\frac{1}{G_N}
	\sum_{g=1}^{G_N}
	\xi_g^2
	\right]^{\frac{1}{2}}
	-
	\left[
	\frac{1}{G_N}
	\sum_{g=1}^{G_N}
	\eta_g^2
	\right]^{\frac{1}{2}}
	\right|
	\leq
	\left[
	\frac{1}{G_N}
	\sum_{g=1}^{G_N}
	\left(\xi_g - \eta_g \right)^2
	\right]^{\frac{1}{2}}
	\\
	&
	\quad
	=
	\left[
	\frac{1}{G_N}
	\sum_{g=1}^{G_N}
	n_g^2
	\left[
	\left(
	\widehat{\varepsilon}_g + \frac{B_{n_g,g}}{n_g^b}-\widetilde{\theta}_{IV}
	\right)^{\prime}
	\Omega_{n_g,g}^{-1}
	\left(
	\widehat{\varepsilon}_g + \frac{B_{n_g,g}}{n_g^b}-\widetilde{\theta}_{IV}
	\right)
	-
	\widehat{\varepsilon}_g^{\prime} 
	\Omega_{n_g,g}^{-1}
	\widehat{\varepsilon}_g
	\right]^2
	\right]^{\frac{1}{2}}
	\\
	&
	\quad
	\leq
	C
	\left[
	\frac{1}{G_N}
	\sum_{g=1}^{G_N}
	n_g^2
	\left(
	\left\|\widehat{\varepsilon}_g\right\|^2
	\left(n_g^{-2b}+ \left\|\widetilde{\theta}_{IV}\right\|^2\right)
	+
	n_g^{-2b} \left\|\widetilde{\theta}_{IV}\right\|^2
	+
	n_g^{-4b}
	+
	 \left\|\widetilde{\theta}_{IV}\right\|^4
	\right)
	\right]^{\frac{1}{2}}
	\\
	&
	\quad
	=
	O_{\mathbb{P}}
	\left[
	\left(
	\frac{G_N}{N}
	\right)^{2b-1}
	+
	\frac{1}{G_N}
	+
	\left(
	\frac{G_N}{N}
	\right)^{2b-1}
	\left(
	\frac{1}{G_N}
	+
	\left(
	\frac{G_N}{N}
	\right)^{2b-1}
	\right)
	+
	\left(
	\frac{G_N}{N}
	\right)^{4b-2}
	+
	\frac{1}{G_N^2}
	\right]
	=
	o_{\mathbb{P}} (1).
\end{align*}
Now, Assumption E ensures that:
\begin{align*}
	\frac{1}{G_N}
	\sum_{g=1}^{G_N}
	\eta_g^2
	=
	s_N^2 + O_{\mathbb{P}} \left(\frac{1}{\sqrt{G_N}}\right),
	\text{ implying }
	\frac{s_N}{\widehat{s}} = 1 + o_{\mathbb{P}} (1).
\end{align*}

\subparagraph{Asymptotic normality of the test statistic.}
Since $
	\sqrt{G_N}
	\frac{\overline{\xi}}{s_N}
	\stackrel{d}{\longrightarrow}
	\mathcal{N}(0,1)
$, it follows that
\begin{align*}
	\sqrt{G_N}
	\frac{\overline{\xi}}{\widehat{s}}
	\stackrel{d}{\longrightarrow}
	\mathcal{N}(0,1),
\end{align*}
and the test has an asymptotic level equal to $\alpha$.

\paragraph{The heterogeneity alternative.}
We first derive the order of the items in \eqref{Prelims}. 
It holds, by \eqref{Hatthetaom},
\begin{align*}
	\left(\frac{G_N}{N}\right)^{\frac{1}{2}}
	\left\|
	\left(\frac{N}{G_N}\right)^{\frac{1}{2}}
	\left(\widehat{\theta}_{IV}- \widetilde{\theta}_{IV}\right)
	\right\|
	=
	o_{\mathbb{P}} (1).
\end{align*}
Assumptions E and He give, together with the Markov Inequality and since the subsamples are balanced:
\begin{align*}
	&
	\left(\frac{G_N}{N}\right)^{\frac{1}{2}}
	\left|
	\left[
	\frac{1}{G_N}
	\sum_{g=1}^{G_N}
	\left\| 
	\sqrt{n_g}
	\left(
	\widehat{\theta}_g -  \widetilde{\theta}_{IV}
	\right)
	\right\|^8
	\right]^{\frac{1}{8}}
	-
	\left[
	\frac{1}{G_N}
	\sum_{g=1}^{G_N}
	\left\| 
	\sqrt{n_g}
	\left(\theta_g-\theta_{IV}\right)
	\right\|^8
	\right]^{\frac{1}{8}}
	\right|
	\\
	&
	\quad
	\leq
	C
	\left[
	\frac{1}{G_N}
	\sum_{g=1}^{G_N}
	\left\| 
	\widehat{\varepsilon}_g + \frac{B_{n_g,g} (\theta_g)}{n_g^b}
	\right\|^8
	\right]^{\frac{1}{8}}
	+
	C
	\left\|
	\widetilde{\theta}_{IV}
	-
	\theta_{IV}
	\right\|
	=
	o_{\mathbb{P}} (1).
\end{align*}
Hence Assumption He and the Law of Large Numbers give
\begin{align*}
	& \left(\frac{G_N}{N}\right)^{\frac{1}{2}}
	\left[
	\frac{1}{G_N}
	\sum_{g=1}^{G_N}
	\left\| 
	\sqrt{n_g}
	\left(
	\widehat{\theta}_g -  \widetilde{\theta}_{IV}
	\right)
	\right\|^8
	\right]^{\frac{1}{8}}
	\asymp
	\mathbb{E}^{\frac{1}{8}}
	\left[
	\left\|\theta_g -\theta_{IV}\right\|^8
	\right]
	+
	o_{\mathbb{P}} (1)
	\\
	&
	\quad
	\geq
	\mathbb{E}^{\frac{1}{2}}
	\left[
	\left\|\theta_g -\theta_{IV}\right\|^2
	\right]
	+
	o_{\mathbb{P}} (1)
	\geq
	C \| \Sigma \|^{1/2}
	+
	o_{\mathbb{P}} (1).
\end{align*}

Hence \eqref{Prelims} gives
\begin{align*}
	\frac{G_N}{N}
	\left|
	\overline{\widehat{\xi}} - \overline{\xi}
	\right|
	=
	o_{\mathbb{P}} (1),
	\quad
	\frac{G_N}{N}
	\left|
	\widehat{s} - \widetilde{s}
	\right|
	=
	o_{\mathbb{P}} (1).
\end{align*}

\subparagraph{The sample mean $\overline{\widehat{\xi}}$.}
Assumption E and the Markov Inequality imply:
\begin{align*}
	&
	\frac{G_N}{N}
	\left|
	\overline{\xi} 
	- 
	\frac{1}{G_N}
	\sum_{g=1}^{G_N}
	n_g \left(\theta_g - \theta_{IV}\right)^{\prime}
	\Omega_{n_{g},g}^{-1}(\theta_g)  \left(\theta_g - \theta_{IV}\right)
	\right|
	\\
	&
	\quad
	\leq
	\frac{C}{G_N}
\sum_{g=1}^{G_N}
\left(
\left\|
\widehat{\varepsilon}_g
\right\|^2
+
\frac{\left\|B_{n_g,g}\right\|^2}{n_g^{2b}}
\right)
+
C
\frac{G_N}{N}
\left(
\left\|\widetilde{\theta}_{IV}- \theta_{IV}\right\|^2
+
D
\right)
=
o_{\mathbb{P}} (1),
\end{align*}
with, by Assumptions He and E,
\begin{align*}
	&
	\frac{G_N}{N}
		\frac{1}{G_N}
	\sum_{g=1}^{G_N}
	n_g \left(\theta_g - \theta_{IV}\right)^{\prime}
	\Omega_{n_{g},g}^{-1} \left(\theta_g - \theta_{IV}\right)
	\\
	& \quad
	=
	\frac{G_N}{N}
	\frac{1}{G_N}
	\sum_{g=1}^{G_N}
	\mathbb{E}\left[
	n_g \left(\theta_g - \theta_{IV}\right)^{\prime}
	\Omega_{n_{g},g}^{-1} \left(\theta_g - \theta_{IV}\right)\right]
	\\
	& \quad
	+
	\overbrace{O_{\mathbb{P}}
	\left[
	\left(
	\frac{1}{G_N^2}
	\sum_{g=1}^{G_N}
	\mathrm{Var}\left(
	\left(\theta_g - \theta_{IV}\right)^{\prime}
	\Omega_{n_{g},g}^{-1} \left(\theta_g - \theta_{IV}\right)
	\right)
	\right)^{\frac{1}{2}}
	\right]}^{=
	o_{\mathbb{P}}
	(1)},
	\\
	&
	\text{ where }
	\frac{G_N}{N}
	\frac{1}{G_N}
	\sum_{g=1}^{G_N}
	\mathbb{E}\left[
	n_g \left(\theta_g - \theta_{IV}\right)^{\prime}
	\Omega_{n_{g},g}^{-1} \left(\theta_g - \theta_{IV}\right)\right]
	\geq
	\frac{C}{G_N}
	\sum_{g=1}^{G_N}
	\mathbb{E}\left[
  \left\|
  \theta_g - \theta_{IV}\right\|^2\right]
	\geq C \| \Sigma \|
.
\end{align*}
Hence
\begin{align*}
	\frac{G_N}{N}\overline{\widehat{\xi}}
	\geq C \| \Sigma \| + o_{\mathbb{P}} (1).
\end{align*}

\subparagraph{The sample variance $\widehat{s}$.}
Arguing as above gives
\begin{align*}
	&
	\left(\frac{G_N}{N}\right)^2
	\widehat{s}^2
	=
	\left(\frac{G_N}{N}\right)^2
	\frac{1}{G_N}
	\sum_{g=1}^{G_N}
	\mathbb{E}\left[
	\left(n_g \left(\theta_g - \theta_{IV}\right)^{\prime}
	\Omega_{n_{g},g}^{-1} \left(\theta_g - \theta_{IV}\right)
	\right)^2\right]
	\\
	&
	\quad
	-
	\left(\frac{G_N}{N}\right)^2
	\left(
	\frac{1}{G_N}
	\sum_{g=1}^{G_N}
	\mathbb{E}\left[
	n_g \left(\theta_g - \theta_{IV}\right)^{\prime}
	\Omega_{n_{g},g}^{-1} \left(\theta_g - \theta_{IV}\right)\right]
	\right)^2
	+
	o_{\mathbb{P}} (1)
	\\
	&
	\quad 
	\leq 
	C \left(\frac{G_N}{N}\right)^2 \left(\frac{N}{G_N}\right)^2 \| \Sigma\|^2 
	+
	o_{\mathbb{P}} (1)
	\leq
	C \| \Sigma\|^2 
	+
	o_{\mathbb{P}} (1).
\end{align*}

\subparagraph{Consistency of the test.}
From the above, there is a $C>0$ such that:
\begin{align*}
	\frac{\overline{\widehat{\xi}}}{\widehat{s}}
	\geq
	\frac{C \| \Sigma\|}{\sqrt{ \| \Sigma\|^2+o_{\mathbb{P}} (1)}}
	\text{, implying that, for any given $\Sigma \neq 0$, }
	\sqrt{G_N} \frac{\overline{\widehat{\xi}}}{\widehat{s}}
	\stackrel{\mathbb{P}}{\longrightarrow}
	+
	\infty.
\end{align*}
This establishes the consistency of the test and ends the proof of the theorem. \hfill $\Box$

\end{document}